\documentclass[]{aa}
\usepackage{graphicx,txfonts,amssymb,natbib,epsfig}
\authorrunning{M. Meneghetti et al.}
\titlerunning{Strong lensing by clusters in the {\sc MareNostrum Universe}}

\begin{document}

\title{Strong lensing in the {\sc MareNostrum Universe}: biases in the cluster lens population}

 \author{M. Meneghetti\inst{1,3}\thanks{E-mail: massimo.meneghetti@oabo.inaf.it}, C. Fedeli\inst{2,3},  F. Pace\inst{4}, S. Gottl\"ober\inst{5}, G. Yepes\inst{6}}

\institute {
  $^1$ INAF-Osservatorio Astronomico di Bologna, Via Ranzani 1, I-40127 Bologna, Italy\\
  $^2$ Dipartimento di Astronomia, Universit\`a di Bologna,
  Via Ranzani 1, I-40127 Bologna, Italy\\
  $^3$ INFN, Sezione di Bologna, Viale Berti Pichat 6/2, I-40127 Bologna, Italy\\
  $^4$ ITA, Zentrum f\"ur Astronomie, Universit\"at
  Heidelberg, Albert \"Uberle Str. 2, D-69120 Heidelberg, Germany \\
  $^5$ Astrophysikalisches Institut Potsdam, An der Sternwarte 16, D-14482 Potsdam, Germany \\
  $^6$ Grupo de Astrof\'isica, Universidad Aut\'onoma de Madrid, Madrid E-28049, Spain}

\date{\emph{Astronomy \& Astrophysics, submitted}}

\abstract{Strong lensing is one of the most direct probes of the mass distribution in the inner regions of galaxy clusters. It can be used to constrain the density profiles and to measure the mass of the lenses. Moreover, the abundance of strong lensing events can be used to constrain the structure formation and the cosmological parameters through the so-called "arc-statistics" approach. However, several issues related to the usage of strong lensing clusters in cosmological applications are still controversial, leading to the suspect that several biases may affect this very peculiar class of objects.}{With this study we aim at better understanding the properties of galaxy clusters which can potentially act as strong lenses.}{We do so by investigating the properties of a large sample of galaxy clusters extracted from the N-body/hydrodynamical simulation {\sc MareNostrum Universe}. We perform ray-tracing simulations with each of them identifying those objects which are capable to produce strong lensing effects. We explore the correlation between the cross section for lensing and many properties of clusters, like the mass, the three-dimensional and projected shapes, their concentrations, the X-ray luminosity and the dynamical activity.}{We quantify the minimal cluster mass required for producing both multiple images and large distortions. While we do not measure a significant excess of triaxiality in strong lensing clusters, we find that the probability of strong alignments between the major axes of the lenses and the line of sight is a growing function of the lensing cross section. In projection, the  strong lenses appear rounder within $R_{200}$, but we find that their cores tend to be more elliptical as the lensing cross section increases. As a result of the orientation bias, we also find that the cluster concentrations estimated from the projected density profiles tend to be biased high. The X-ray luminosity of strong lensing clusters tend to be higher than that of normal lenses of similar mass and redshift. This is particular significant for the least massive lenses. Finally, we find that the strongest lenses generally exhibit an excess of kinetic energy within the virial radius, thus indicating that they are more dynamically active than usual clusters.}
{We conclude that strong lensing clusters are a very peculiar class of objects, affected by many selection biases which need to be properly modeled when using them to study the inner structure of galaxy clusters or to constrain the cosmological parameters.}


\maketitle

\section{Introduction}

Gravitational lensing is one of the most powerful tools for studying the formation of cosmic structures in the universe. The light from distant sources, traveling in space and time,  is deflected by the matter along its path before being collected by the observers.  Thus, we measure an integrated effect which contains a wealth of information about the cosmic structures at different epochs. 

Depending on the impact parameter on the intervening matter and on the mass of the deflectors encountered by the light along its path, gravitational lensing manifests itself in the weak and in the strong regimes. In the weak lensing regime the shapes of distant galaxies, which happen to be at large angular distances  from the largest mass concentrations on the sky, are slightly changed, such that this effect can be revealed only though statistical measurements. Nevertheless, these tiny distortions can be used for tracing the large scale structure of the universe both in two and in three dimensions \citep[see e.g.][for some  recent results]{2008A&A...479....9F,2007MNRAS.381..702B,2007ApJS..172..239M}, from which important cosmological constraints can be derived \citep{BA01.1}. This is a field of research which will have extraordinary improvements in the next decades, thanks to some upcoming missions 
\citep{2006AAS...209.8610W,2007AAS...210.5102K,2006AAS...209.9809J,2008arXiv0807.4036R}. 
Weak lensing allows to reconstruct the mass distribution up to the outskirts of galaxy clusters \citep[see e.g.][]{2006ApJ...653..954D,2006A&A...451..395C,2007MNRAS.379..317H}. 

Strong lensing is an highly non-linear and relatively rare effect which is observable in the central regions of galaxies and clusters. In this regime, the background sources can be multiply imaged and/or highly distorted to form very elongated images, the so called "gravitational arcs". They are powerful cosmological probes for many reasons. First, such events can be used to investigate the inner regions of the lenses. Thus, they can be used to test the predictions of the Cold-Dark-Matter paradigm on the inner structure of dark matter halos  \citep{ME01.1,2004ApJ...604L...5M,2004ApJ...604...88S,BA04.1,2007MNRAS.381..171M,2008A&A...489...23L}. Second, they can be used to recover the mass distribution in the centre of the lenses, providing complementary informations to those obtained from weak lensing \citep{2005A&A...437...49B,2005MNRAS.362.1247D,2006A&A...458..349C,2007ApJ...668..643L,2009A&A...500..681M}. Lensing masses can then be used for measuring the cluster mass function. Third, the position and the distortions of the strongly lensed images as a function of the source redshift reflect the geometry of the universe \citep{2004A&A...417L..33S,2005MNRAS.362.1301M}. Finally, the probability of observing  strong lensing events is deeply connected to the abundance, the mass and the formation epoch (through the concentration) of the lenses. This makes statistical lensing a potentially powerful tool to study the structure formation \citep{BA98.2,LI05.1,ME05.1}.
 
In this paper, we focus on the properties of the most massive and therefore most efficient strong lenses in the universe: the galaxy clusters. In the framework of the hierarchical scenario of structure formation, these are the youngest bound systems in the sky.  About $85\%$ of their mass is believed to be in the form of cold-dark-matter \citep{2007ApJ...664..117G}. The remaining $15\%$ is made of a diffuse gas component, the Intra-Cluster-Medium, and of other baryons in the form of stars, the vast majority of which is inside the cluster galaxies. Being relatively young structures, the interaction between the baryons and the dark matter is less strong than in older systems like galaxies. For this reason, clusters are important laboratories for studying the properties of the dark matter \citep{2004ApJ...606..819M}. 
However, there are several issues that we need to take into account when studying these systems. In particular, clusters where gravitational arcs are observed are a limited fraction of the total number \citep{2005ApJ...627...32S,LU99.1,GL03.1,ZA03.1,2008AJ....135..664H} and therefore a particular class of objects. Broadly speaking they are the most massive clusters, but there are several other properties that boost the cluster ability to produce strong lensing events. For example, we know that substructures, asymmetries and projected ellipticity of the lenses are all contributing to the strong lensing cross section of clusters \citep{ME03.1,ME07.1}. Both observations and simulations agree that strong lenses have high concentrations \citep{HE07.1,2007A&A...473..715F,2003ApJ...598..804K,2003A&A...403...11G,2008ApJ...685L...9B}. Triaxiality is also relevant, because clusters seen along their major axis are more efficient lenses \citep{OG03.1}. Although cluster galaxies statistically do not change the distributions of the arc properties \citep{ME00.1,FL00.1,2008MNRAS.386.1845H}, cD galaxies sitting at the bottom of the cluster potential well increase the ability for producing long and thin arcs by $\sim 30-50\%$ \citep{ME03.2}. The gas physics, in particular cooling, could also affect the strong lensing properties of clusters \citep{PU05.1,2008ApJ...676..753W,2010arXiv1001.2281M}. Finally, \cite{TO04.1} showed that the cluster ability to produce gravitational arcs can also be enhanced by the dynamical activity in the lens. By studying with high time resolution how the lensing cross section changes during an edge-on collision between the main cluster clump and a substructure, these authors found that the strong lensing efficiency is boosted by a factor of 10 during the merging phase. Later, \cite{FE06.1}, using semi-analytic methods, showed that the arc optical depth produced by clusters with moderated
and high redshifts is more than doubled by mergers.      
   
Although an extensive work has been done in the past decade, a better characterization of the strong lens cluster population is mandatory. Given the complexity of clusters and the great importance that many of their properties have for strong lensing, the only reliable way to do that is through the ray-tracing analysis of a large number of simulated clusters. A first important work in this framework was done by \cite{HE07.1}, who analyzed a sample of 878 clusters from an N-Body pure dark-matter cosmological simulation. Important properties like concentrations, axis ratios and substructures were discussed. In this work, we aim at extending the analysis of \cite{HE07.1} in three ways. First, we include a much larger number of clusters (now 49366 systems), taken from a larger cosmological volume ($500\,h^{-1}$Mpc vs $320\,h^{-1}$Mpc). Second, the clusters used here are obtained from an N-body-hydrodynamical simulation where the evolution of the gas component is also considered. Thus, we can correlate the lensing properties of clusters with some important X-ray observables. Third, we study in detail the possible correlations between cluster dynamics and strong lensing efficiency, which was made only through analytical models so far.

The plan of the paper is as follows. In Sect. 2 we summarize the main characteristics of the cosmological simulation {\sc MareNostrum Universe}. In Sect. 3 we discuss the simulation methods and we define several lensing quantities useful for the following analysis. In Sect. 4 we discuss the correlation between lens masses and strong lensing ability. Sect. 5 is dedicated to the statistical analysis of the shapes and orientations of strong lensing clusters. We discuss the biases in the concentrations in Sect. 6. In Sect. 7 we focus on the X-ray properties of strong lensing clusters. Finally, in Sect. 8 we correlate the strong lensing efficiency with the dynamical state of the lenses. We summarize the main results and the conclusions of this study in Sect. 9. 

\section{The {\sc MareNostrum Universe} simulation}

The {\sc MareNostrum Universe}  \citep{2007ApJ...664..117G} is a large scale cosmological non-radiative SPH
simulation performed with the {\sc Gadget2} code \citep{SP05.1}. We briefly summarize the relevant characteristics here and we refer the reader to the paper by \cite{2007ApJ...664..117G} for a more detailed description of the simulation. This was run in 2005, during the testing period of the MareNostrum supercomputer using the WMAP1 normalisation, namely $\Omega_{\mathrm{m},0} = 0.3$, $\Omega_{\Lambda,0} = 0.7$ and $\sigma_8 = 0.9$ with a scale invariant primordial power spectrum.. After the
release of the 3-year WMAP data the simulation has been repeated {  at lower resolution} with
the predicted low normalization \citep{wmap3} and with a higher
normalization of $\sigma_8 = 0.8$ which is better in agreement with
the 5-year WMAP data \citep{wmap5}.  Comparing predictions from these
numerical simulations with recent observational estimates of the
cluster X-ray temperature functions \cite{YSGS07} argue that the low
normalization cosmological model inferred from the 3 year WMAP data
results is barely compatible with the present epoch X-ray cluster
abundances. {Therefore, the original WMAP1 normalized version of the simulation is used in this paper.}The simulation consists of a comoving box size of $500\;h^{-1}$ Mpc containing $1024^3$ dark matter particles and $1024^3$ gas particles. The mass of each dark matter particle equals $8.24 \times 10^{9} M_\odot h^{-1}$, and that of each gas particle, for which only adiabatic physics is implemented, is $1.45 \times 10^{9} M_\odot h^{-1}$.   The baryon density parameter is set to $\Omega_{\mathrm{b},0} = 0.045$. The spatial force resolution is set to an equivalent Plummer
gravitational softening of $15$~h$^{-1}$~kpc, and the SPH smoothing
length was set to the 40th neighbour to each particle.

In order to find all structures and substructures within the
distribution of 2 billion particles and to determine their properties
we use a hierarchical friends-of-friends (FOF) algorithm
\citep{Klypin99}. With a basic linking length set to 0.17 times the mean
interparticle distance, we extract the FOF objects at all redshifts.
The final catalog of identified objects contains more than 2 million
objects with more than 20 DM particles at $z=0$. { The same objects and their progenitors are contained in the catalogs corresponding to higher redshift outputs of the simulation. In this sense a correlation  exists between different redshift slices.}
In a second step we divide the linking length by $2^n$
($n=1,3$) to find substructures of the clusters. In particular we
use  $n=2$ to identify the highest density peak
which we associate with the center of the cluster.

All the FOF groups with mass larger than $10^{13}\,h^{-1} M_\odot$ are then stored into sub-boxes of cubic shape with side length $5 \,h^{-1}$Mpc for the subsequent lensing analysis. 

\section{Lensing properties}

\subsection{Ray-tracing}\label{sct:ray}
In this section, we illustrate the techniques used to derive the strong lensing properties of the clusters in the {\sc MareNostrum Universe} cosmological volume.
The deflection angle maps are calculated as explained in several previous papers \citep[see e.g.][]{ME00.1,ME05.1}. The particles in each cube are used to
produce a three-dimensional density field, by interpolating their
position on a grid of $512^3$ cells using the {\em Triangular Shaped
Cloud} method \citep{HO88.1}. Then, we project the three-dimensional
density field along the coordinate axes, obtaining three surface
density maps $\Sigma_{i,j}$, used as lens planes in the following
lensing simulations. 

The following step consists of tracing bundles of light rays through a regular grid covering the central part of each lens plane. We choose to set the size of this region as $1.5\times 1.5\,h^{-2}$ Mpc$^2$ comoving. This choice is driven by the necessity to
study in detail the central region of the clusters, where critical
curves form. However, we do this by taking into account the contribution from the surrounding mass distribution to the deflection angle of each ray.

We first define a grid of $256\times256$
``test'' rays, for each of which the deflection angle is calculated by
directly summing the contributions from all cells on the surface
density map $\Sigma_{i,j}$,
\begin{equation}
  \vec \alpha_{h,k}=\frac{4G}{c^2}\sum_{i,j} \Sigma_{i,j} A
  \frac{\vec x_{h,k}-\vec x_{i,j}}{|\vec x_{h,k}-\vec x_{i,j}|^2}\;,
\end{equation}  
where $A$ is the area of one pixel on the surface density map and
$\vec x_{h,k}$ and $\vec x_{i,j}$ are the positions on the lens plane
of the ``test'' ray ($h,k$) and of the surface density element
($i,j$). Following \cite{WA98.2}, we avoid the divergence when the
distance between a light ray and the density grid-point is zero by
shifting the ``test'' ray grid by half-cells in both directions with
respect to the grid on which the surface density is given.
We then define an higher resolution grid of rays covering the same region. We determine the deflection angle of each new ray  by bi-cubic interpolation between the four nearest test rays.
The grid size is chosen such that the resolution of the deflection angle map is fixed at $0.2''$. This results into $\sim 2400 \times 2400$ rays traced through the central region of deflectors at $z_{\rm l} = 0.3$. The number of grid point obviously increases at lower redshift due to the combined increase of the physical size of the grid and to the decrement of the angular diameter distance. Conversely, the number of pixels decreases at higher redshifts. We selected an upper bound of $3500 \times 3500$ grid points because the time consumption and the needed memory for the code become too demanding, and a lower bound of $650 \times 650$ grid points in order to be able to fairly capture the relevant  structures in the deflector.

\subsection{Strong lensing clusters}
For a fixed source redshift, a cluster can produce strong lensing events if it develops critical lines on the lens plane. These lines correspond to the caustics on the source plane. Only sources within the caustics have multiple images. Only sources which happen to lay close to the caustics are strongly distorted and magnified.

A cluster can produce both tangential and radial critical lines, where the tangential and the radial magnifications diverge, respectively. The critical lines form where 
\begin{eqnarray}
	\kappa(\vec x)\pm \gamma(\vec x)=1 \;,
\end{eqnarray}
where $\kappa(\vec x)$ and $\vec\gamma(\vec x)=[\gamma_1,\gamma_2]$ are the convergence and the shear at the the position $\vec x$ on the lens plane. Both the convergence and the shear are linear combinations of the spatial derivatives of the deflection angle components, $\vec{\alpha}=[\alpha_1,\alpha_2]$:
\begin{eqnarray}
	\kappa& = &\frac{1}{2}\left(\frac{\partial \alpha_1}{\partial x_1}+\frac{\partial\alpha_2}{\partial x_2}\right) \;, \\
	\gamma_1& = &\frac{1}{2}\left(\frac{\partial \alpha_1}{\partial x_1}-\frac{\partial\alpha_2}{\partial x_2}\right) \;,	\\
	\gamma_2& = &\frac{\partial \alpha_1}{\partial x_2}=\frac{\partial \alpha_2}{\partial x_1} \;.
\end{eqnarray}
It can be shown that the convergence is the surface density divided by a critical surface density:
\begin{eqnarray}
	\kappa=\frac{\Sigma}{\Sigma_{\rm crit}} \;,
\end{eqnarray} 
which depends on the angular diameter distances between observer and lens, $D_{\rm l}$, between the lens and the source, $D_{\rm ls}$, and between the observer and the source, $D_{\rm s}$:
\begin{equation}
	\Sigma_{\rm crit}=\frac{c^2}{4\pi G}\frac{D_{\rm s}}{D_{ls}D_{s}} \;.
\end{equation}
It is clear that, in order to be a strong lens, the cluster convergence and shear must be large enough, such that their sum is larger than unity somewhere.

  In order to focus on the subsample of strong lenses, we start by selecting those halos that are capable of developing critical lines. For doing this, we use the  $256 \times 256$ test rays first. We numerically calculate the spatial derivatives of the deflection angles to compute $\kappa$ and $\gamma$ and to determine the positions of the critical points. If at least a critical point is found in the three cluster projections using these maps, we consider the cluster for further, more detailed analysis. We are aware that, using this selection criterium, all lenses whose critical lines have sizes smaller than $\sim 20 h^{-1}$ kpc comoving, corresponding to the spatial resolution of the coarse grid, are not included in our analysis. Because of this limitations our results should be used with caution in reference to small strong lensing systems. Instead, such objects would never produce significant image splittings and large distortions or giant arcs, which are the most relevant strong lensing features in this work. 

As a result of this preliminar selection, it turns out that $49,366$ clusters produce critical curves for sources at redshift $z_\mathrm{s} = 2$ in at least one of their projections. For these objects, we repeat the calculation of the deflection angles on grids with higher spatial resolution on all the three projections, obtaining $148,098$ deflection angle maps. 

\subsection{Cross-sections for giant arcs}
\label{sect:crsec}

Once the high-resolution deflection angle maps for the three projections of each numerical cluster 
are computed, the strong lensing efficiency for long and thin arcs was evaluated by using the fast, semi-analytyc algorithm presented in \cite{FE06.1}. The reader is referred to the quoted paper for details, while here we give just a quick overview of the method. The lensing efficiency is quantified by the lensing cross section for highly distorted arcs $\sigma_{d_0}$. This is defined as the area of the region surrounding the caustics within which a source is mapped on the lens plane as an image with a minimal length-to-width ratio $d_0$. The size of the lensing cross section is related to the expected number of arcs with a minimal distortion observed behind the cluster.  In fact, the number of arcs above a minimal surface brightness $S_{0}$ expected from a cluster with cross section $\sigma$ for sources at redshift $z_{\rm s}$ is 
\begin{equation}
	N_{\rm arcs}(S_0)=\int_{z_{\rm l}}^{\infty} \int_{S_0}^{\infty} \sigma(z_{\rm s}) n_{\rm s}(S,z_{\rm s}) dS d z_{\rm s}
	\label{eq:narcs}
\end{equation}
where $n_{s}(S,z_{\rm s})$ is the number density of sources with surface-brightness $S$ and redshift $z_{\rm s}$. Note that lensing does not change the source surface brightness, thus $N_{arcs}$ does not depend on the magnification, {  at least if the PSF size is smaller that the size of arcs. This generally applies for extended arcs like those considered here.} 

When sources are much smaller than the characteristic length over which the lensing properties of the deflector change significantly, they can be considered as pointlike. In this case the lens mapping can be linearised and the length-to-width ratio of the distorted images, $d$, is simply given by the ratio of the eigenvalues of the Jacobian matrix at image position. Hence, in this case the cross section for arcs with length-to-width ratio larger than some threshold $d_0$ is by definition the area of the lens plane where the eigenvalue ratio is larger than $d_0$, mapped back to the source plane. This framework can be also easily modified to account for the extended size of real sources, by convolving the lensing properties over the typical source domain, assumed here to be circular with angular radius of $0.5''$.   Several studies have shown that the properties of the sources  are relevant for determining the shape of gravitational arcs \citep[see e.g.][]{2008A&A...482..403M,2009arXiv0910.4013G}. Our method cannot take into account all the effects due to source morphologies and luminosity profiles, however the intrinsic ellipticity of real sources is accounted for according to the elegant algorithm proposed by \cite{KE01.2}.

Because of the huge number of cross sections computed in this work, we focused only on a single value for the length-to-width threshold, namely $d_0 = 7.5$. While a distribution of thresholds would be preferable, we expect the change in $d_0$ to produce only a shift in the normalisation of cross sections, by leaving every qualitative conclusion unchanged. As mentioned above, we consider one  source redshift, $z_\mathrm{s} = 2$ only. 

\begin{figure}[lt!]
\begin{center}
  \includegraphics[width=\hsize]{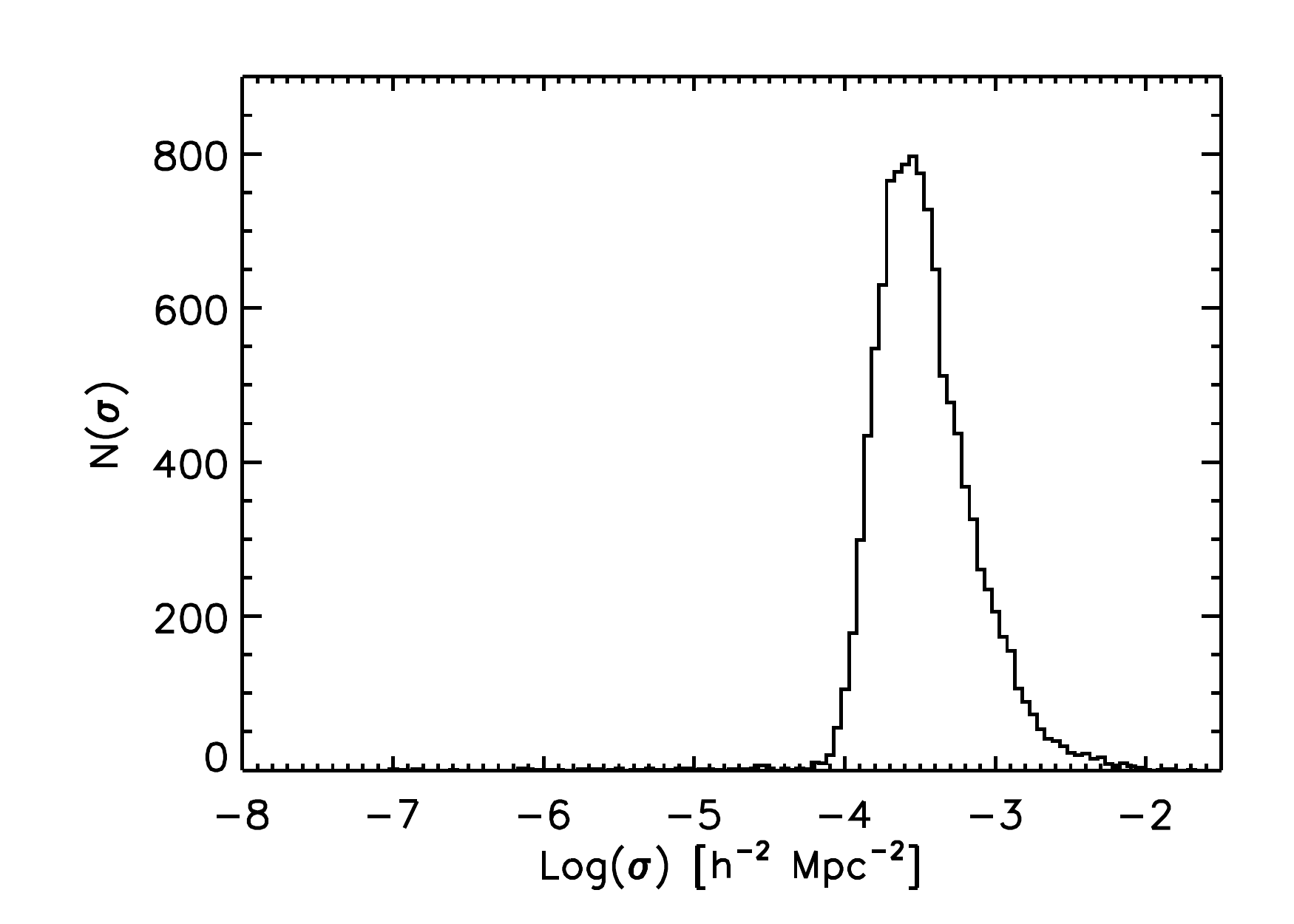}
\end{center}
\caption{The distribution of the lensing cross sections for giant arcs of all the strong lensing clusters between $z_{\rm l}=0$ and $z_{\rm l}=2$}
\label{fig:csdist}
\end{figure}

We computed the strong lensing cross sections for each of the $49,366$ high resolution deflection angle maps, produced as described in section \ref{sct:ray}. A large number of these however is vanishing, due to the fact that, even though the deflector is able to produce critical curves, these are small compared to the typical source size to efficiently distort images. It turns out that only $6375$ clusters have at least one projection with non vanishing cross section for $z_\mathrm{s} = 2$. In total, the cluster projections capable of large distortions are $11,347$. The lensing cross sections range from a minimal value of $9.5\times 10^{-8}\;h^{-2}$Mpc$^2$ to a maximal value of $1.9\times 10^{-2}\;h^{-2}$Mpc$^2$ (the properties of this super-lens are shown in Appendix \ref{sect:app1}). However, the vast majority of the lenses capable of producing giant arcs have cross sections larger than $\sim10^{-4}\;h^{-2}$Mpc$^2$, as it can be seen in Fig.~\ref{fig:csdist}, where we show the distribution of the lensing cross sections for giant arcs among the clusters analyzed here.  

\begin{figure*}[lt!]
\begin{center}
  \includegraphics[width=0.49\hsize]{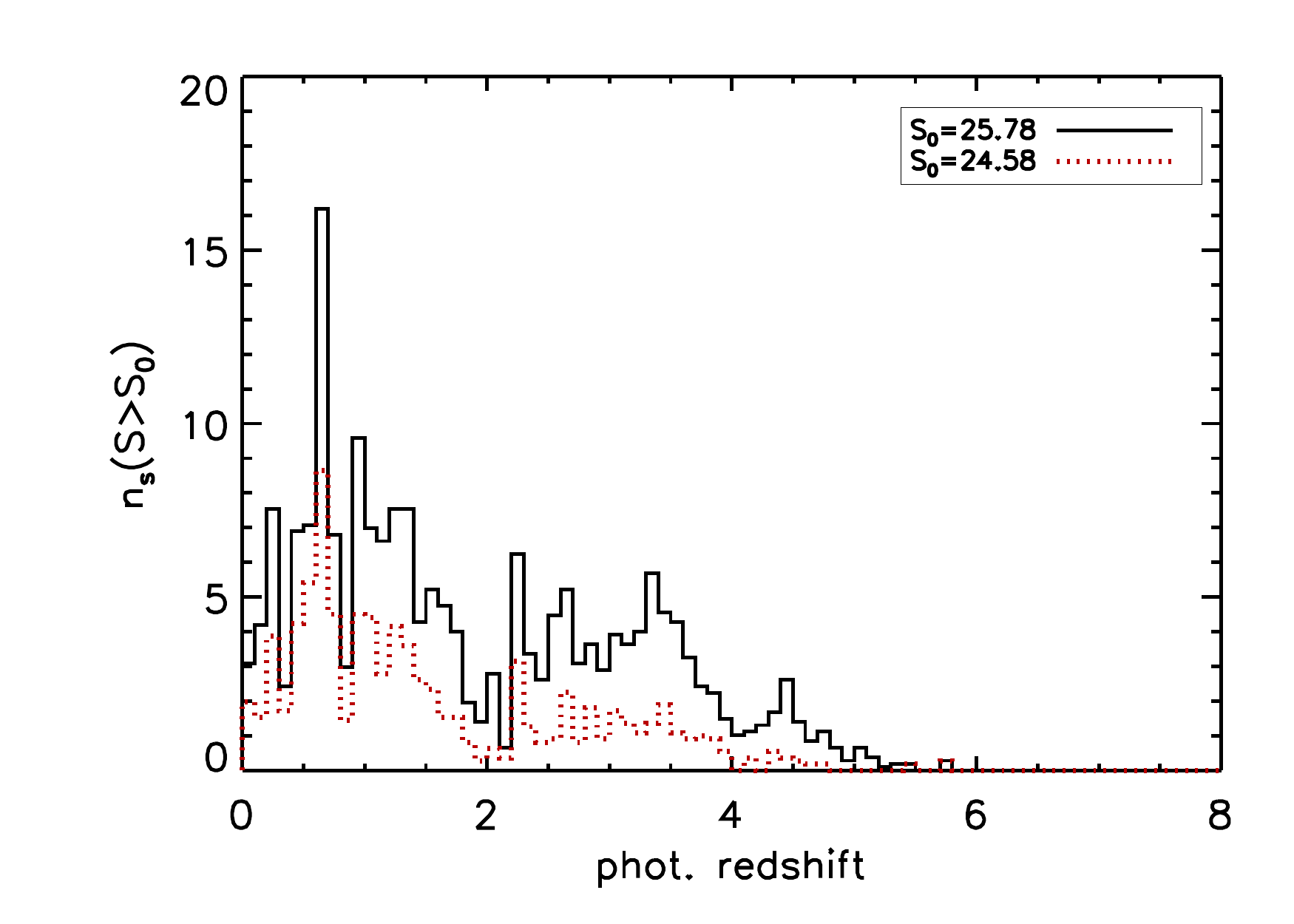}
   \includegraphics[width=0.49\hsize]{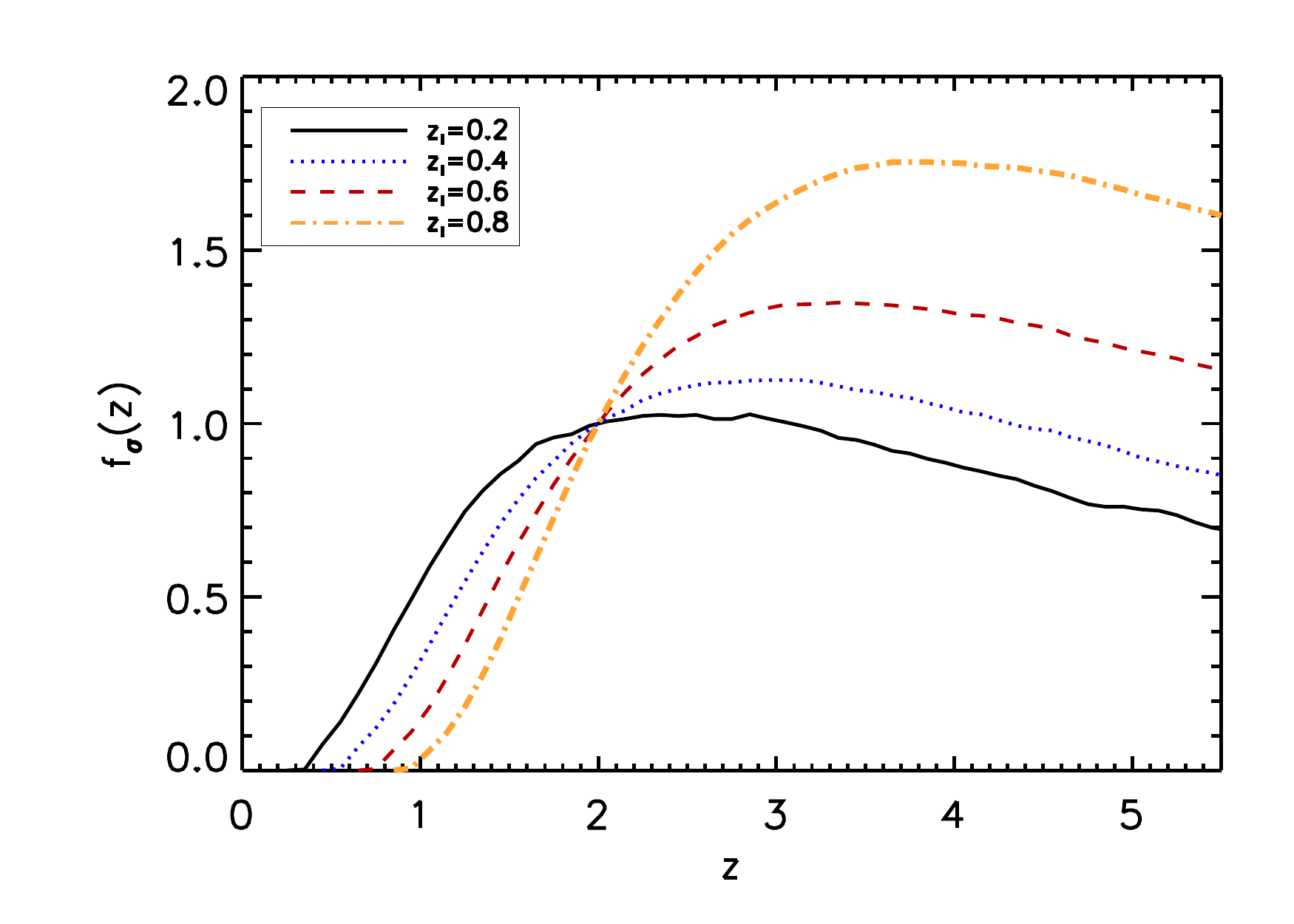}
\end{center}
\caption{Left panel: Galaxy number counts per square arcmin in redshift bins as obtained by simulating a deep observation of 8000s with HST/ACS in the F775W filter. The two histograms refer to two different SExtractor detection thresholds, namely $S_0=25.78$ and $S_0=24.58$ mag arcsec$^{-2}$, which correspond to 1$\sigma$ and 3$\sigma$ above the background r.m.s. See text for more details. Right panel: scaling functions for cross sections of lenses with mass $M=10^{15}\;h^{-1}M_{\odot}$ at four different redshifts: 0.2 (solid line), 0.4 (dotted line), 0.6 (dashed line), and 0.8 (dot-dashed line).}
\label{fig:udfzdist}
\end{figure*}

Given the large number of clusters analyzed here and the computational time required to analyze them it was not possible to calculate the lensing cross sections for several source redshifts. Thus it is not easy to convert the lensing cross section into a number of arcs using the Eq.~\ref{eq:narcs}. Nevertheless, we can estimate such number using some approximation. If we assume that the lensing cross section evolves with redshift as $\sigma(z_{\rm s})= \sigma \times f_{\sigma}(z_{\rm s})$, where $\sigma$ is the lensing cross section for sources at $z_{s}=2$ and $f_\sigma=\sigma(z_{\rm s})/\sigma$ is a scaling function, then the number of arcs detectable behind a cluster can be expressed as follows:
\begin{eqnarray}
	N_{\rm arcs}(S_0) & = & \sigma \times \int_{z_{\rm l}}^{\infty} \int_{S_0}^{\infty} f_\sigma(z_{\rm s}) n_{\rm s}(S,z_{\rm s}) dS d z_{\rm s} \nonumber \\
	& = & \sigma \times n_{\rm eff}(S_0) \;.
		\label{eq:narcse}
\end{eqnarray}  
In the last equation, we have introduced the {\it effective} source number density. Apart from the dependency on the scaling of the lensing cross section with the source redshift, which will be discussed below, the effective source number density is set by the minimal surface brightness (i.e. flux per square arcsec) of detectable arcs. Thus, it is  determined by the characteristics of the observation, i.e. by the throughput of the instrument and by the level of the background. Using the optical simulator {\tt SkyLens} \citep[][Meneghetti et al. 2009, in prep.]{2008A&A...482..403M}, we have simulated a deep exposure of 8000s with the Advanced Camera for Surveys on board the Hubble Space Telescope in the F775W filter. This code uses the morphologies, the luminosities, and the redshifts of the galaxies in the Hubble Ultra-Deep-Field \citep{2006AJ....132.1729B} to produce extremely realistic images of the sky including several observational noises. Setting the background level to 22.4 mag arcsec$^{-2}$, the 1$\sigma$ and the 3$\sigma$ detection thresholds above the background r.m.s. are $S_0 \sim 25.78$ mag arcsec$^{-2}$ and $S_0 \sim 24.58$ mag arcsec$^{-2}$, respectively. The galaxy number counts per square arcmin above the detection thresholds are shown in the left panel of Fig.~\ref{fig:udfzdist} as a function of redshift. Here, we use the photometric redshifts of the  HUDF galaxies from the public catalog by \cite{2006AJ....132..926C}.   
 
The scaling function $f_\sigma$ is expected to depend on several properties of the lenses, like their redshifts, density profiles, ellipticity, and substructures. Therefore, adopting a universal scaling law is certainly a gross approximation. On the other hand, it is useful also to have a rough estimate of the effective number density of sources in order to link a quantity like the cross section, which is not directly measurable, to something which can be observed, like the number of arcs behind a cluster. To estimate the scaling function $f_\sigma$ for a cluster at reshift $z_{\rm l}$, we use a toy lens model with an NFW density profile and fixed projected ellipticity $\epsilon=0.2$. The ellipticity is introduced in the lensing potential as discussed in \cite{ME03.1}. Using the same algorithm used to analyze the deflection angle maps of numerically simulated clusters, we measure how the lensing cross section grows as a function of the source redshift. In the right panel of Fig.~\ref{fig:udfzdist}, we show the scaling functions for a cluster with mass $10^{15}\;h^{-1}M_\odot$ at several redshifts between $z=0.2$ and $z=0.8$. 

\begin{figure}[lt!]
\begin{center}
  \includegraphics[width=\hsize]{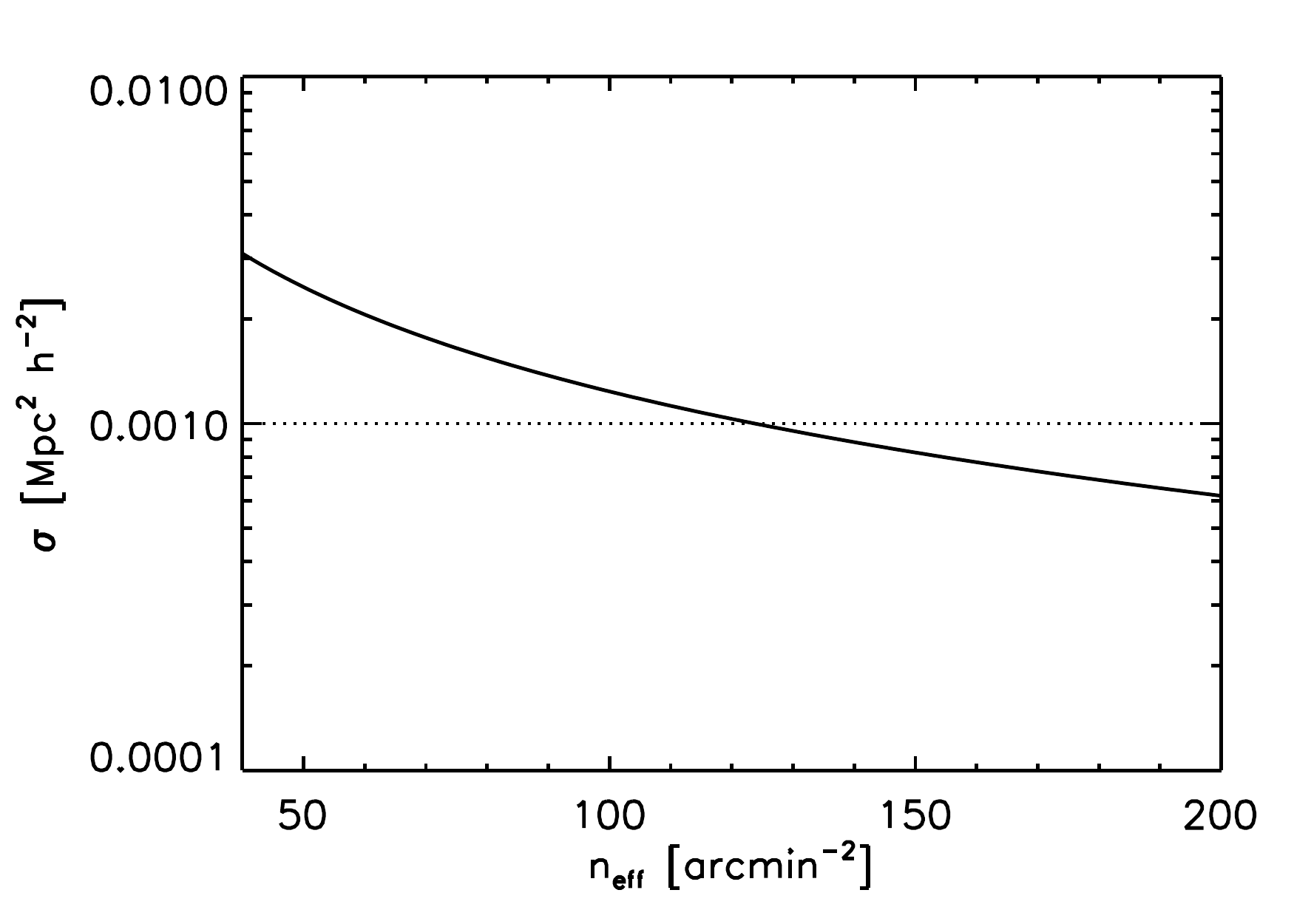}
\end{center}
\caption{The lensing cross section required for an expectation value of one giant arc behind a cluster as a function of the effective number of background sources.}
\label{fig:sigd}
\end{figure}

The effective number counts derived  as explained above are reported in Tab.~\ref{tab:neff} for different cluster masses and redshifts. First, we note that the dependence on the mass is weak, which allows to extend the validity of these calculations to a broad range of masses. Second, the rise of the scaling function for increasing lens redshift compensate for the lower number of galaxies at high redshift. Thus the effective source number counts do not drop, but tend to increase as the lens redshift increases. Using the Eq.~\ref{eq:narcse}, we can finally link the number of arcs expected for a given lensing cross section $\sigma$ to the effective number density of background sources, i.e. to the depth of the observation. For example, for a cluster with cross section $10^{-3}\;h^{-2}$ Mpc$^{2}$ the expected number of giant arcs varies from 0.3 to 1.6 for $n_{eff}$ in the range $[40-200]$. Conversely, in Fig.~\ref{fig:sigd} we show the lensing cross section required for $N_{\rm arcs}=1$ as a function of the effective number density of background sources. Even for very high effective number counts (or equivalently very deep exposures), the lensing cross section needs to be very large in order to expect  at least one arc behind a galaxy cluster. For example, for a cross section of $\sigma=10^{-3}h^{-2}$ Mpc$^2$ the effective number density of background sources needs to be $\sim 130$ galaxies per square arcmin. Such number density needs to be doubled in order to expect to observe two arcs, and so forth. 

{  In the rest of the paper, we will use the lensing cross section to discriminate between lenses of different strengths. Observationally, the lensing cross section is not a directly measurable  quantity. A possible method to estimate the lensing cross section is through the detailed parametric reconstruction of the lens potential (Meneghetti et al. in prep.). Indeed,  the deflection field can be readily derived from the lensing potential and used to measure the lensing cross section using the same method adopted in our simulations. Other methods to define the strength of the lenses may be based on quantities which are more directly measurable, like the angular separations of multiple images, which can be used to estimate the size of the critical lines, etc. However, in this paper, given the huge size of the cluster sample considered, we could not explore this other possibility. }

\begin{table}[htdp]
\caption{Effective galaxy number counts per square arcmin behind clusters with different masses and redshifts. The counts refer to an observation with HST/ACS in the F775W filter with an exposure time of $8000$s. In each column, the biggest and the smallest number correspond to detections at 1$\sigma$ and at 3$\sigma$ above the background r.m.s.}
\begin{center}
\begin{tabular}{|l|c|c|c|c|}
\hline
\hline
 & $z_{\rm l}=0.2$ & $z_{\rm l}=0.4$ & $z_{\rm l}=0.6$ & $z_{\rm l}=0.8$ \\
\hline
\hline
$M=1\times 10^{15}\;h^{-1}M_\odot$ & 154/61 & 140/52 & 157/55 & 179/59 \\
$M=7\times 10^{14}\;h^{-1}M_\odot$ & 143/57 & 143/53 & 151/52 & 180/60 \\ 
\hline\hline
\end{tabular}
\end{center}
\label{tab:neff}
\end{table}%


\section{Cluster masses}

The easiest way to characterize a cluster lens is through its mass.  In the following, we refer to the cluster mass as the mass contained in spheres of radius $R_{\rm vir}$. Such virial radius encloses a density of $\Delta_{\rm vir}(z)$ times the closure density of the Universe at the redshift of the cluster, 
\begin{equation}
\rho_c(z)=\frac{3 H_0^2}{8\pi G}\left[ \Omega_m(1+z)^3+\Omega_\Lambda \right] \,
\end{equation}
where $H_0$ is the present value of the Hubble constant, $z$ the redshift, $G$ is the gravitational constant. The virial overdensity $\Delta_{\rm vir}$ depends on redshift \citep[see][for definitions]{gunngott72,1998ApJ...495...80B}. The corresponding mass, $M_{\rm vir}$, is given by
\begin{equation}
M_{\rm vir} \equiv \frac{4}{3} \pi R_{\rm vir}^3 \rho_\mathrm{c}(z) \Delta_{\rm vir}(z) \;.
\end{equation}
In alternative to the virial mass, different mass definitions are often adopted in literature, such as the mass corresponding to a constant over-density $\Delta$, with $\Delta=200, 500$, or $2500$ times the critical density. The general definition in this case is
\begin{equation}
	M_\Delta \equiv \frac{4}{3} \pi R_{\Delta}^3 \rho_\mathrm{c}(z) \Delta \;.
\end{equation}

\begin{figure}[lt!]
\begin{center}
  \includegraphics[width=\hsize]{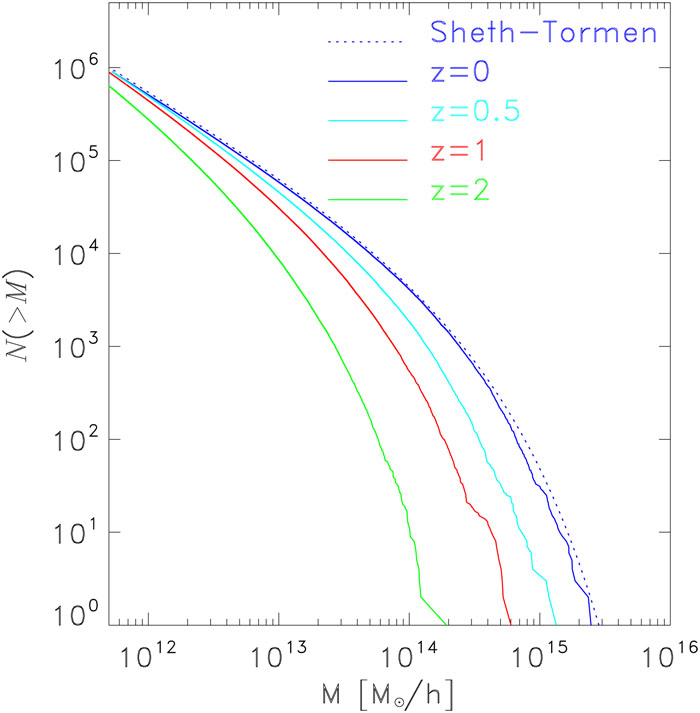}
\end{center}
\caption{The number of halos above a given mass within the {\sc MareNostrum Universe} simulation box at four different redshifts between 0 and 2. The dotted line shows the theoretical expectations from the Sheth \& Tormen mass function.}
\label{fig:massf}
\end{figure}

The mass function of objects in the {\sc MareNostrum Universe} is in very good agreement with the theoretical expectations \citep{SH02.1,2006astro.ph..8289G}. This is shown in Fig.~\ref{fig:massf}, where the number of halos above a minimal virial mass is shown for different redshifts and compared to the predictions of the Sheth \& Tormen mass function at redshift $z=0$.
At this redshift more than 4,000 cluster sized objects with masses
larger than $10^{14}\, h^{-1} M_\odot $ are found. About 58,000 objects have masses
larger than $10^{13}\, h^{-1} M_\odot$. At redshift $z=1$ more than
30,000 objects with masses larger than $10^{13}\, h^{-1} M_\odot$ are
detected. 

\begin{figure*}[lt!]
\begin{center}
  \includegraphics[width=0.49\hsize]{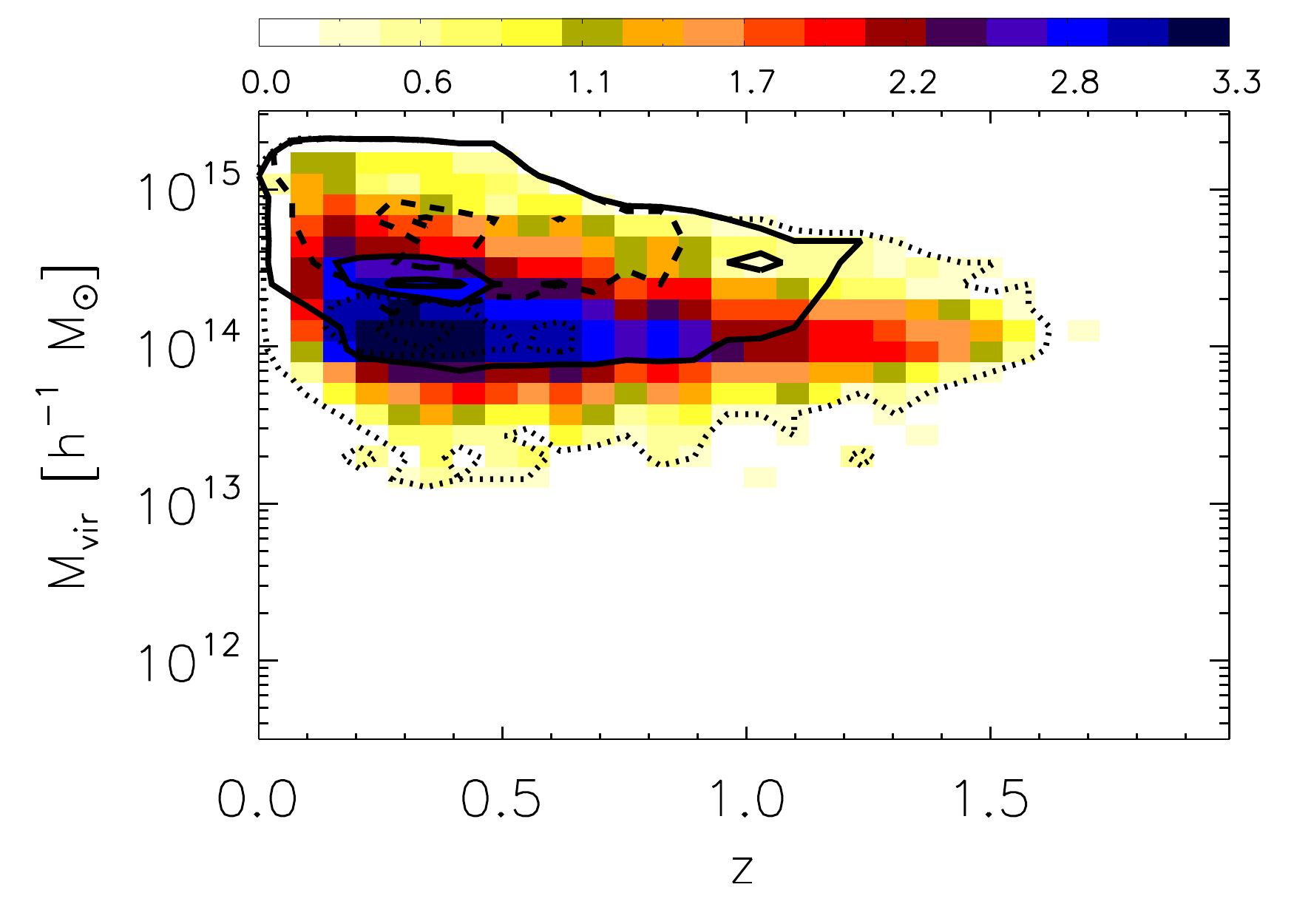}
    \includegraphics[width=0.49\hsize]{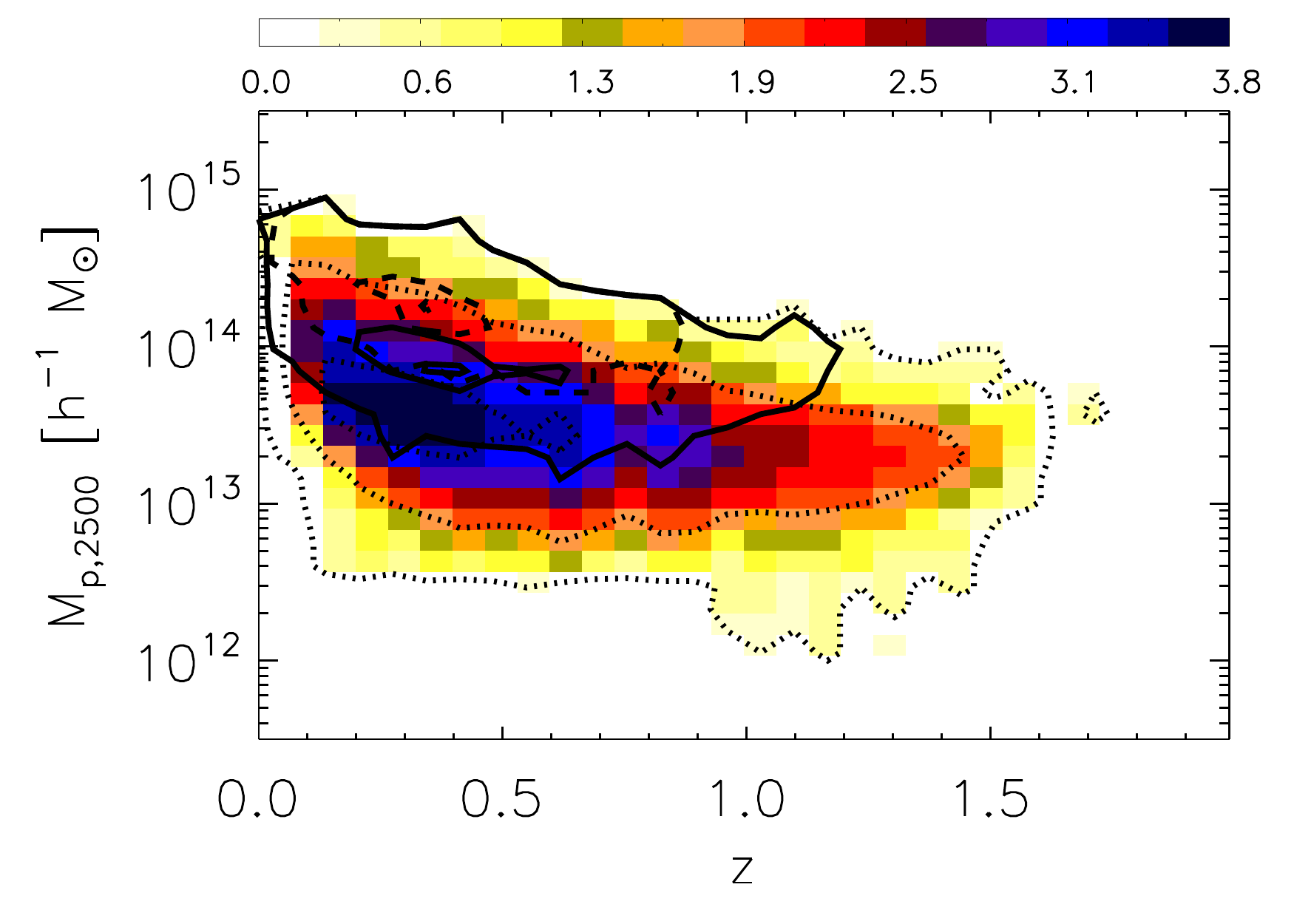}
\end{center}
\caption{The distributions of strong lensing clusters in the $M_{\rm vir}-z$ (left panel) and in the $M_{\rm p,2500}-z$ plane (right panel). The color levels correspond to different number counts of halos {  in the {\sc MareNostrum Universe} ($500^3h^{-3}$Mpc$^3$ comoving)} which are critical for a source redshift of $z_{\rm s} = 2$. The inner dotted contours show the levels corresponding to $90\%$ and $50\%$ of the peak of the distribution. The outer dotted contour encloses $100\%$ of the halos, thus it shows the minimal and the maximal mass of halos producing critical lines for each redshift bin. Similarly, the solid and the dashed contours refer to the distributions of the halos with cross sections for giant arcs larger than $0$ and than $10^{-3}h^{-2}\;$Mpc$^{2}$. The color-bar on the top of each panel show the link between the colors and the $\log$ of the number of cluster per mass and redshift bin.}
\label{fig:minmass}
\end{figure*}

Selecting the clusters via their strong lensing efficiency implies that only the high mass tail of the distribution is properly sampled. Indeed, since the amplitude of the gravitational deflection directly depends on the mass, strong lenses are the most massive objects at each epoch. In particular, we expect a minimal mass below which clusters do not develop critical lines and are unable to produce very distorted images like gravitational arcs. Since clusters must be located at a convenient angular diameter distance between the observer and the sources, the number and the typical mass of strong lensing clusters should vary as a function of both the lens and the source redshifts. Assuming a fixed source redshift of $z_s=2$, the mass distribution of strong lensing clusters {  in a comoving volume of $500^3h^{-3}$Mpc$^3$} at different redshifts is given in the left panel of Fig.~\ref{fig:minmass}. The color levels show the number counts of critical clusters in the $M_{\rm vir}-z$ plane. Lighter (darker) colors correspond to smaller (larger) number counts. The outer dotted contour show the limits of the distribution: no critical clusters have been found outside the region enclosed by this line. The two inner contours correspond to the $50\%$ and to the $90\%$ of the peak of the distribution. Thus, they show how rapidly the critical cluster counts decrease as a function of both mass and redshift.  As the figure shows, the region of the plane where clusters are able to produce critical lines extends down to masses of groups at the most favorable redshifts. However these are very rare objects. At redshifts larger than 1.2, or smaller that 0.2, the mass threshold grows rapidly, while the number counts of critical clusters drop. These lenses are too close to the sources or to the observer to be critical. Requiring that clusters are also able to produce giant arcs produces an additional selection effect. Using the same convention as for the dotted contours, the solid contours refer to arcs with non-vanishing cross-sections for giant arcs.  Note that there are no clusters at $z_{\rm l}>1.3$ that are able to produce large distortions, although they could be still efficient for sources at much larger redshift. The most massive clusters in the box are still able to develop small critical lines up to $z_{\rm l}=1.7$. The bulk of clusters producing giant arcs is concentrated at $0.15<z_{\rm l}<0.8$. 
Finally, the dashed contours show the distribution of the lenses with lensing cross sections above $10^{-3}\;h^{-2}$Mpc$^2$. 
These lenses are likely to be {the most easily targeted for strong lensing studies and contribute significantly to the lensing signal in the universe} given that, as discussed in the previous section, the expected number of strong lensing features produced by these objects is by far larger than for lenses with smaller cross sections. {These, on the other hand, are more abundant, thus they will dominate the total lensing optical depth.} As shown by the contours, these objects are confined in a smaller area on the $M_{\rm vir}-z$ plane. They are typically clusters with masses exceeding few times $10^{14}\;h^{-1}M_\odot$ and with redshift below unity. Most of them are concentrated in a narrow redshift window between $0.2<z<0.6$. 

Although we see an interesting selection effect in the 3D masses, what really matters for strong lensing is the projected mass. In particular, the mass contained in a cylinder around the cluster center, where the critical lines form. For each cluster in our sample, we measure the projected mass within $R_{2500}$, $M_{\rm p,2500}$, for each of the three projections used for ray-tracing. This  is defined as that of a sphere encompassing a mean density of $2500 \times \rho_c$. Typically, it corresponds to a region which is large enough to contain the cluster critical lines.  Here, the projected mass is obtained by integrating all the mass in a cylinder of height $5\,h^{-1}$Mpc. We show  in  the right panel of Fig.~\ref{fig:minmass} the distribution of the strong lensing clusters in the plane $M_{\rm p,2500}-z$. Interestingly, although it appears clear that strong lensing depends on the mass in the central region of the deflectors, the spread in projected mass is larger by about one order-of-magnitude than in 3D. There are clusters which have relatively small mass projected in the core but that are still capable to produce strong lensing effects of different intensity. We interpret this result as due to the importance that other properties of the lenses have for strong lensing, like the amount of substructures and the level of asymmetry and ellipticity in the cluster cores, as shown in \cite{ME07.1}. In several cases, and especially for clusters producing mild strong lensing effects (i.e. clusters with critical curves) the excess of shear produced by a clumpy and asymmetric mass distribution can compensate for the low value of the central convergence.   


\section{Halo triaxiality and orientation}
Since simulated galaxy clusters are triaxial \citep[see e.g.][]{JI00.2,2007ApJ...664..117G}, their projected mass depends on their orientation with respect to the line of sight. In order to evaluate how this impacts on the strong lensing ability of clusters, we measure their triaxial best fit model and discuss the correlation of the orientation with the occurrence of critical lines and giant arcs. To do that, we measure the moment of inertia tensor $I_{ij}$ of each cluster in the sample. The cluster particles within $R_{\rm vir}$ are sorted in a regular cubic grid of $256\times256\times 256$ cells. The mass density in each grid cell is then computed
The inertial tensor components are given by
\begin{equation}
I_{ij}=\sum_{k}m_{k}(R_{k}^2\delta_{ij}-R_{k,i}R_{k,j}) \;,
\end{equation}
where $m_{k}$ is the mass of in the $k$-th selected cell and $R=[R_i]$ is the vector which identifies the cell position with respect to the centre of mass of the system. The triaxial model of the cluster and its principal axes are obtained by diagonalizing the inertial tensor, finding its eigenvalues and eigenvectors.  

\begin{figure}[t!]
\begin{center}
  \includegraphics[width=\hsize]{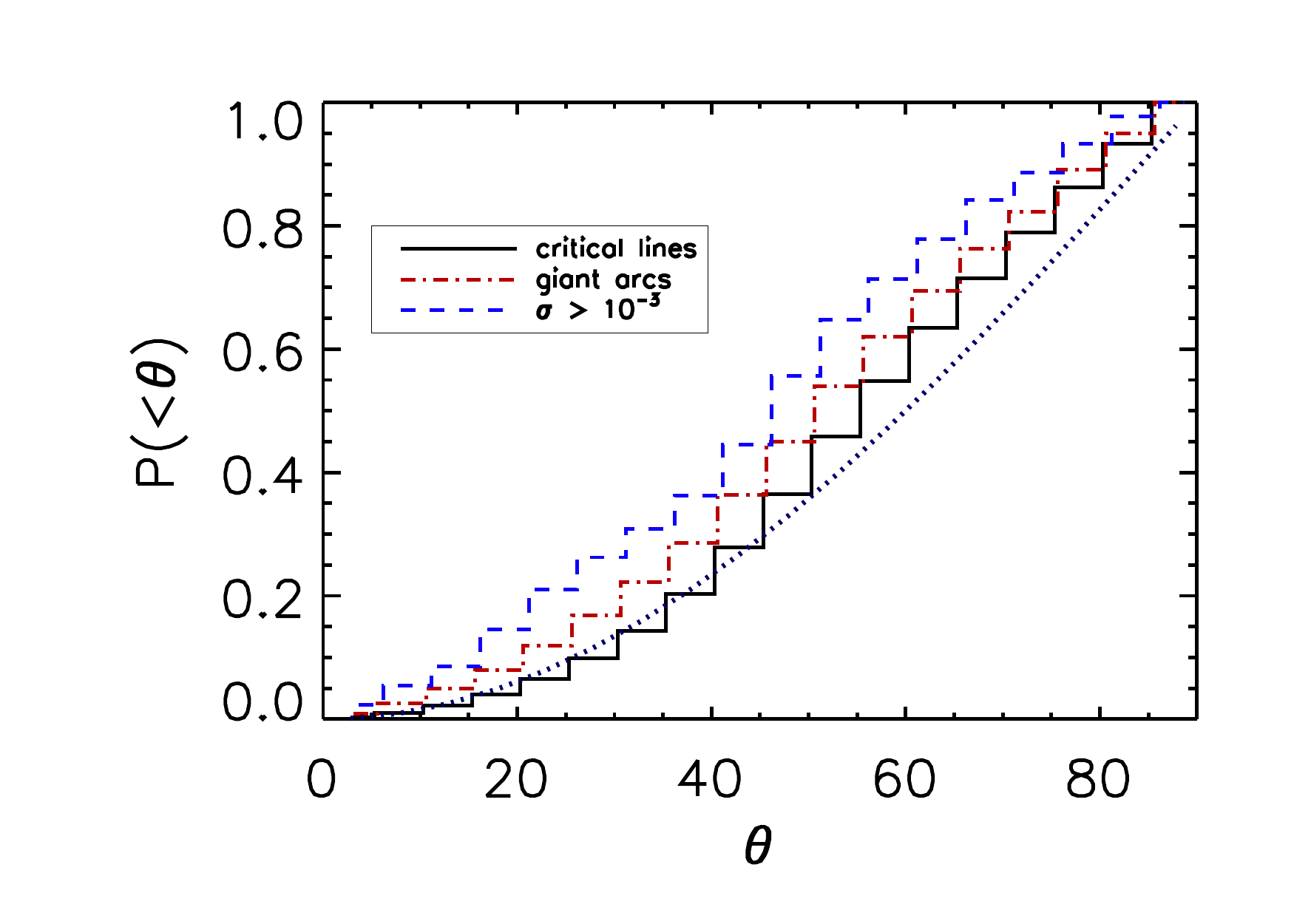}
\end{center}
\caption{The cumulative probability distribution function of the angles between the major axes of the strong lenses in the {\sc MareNostrum Universe} and the line of sight. Shown are the results for the clusters with critical lines, (solid line), for the clusters with cross section for giant arcs larger than zero (dot-dashed line), and for the clusters with cross section for giant arcs larger than $10^{-3}h^{-2}\;$Mpc$^2$ (dashed line). The dotted line shows the expected distribution in case of randomly oriented halos.}
\label{fig:orient}
\end{figure}

The fits show that clusters have prolate triaxial halos and the distributions of the axis ratios of strong lensing clusters is not significantly different from that expected for the general cluster population \citep[see e.g.][]{JI02.1}. This is in agreement with the results of \cite{HE07.1}, who also find that strong lensing clusters are not significantly more triaxial than normal clusters. However, strong lensing clusters seem to be affected by an orientation bias. In Fig.~\ref{fig:orient}, we show the cumulative probability distribution function of the angle between the major axes of the inertial ellipsoid and the line of sight to the cluster. The solid and the dot-dashed lines indicate the results for the whole sample of lensing clusters and for the sub-sample of clusters producing giant arcs. The dashed line shows the distribution of the orientation angles of the most efficient lenses in the sample, i.e. with lensing cross section $\sigma >10^{-3}h^{-2}$Mpc$^{2}$. We also display, using a dotted line, the distribution corresponding to totally randomly oriented lenses. We find that lensing clusters tend to be preferentially aligned with the line of sight. This orientation bias increases with the strength of the lens: the median angle for critical clusters is $\sim 57\deg$, while for the sub-sample of clusters capable of producing giant arcs the median angle is $\sim50 \deg$. The median decreases to $\sim 47 \deg$ for clusters with lensing cross section $\sigma >10^{-3}h^{-2}$Mpc$^{2}$. Note that in the case of random orientation we should expect a median angle of $60 \deg$.  This is an important effect, which can affect the conclusions of many studies aiming at estimating the mass of clusters through strong lensing or at measuring cosmological parameters using the abundance of highly elongated arcs on the sky. In fact, we expect that, due to the orientation bias, 3D strong-lensing masses are biased high, if the approximation of spherical symmetry is  used to convert the measured two-dimensional into three-dimensional mass profiles. Moreover, this alignment bias has to be properly modeled when estimating the lensing optical depth for a population of strong lenses in a given cosmology.  Similar results were found by \cite{HE07.1}. They also find a correlation between strong lensing and orientation of the lenses and they find similar distributions of the orientation angles as those we find here. 

\begin{figure*}[t!]
\begin{center}
  \includegraphics[width=0.49\hsize]{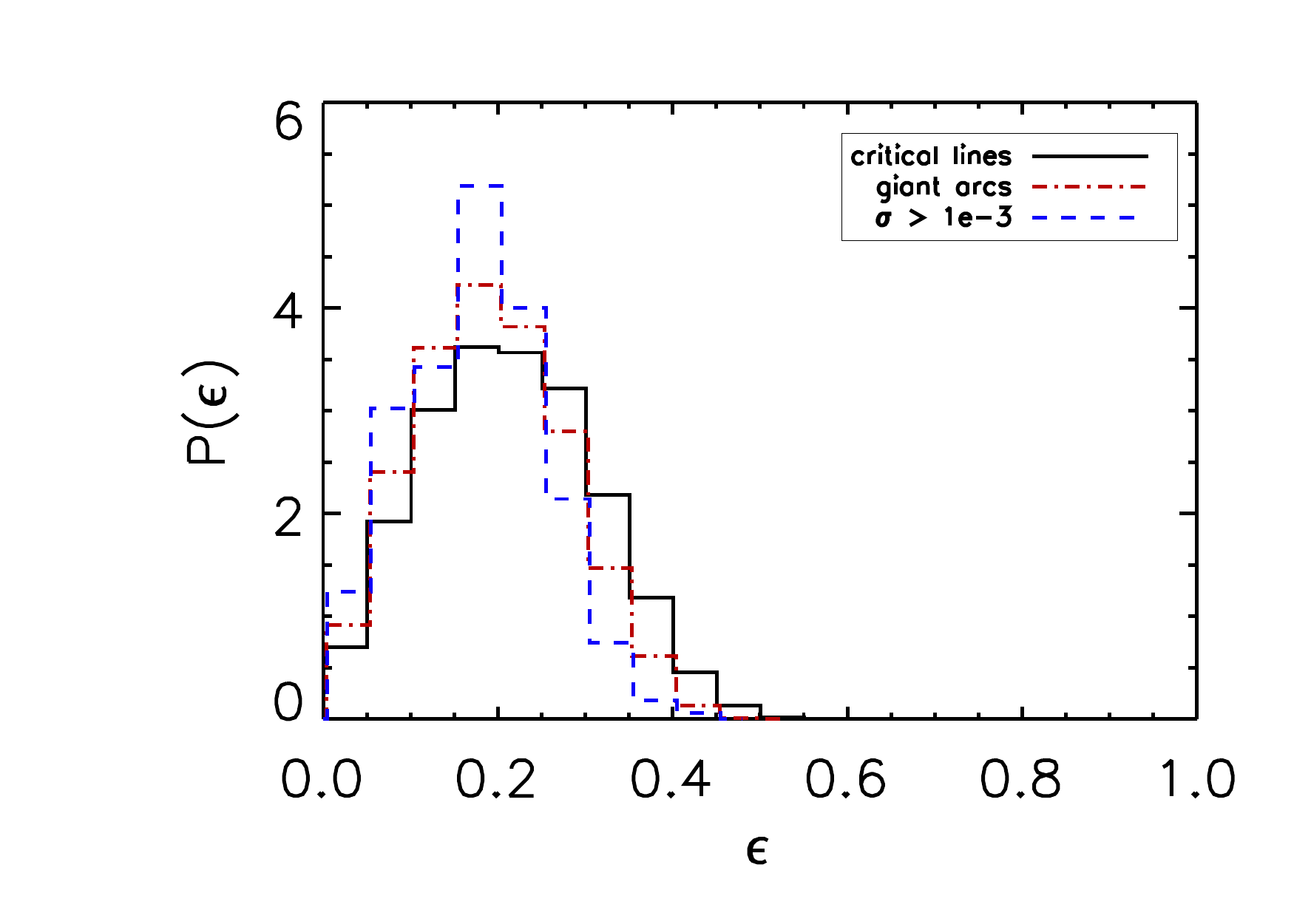}
  \includegraphics[width=0.49\hsize]{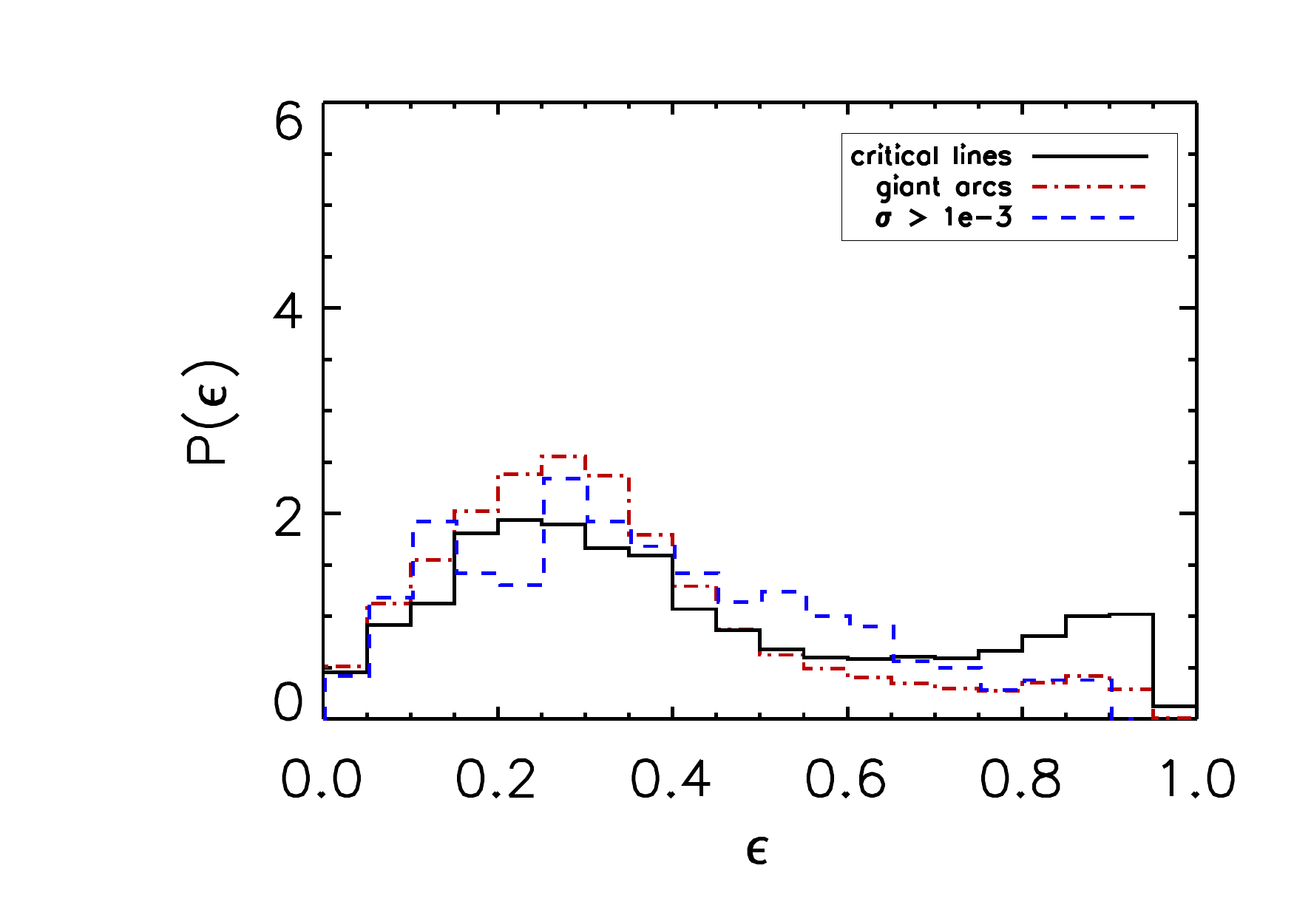}
\end{center}
\caption{The probability density functions of the projected ellipticity of the lensing clusters in the {\sc MareNostrum Universe}. Shown are the results for the clusters with critical lines (solid lines), for the sub-sample of clusters with non-vanishing cross sections for giant arcs (dot-dashed lines), and for cluster with $\sigma >10^{-3}\;h^{-2}$Mpc$^{2}$. The  left and the right panels refer to ellipticities measured at $R_{\rm vir}$ and at $0.1\times R_{\rm vir}$.}
\label{fig:ellipt}
\end{figure*}

Apart from the orientation, the halo triaxiality is important because it determines the projected shape of the lenses.
It has been shown in several papers that, for a fixed mass, the strong lensing cross section is larger for higher ellipticities of the projected mass distribution \citep{ME03.1,ME07.1}. With such a large sample of lensing clusters, we can address the question of what is the distribution of their projected ellipticities. These are measured similarly to the three-dimensional shape of the lenses. We measure and diagonalize the inertial tensor of the cluster mass distribution projected on a regular grid of $256\times256$ cells. We select those cells where the surface density exceeds some thresholds. The thresholds we use are given by the mean surface densities at $R_{\rm vir}$ and at $0.1\times R_{\rm vir}$. Thus, we measure the projected ellipticity both in the outer and in the inner cluster regions. The probability distribution functions of the projected ellipticity are shown in Fig.~\ref{fig:ellipt}. The ellipticity is defined as $\epsilon=(1-b/a)/(1+b/a)$, where $a$ and $b$ are the major and minor axes of the ellipse. The solid, the dot-dashed, and the dashed lines refer to critical clusters (assuming again a source redshift of $z_{\rm s}=2$), to clusters with non-vanishing cross section for giant arcs, and to clusters with large cross section for giant arcs ($>10^{-3}\;h^{-2}$Mpc$^2$), respectively. The left and the right panels show the distributions of the outer and of the inner ellipticities. We find that the projected cores are more elliptical, with distributions which peak at $\epsilon \sim 0.2-0.4$. It is interesting to note that critical clusters have a bimodal ellipticity distribution: a large number of clusters have extremely elongated cores with ellipticities which extend to $\epsilon=0.9$. Since we are fitting each lens with a single ellipse, these are mainly clusters with substructures near the centers which mimic large ellipticities.  We shall recall that the tangential critical lines form where 
\begin{equation}	
	\kappa+\gamma=1	
\end{equation} 
As discussed in \cite{TO04.1}, the shear produced by the substructures enhance the the ability of the clusters to produce strong lensing, because it makes critical even those lenses where the convergence is not large enough to ensure it ($\kappa \ge 1$ at some point). However, although the additional shear allows several clusters to have critical lines, several of them still are unable to produce large distortions. Indeed, the second peak in the PDFs at large ellipticities is less prominent for clusters producing giant arcs, and even less for clusters with large lensing cross sections. For this class of lenses, it is more important to have a large amount of mass projected onto cluster core. Thus, either they have substructures closer to center, or they are more strongly aligned with the line of sight, thus appearing a bit rounder than other critical lenses. Note that, as long as the lensing cross section grows, the ellipticity distribution becomes unimodal but its peak shifts towards larger ellipticities, also suggesting that most of these cluster have substructures close to their centers. 

At large radii, we find that the ellipticities are smaller and the projected ellipticity becomes smaller as the strength of the lens increases. This is clearly related to the orientation bias discussed above. The strongest lenses typically are elongated along the line of sight, thus they appear rounder on the sky. 

\section{Concentrations}
Several previous studies have discussed the importance of the halo concentration for lensing. Using simulations, \cite{HE07.1} find that concentrations of lensing clusters are on average $\sim 18\%$ larger than the typical clusters in the universe. \cite{2008ApJ...685L...9B} report very high level of mass concentrations ($c \sim 10$) in a sample of four well known strong lensing clusters. \cite{FE07.2} show that, at a given mass, the strong lenses are $\sim 10\%$ to $\sim25\%$ more concentrated than the average. Here, we discuss the concentrations of the clusters in the {\sc MareNostrum Universe}.

We measure the concentrations by fitting the density profiles of the clusters in our sample with the Navarro-Frenk-White \citep{NA97.1} formula,
\begin{equation}
	\rho(r)=\frac{\rho_s}{r/r_s(1-r/r_s)^2} \;,
\end{equation} 
where $\rho_s$ is a characteristic density and $r_s$ is the scale radius. The concentration is defined as $c=R_{200}/r_s$.  

\begin{figure*}[t!]
\begin{center}
\includegraphics[width=0.49\hsize]{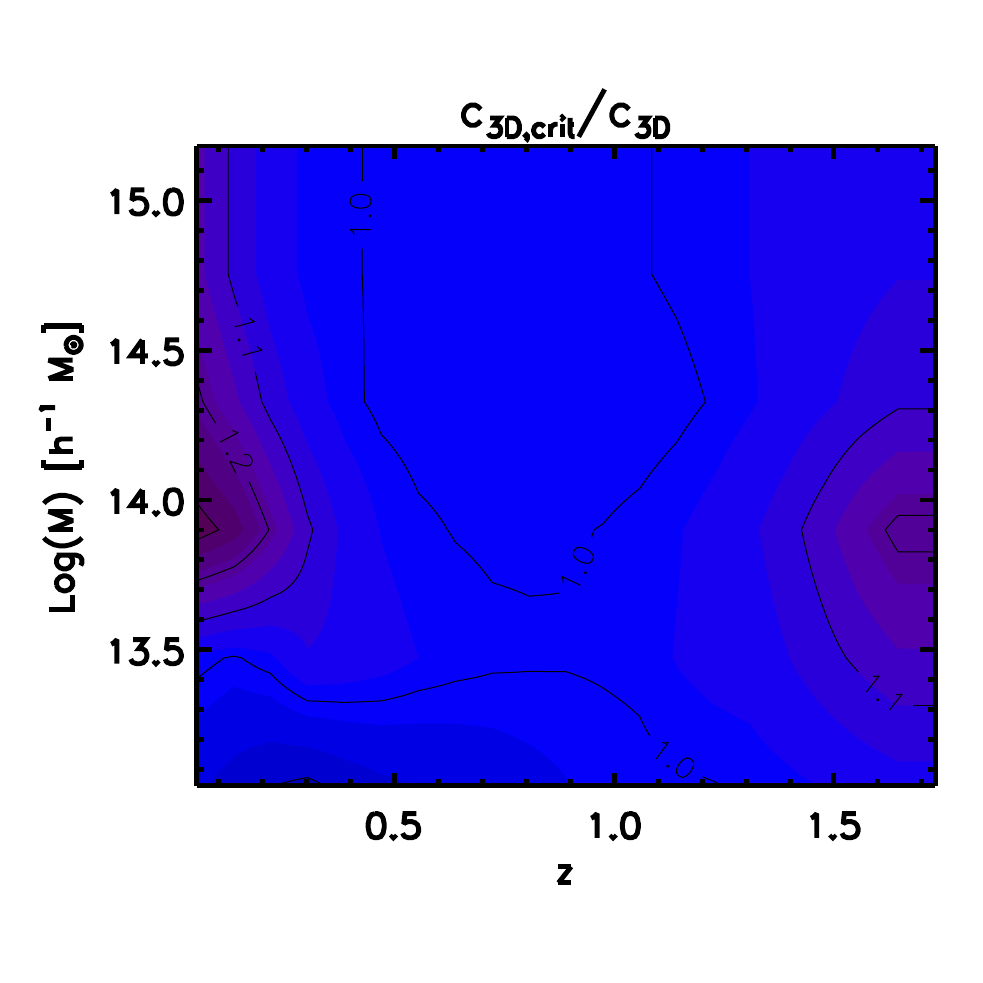}
\includegraphics[width=0.49\hsize]{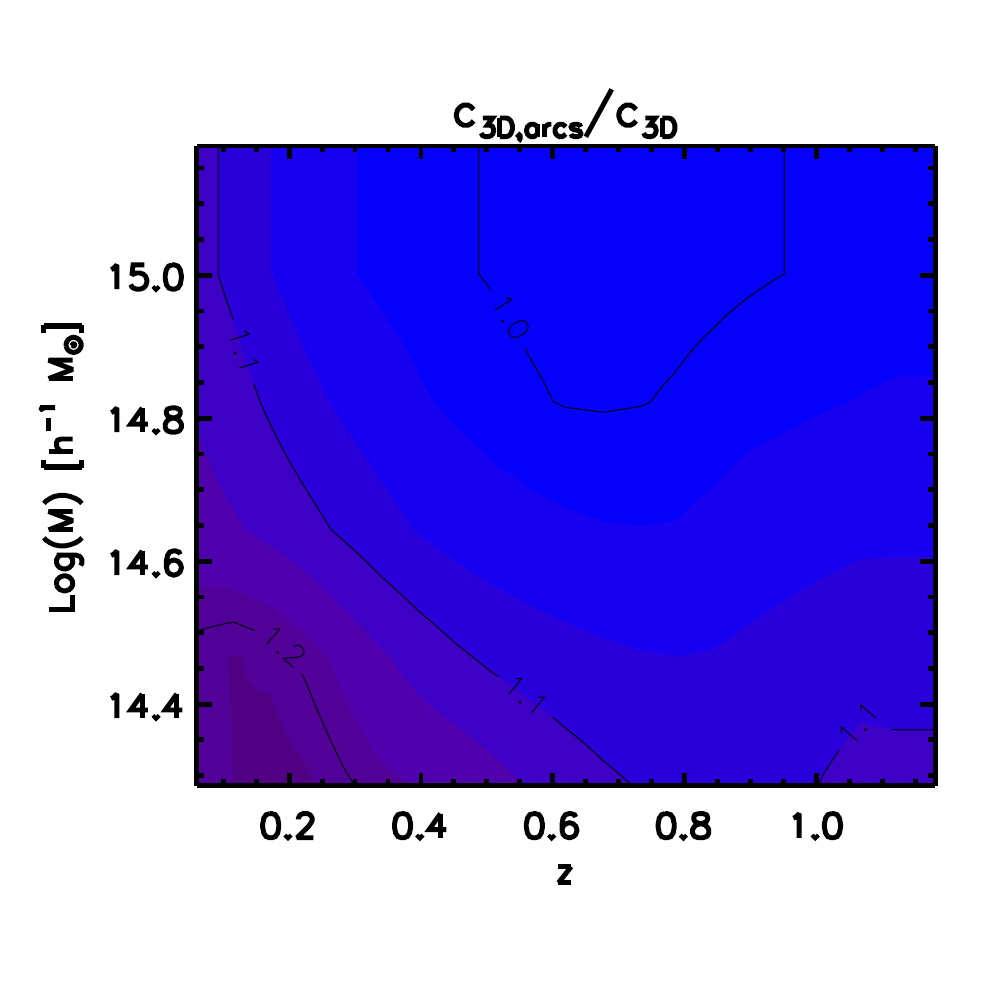}
\end{center}
\caption{NFW concentrations as a function of the halo mass and redshift. Left and right panels refer to clusters with critical lines and to clusters with non-vanishing lensing cross section for giant arcs, respectively. The concentrations  are normalized to those measured on the whole sample of clusters in the simulation box, regardless of their ability to produce strong lensing effects. Different colors are used to encode the different values of the normalized concentrations. The labelled contours provide the link between the color scale and the concentration values.}
\label{fig:3dc}
\end{figure*}

\begin{figure*}[t!]
\begin{center}
\includegraphics[width=0.49\hsize]{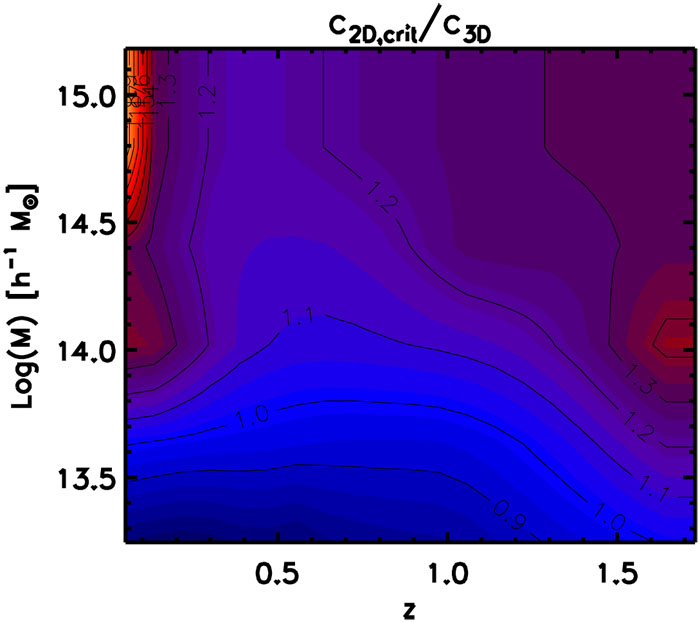}
\includegraphics[width=0.49\hsize]{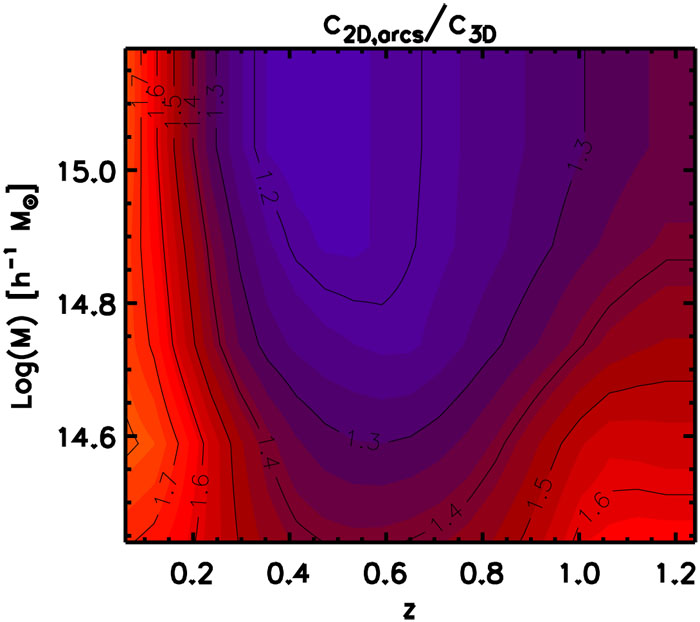}
\includegraphics[width=0.49\hsize]{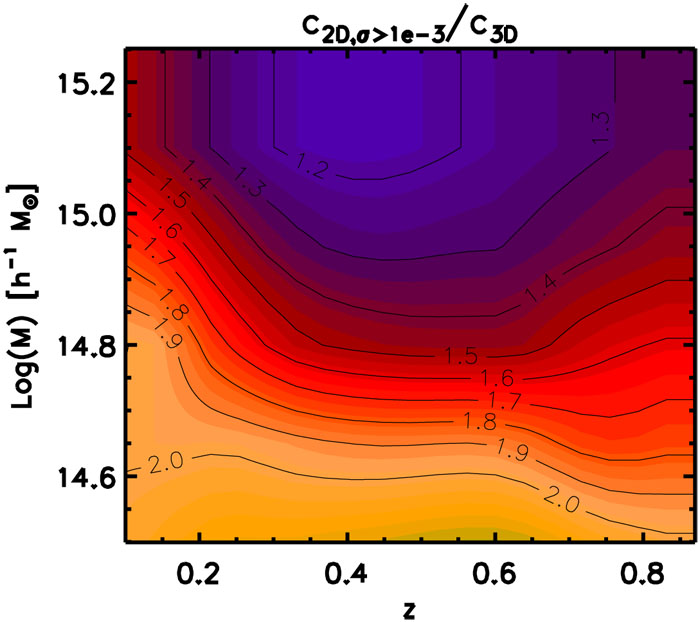}
\includegraphics[width=0.49\hsize]{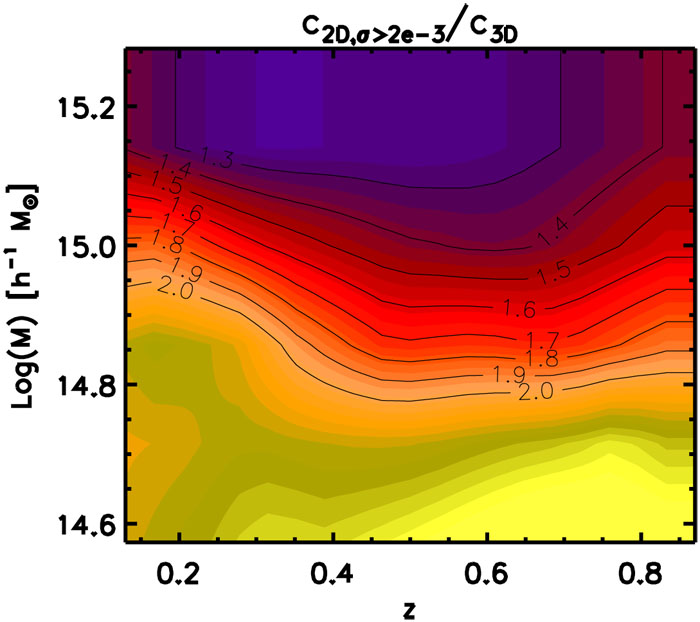}
\end{center}
\caption{Same as in Fig.~\ref{fig:3dc}, but showing 2D-concentrations of lensing clusters. The 2D-concentrations have been normalized to the 3D-concentrations of the whole cluster sample. Starting from the top-left panel, we show the 2D-concentrations as a function of mass and redshift for 1) clusters with critical lines; 2) clusters with non-vanishing lensing cross sections for giant arcs; 3) clusters with cross section $\sigma>10^{-3}\;h^{-2}$Mpc$^2$; 4) clusters with cross section $\sigma>2 \times 10^{-3}\;h^{-2}$Mpc$^2$.  }
\label{fig:conc2}
\end{figure*}

Instead of fitting the density profiles of individual halos, which are noisy, we prefer to fit the stacked profiles of clusters with similar redshifts and masses. The mass bins are equally spaced in logarithmic scale. We stack the profiles of all clusters in the mass bins and perform the NFW fit. Again, we select those objects which exhibit critical lines for sources at $z_{\rm s}=2$, and, among them, those halos which are also able to produce large distortions, with lensing cross section above some minimal value. The concentrations of the clusters in these two sub-samples are compared with those of general clusters, regardless of their ability to behave as strong lenses.  The results are shown in  Fig.~\ref{fig:3dc}. The left and the right panel refer to critical and to large distortion clusters, respectively. The color intensity reflects the amplitude of the concentrations. The concentrations are normalized to those of general clusters of similar mass and redshifts. The labels in the overlaid contours indicate the numerical value of the normalized concentration at the corresponding color level. Strong lensing clusters at moderate redshifts have concentrations similar to those of the general cluster population. Only small mass lenses have a relatively small concentration bias ($\lesssim 20\%$). The bias become more significant at small and high redshifts, where it affects also the largest masses. Due to their short distance to the observer or to the sources, these clusters need to be very concentrated in order to focus the light from distant sources. Note that the bias is mass dependent. As the mass decreases, the bias is stronger. This is a clear selection effect: if we require a cluster to be critical or even to produce large arcs, only the most concentrated halos in the smallest mass bins are able to satisfy the requirement. 

As mentioned above, lensing probes the projected mass distribution of clusters. The concentrations are typically measured by fitting multiple image systems and arcs with combinations of projected parametric models. Then, the three dimensional density profiles are determined by assuming spherical symmetry. As we discussed earlier, clusters have triaxial shapes, thus the assumption of spherical symmetry is generally wrong. Moreover, as we have shown in the previous section, strong lensing clusters tend to be seen along their major axes. For these reasons, the concentrations measured in 2D through strong lensing are expected to be more biased compared to 3D concentrations. This effect is also discussed in \cite{HE07.1}, where a comparison of 2D vs 3D concentrations of individual clusters led to the conclusion that the former are typically $\sim 20\%$ larger than the latter.  To verify this result, we proceed to fit the surface density profiles of our strong lensing clusters in their projections. Again, to do this, we stack the profiles in mass and redshift bins. The fitting formula is given by the truncated NFW surface density profile \citep{ME00.1},
\begin{equation}
  \Sigma _{_{\rm NFW}}(\xi)=2\int_0^{\zeta_{\rm max}}
  \rho_{_{\rm NFW}}(r)\,{\rm d}\zeta \ ,
\label{prnf}
\end{equation} 
where $\zeta$ is the coordinate along the line of sight and $\xi$ is the
component of $r$ perpendicular to $\zeta$. The maximum of $\zeta$ is given by
half the size of the sub-box containing each cluster, i.e. $2.5 \;h^{-1}$Mpc comoving.
Using the dimensionless coordinate on the projection plane $x \equiv
\xi/r_s$ and defining the quantities $u \equiv \mbox{arcsinh} (\zeta/\xi)$, and $\kappa_{\rm s}\equiv \rho_{\rm s}r_{s}\Sigma_{\rm
cr}^{-1}$, the previous equation can be written as
\begin{equation}
  \Sigma_{_{\rm NFW}}(x)=2\kappa_{\rm s} \Sigma_{\rm cr} f(x) \ ,
\end{equation}
where 
\begin{eqnarray}
  f(x) &=& -\frac{2}{(x^2-1)^{3/2}}
  \arctan\left[\frac{x-1}{\sqrt{x^2-1}}
  \tanh\left(\frac{u}{2}\right)\right] \nonumber \\
  &+& \left.\frac{1}{x^2-1}\frac{x\sinh u}{1+x\cosh u}
  \right|_0^{u_{\rm max}} 
\label{app_f1}
\end{eqnarray}
if $ x>1$;
\begin{equation}
  f(x)=\frac{2\cosh(\frac{u}{2})\sinh(\frac{u}{2})}{3(1+\cosh u)^2}+
  \left.
  \frac{4\cosh (\frac{u}{2})^3\sinh(\frac{u}{2})}{3(1+\cosh u)^2}
  \right|_0^{u_{\rm max}} 
\label{app_f2}
\end{equation}
if $x=1$; and
\begin{eqnarray}
  f(x) &=& \frac{2}{{(1-x^2)}^{3/2}}
  \mbox{arctanh}\left[\frac{1-x}{\sqrt{1-x^2}}
  \tanh\left(\frac{u}{2}\right)\right] \nonumber \\
  &+& \left.\frac{1}{1-x^2} \frac{x \sinh u}{1+x \cosh u}
  \right|_0^{u_{\rm max}}
\label{app_f3}
\end{eqnarray}
if $x<1$. 
In the previous formulae, $u_{\rm max}=\mbox{arcsinh}
(\zeta_{\rm max}/\xi)$.

The resulting 2D concentrations, for several classes of strong lenses,  as a function of mass and redshift are shown in  Fig.~\ref{fig:conc2}. As done in Fig.~\ref{fig:3dc}, the 2D-concentrations are normalized to the 3D-concentrations of general clusters of similar masses and redshifts. critical and large distortion clusters separately and we normalize the measured concentrations with those derived fitting the 3D density profiles of the whole sample of clusters in the cosmological box. Starting from the top-left panel we show the results for clusters with critical lines and for clusters with lensing cross sections for giant arcs larger than $0$, $10^{-3}\;h^{-2}$Mpc$^2$,  and $2\times 10^{-3}\;h^{-2}$Mpc$^2$, respectively. As expected, the bias grows compared to the 3D case, and the amount by which it increases depends on the class of lensing clusters we are considering. For critical clusters, the ratios between 2D concentrations and the corresponding 3D concentrations are of the order of $\sim 1.2$ for intermediate redshift clusters but they become larger than $1.3$ at low and at high redshifts. For clusters which are able to produce giant arcs, the 2D-concentration bias is significantly larger. As discussed for the 3D concentrations the amplitude of the bias depends on both redshift and mass: smaller masses at short distances from the observer or from the sources have the largest biases. Moreover, increasing the lensing cross section for giant arcs, the concentration bias grows dramatically. For example, massive clusters ($M\sim 10^{15}\;h^{-1}M_\odot$) at the most efficient redshifts for strongly lensing sources at $z_{\rm s }=2$ ($z_{\rm l}\sim 0.4$) with lensing cross sections larger than $2\times 10^{-3}\;h^{-2}$Mpc$^2$ have 2D-concentrations which are typically larger by $\sim 50\%$ than the 3D-concentration of general clusters. For smaller masses and redshifts, the 2D-concentrations can be higher than expected in 3D by more than a factor of two. 
This is a consequence of the orientation bias discussed in the previous section. In order to be able to produce large and very elongated arcs, clusters laying too close to the observer or to the source must be optimally oriented and extremely concentrated. Due to triaxiality, the concentrations measured from the 2D mass distributions of these clusters are much bigger than the corresponent 3D concentrations \citep{2005ApJ...632..841O,2005A&A...443..793G}.  
Note that, as discussed in Sect.~\ref{sect:crsec}, a lensing cross section of $\sigma \sim 10^{-3}\;h^{-2}$Mpc$^2$ corresponds to an expectation value of $\sim 1$ giant arc in a deep HST observation. Clusters like $A1689$, which contains about $10$ arcs with length-to-width ratio larger than 7.5 \citep{2005ApJ...627...32S}, are thus expected to have extremely large lensing cross sections. If the properties of real clusters are well reproduced by the clusters in our simulations, these very efficient strong lenses are likely  to have extremely biased 2D-concentrations, as recently reported by \cite{2008ApJ...685L...9B} \cite[see also][]{2009ApJ...699.1038O}. Our findings are in agreement with the results recently published by  \cite{2009MNRAS.392..930O}, who use semi-analytic models of triaxial halos to estimate that the projected mass distributions of strong lensing clusters have $\sim 40 - 60\%$ larger concentrations compared with typical clusters with similar redshifts and masses \citep[see also][]{2010arXiv1001.1696S}.

\section{X-ray luminosities}

Gas physics is known to be potentially very important for strong lensing \citep[see e.g.][]{PU05.1,2008MNRAS.386.1845H}. Several processes taking place in the Intra-Cluster-Medium (ICM), like cooling, heating, energy feedback from AGNs and supernovae, thermal conduction, etc. can also affect the distribution of the dark-matter in clusters, influencing the shape of the density profiles \citep{DO04.1,2009arXiv0904.0253P,YSGS07} as well as the triaxiality of the dark matter halos.  {  Phenomena like cooling, star formation, and energy feedback change the thermal properties of the ICM, thus influencing the X-ray emissivity. In fact, several numerical studies report that X-ray luminosities in non-radiative simulations are  higher than in simulations where cooling and feedback are active, while the $L_X-M$ relation derived from the same simulations is less steep than observed \citep[e.g.][]{2010arXiv1002.4539S,2008MNRAS.387.1179M}. The extreme complexity of the processes involved presents a serious challenge for simulating them accurately in a cosmological setting \citep[e.g.][]{2004MNRAS.348.1078B}. Nevertheless, also a non-radiative simulation like the {\sc MareNostrum Universe} can provide useful qualitative information on the possible correlation between strong lensing and X-ray emission by galaxy clusters. Here, we focus in particular on the X-ray luminosity, which is often used to select clusters for strong lensing surveys \citep{LU99.1}.}

The X-ray bolometric luminosity is calculated from the temperature and internal energy of each gas particle in the simulated clusters. In short, the X-ray luminosity is the sum of the contributions to the emissivity from each gas particle, $L_{X}=\sum_i \varepsilon_i$, where the sum extends over all the particles within $R_{\rm vir}$. The emissivity of each gas element can be written as
\begin{equation}
\varepsilon_i=n_{e,i}n_{H,i}\Lambda(T_i,Z_i)dV_i \;,
\end{equation} 
where $n_{e,i}$ and $n_{H,i}$ are the number densities of electrons and of hydrogen atoms, respectively,  associated to the i-th gas particle of given density $\rho_i$, temperature $T_i$, and metallicity $Z_i$. The cooling function $\Lambda(T,Z)$ is calculated by using a \texttt{MEKAL} plasma model \citep{ME85.1,ME86.1,LI95.2} implemented in the \texttt{XSPEC} software package \citep{AR96.1}. For the metallicity, we adopt the typical value $Z = 0.3Z_\odot$ (\citealt{FU98.1,SC99.2}, see also \citealt{BA03.4}). Finally, $dV_i=m_i/\rho_i$ is the volume of the $i$-th gas particle of mass $m_i$.

\begin{figure}[lt!]
\begin{center}
  \includegraphics[width=1.0\hsize]{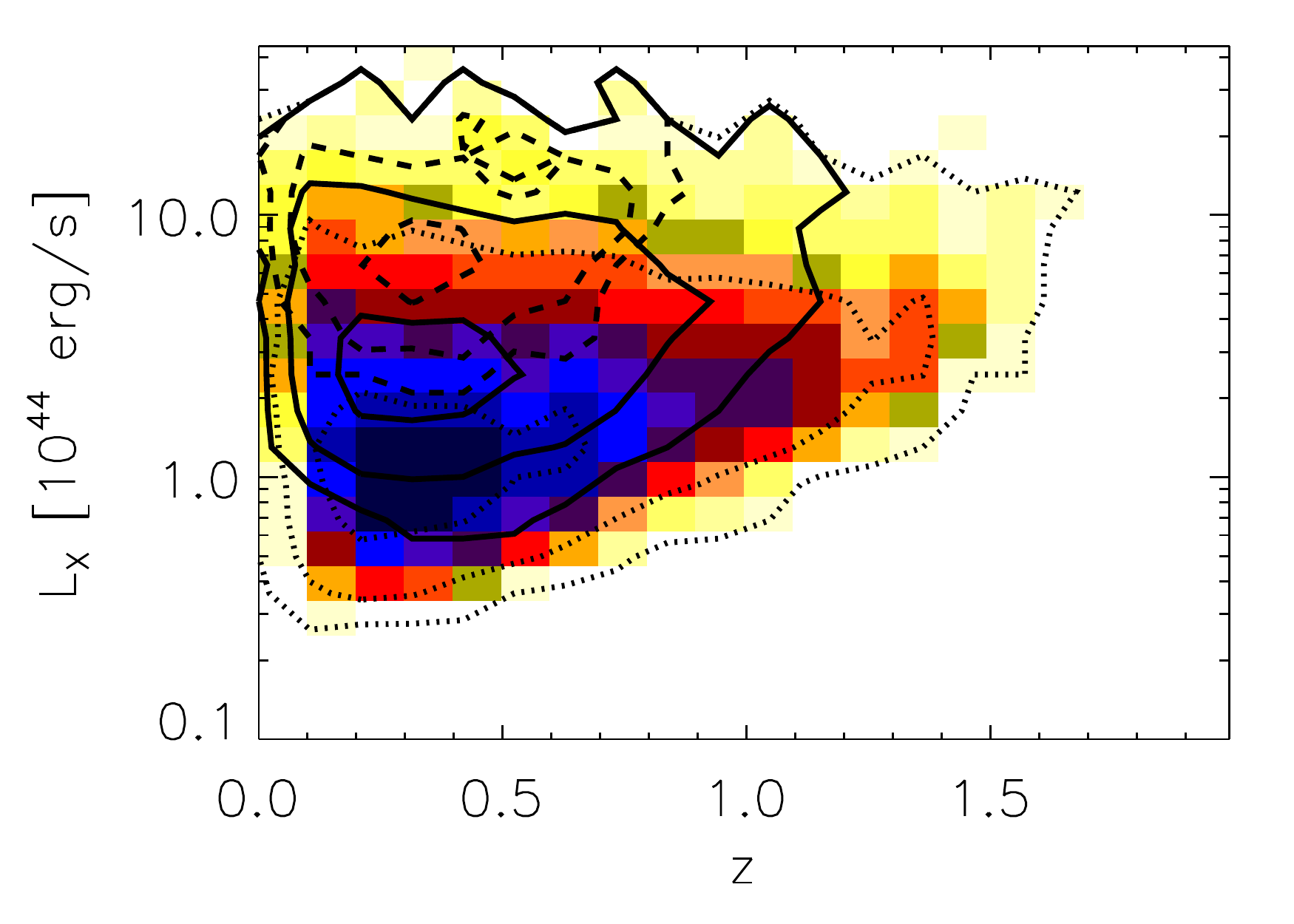}
\end{center}
\caption{The distribution of clusters with critical lines (dotted contours and color levels), with non-vanishing cross section for giant arcs, and with large lensing cross sections for giant arcs (solid contours), $\sigma >10^{-3}\;h^{-2}$Mpc$^{2}$ (dashed contours) in the $L_X-z$ plane. The contour description is the same as in Fig.~\ref{fig:minmass}}
\label{fig:lumz}
\end{figure}

\begin{figure*}[lt!]
\begin{center}
  \includegraphics[width=1.0\hsize]{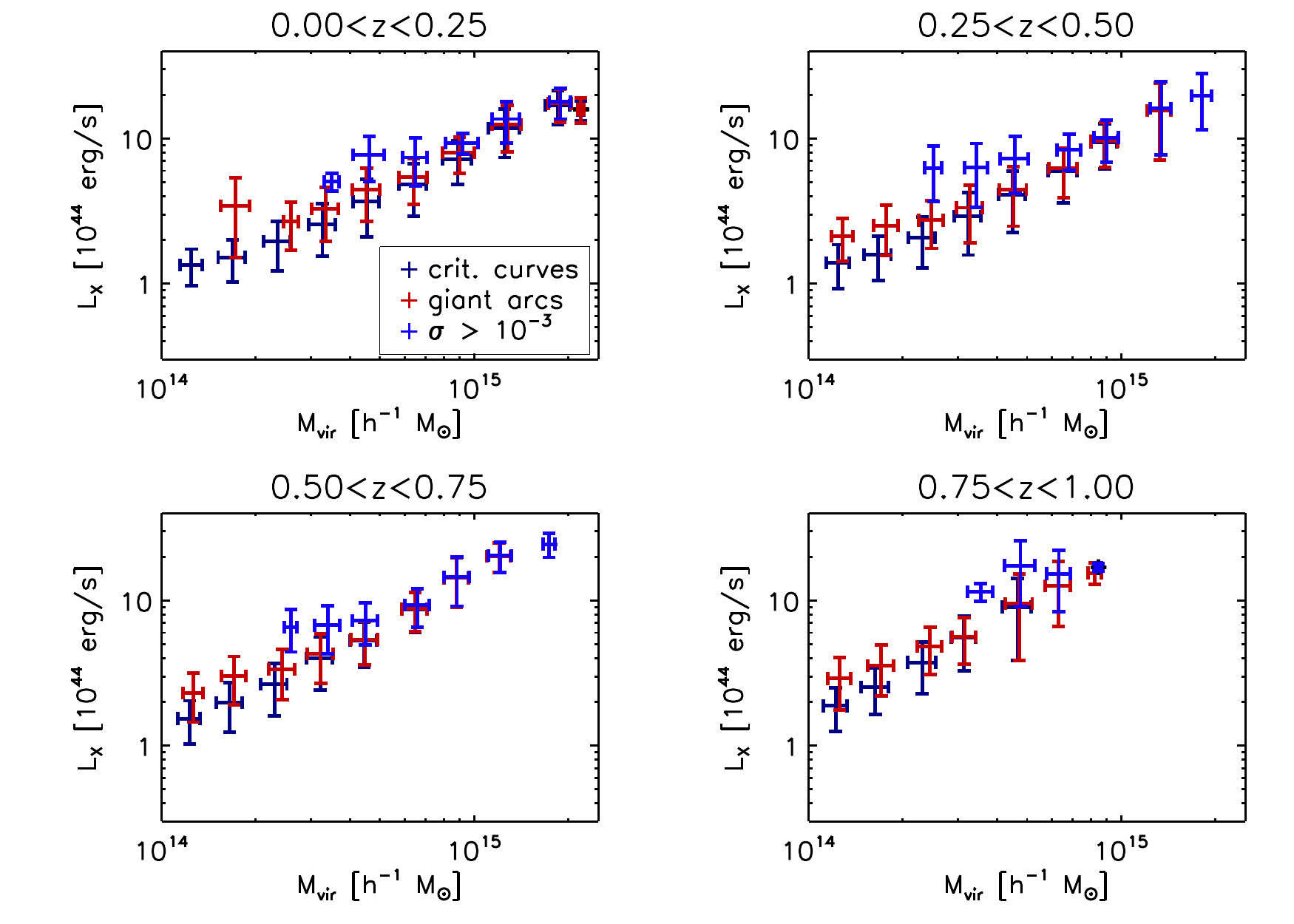}
\end{center}
\caption{The relation between the X-ray bolometric luminosity and the cluster mass. Results are shown in four different redshift bins, as indicated on the top of each panel. Black, red, and blue data points (and errorbars) refer to the sub-samples of critical and large distortion lenses.}
\label{fig:lumM}
\end{figure*}

The distribution of strong lensing clusters in the $L_X-z$ plane is shown in Fig.~\ref{fig:lumz}, where we use the same notation as in Fig.~\ref{fig:minmass}. {  Again the counts correspond to the comoving volume of the {\sc MareNostrum Universe}.} Not surprisingly, given that the X-ray luminosity scales with the cluster mass \citep[e.g][]{1986MNRAS.222..323K},  the  distribution of the strong lenses in the $L_X-z$ plane is very similar to that in the $M-z$ plane. As found for the masses, at each redshift there exists a minimal X-ray luminosity below which no critical lenses are found. The ``critical" X-ray luminosity reaches a minimum between $z \sim 0.3$ and $z\sim 0.5$. Increasing the minimal lensing cross section the distributions of clusters producing giant arcs moves upwards and shrink along the redshift axis.


We now explore more in details the scaling of  the X-ray luminosity with the mass for strong lensing clusters. In Fig.~\ref{fig:lumM} we show the $L_X-M$ relation for four redshift bins, namely $0<z\leq0.25$, $0.25<z\leq0.5$, $0.5<z\leq0.75$, and $0.75<z\leq1$. Numerical simulations are known to be poor at describing the X-ray properties of the cosmic structures on the scales of groups \citep{2008SSRv..134..269B}.  Thus, we limit our analysis to clusters of masses $M_{\rm  vir}>10^{14}h^{-1}M_\odot$. The $L_X-M$ relations found for clusters with critical lines, with non-vanishing cross sections for giant arcs, and with lensing cross sections $\sigma > 10^{-3}\;h^{-2}$Mpc$^2$ are shown in black, red, and blue, respectively. We find that, increasing the strong lensing efficiency, the slope of the $L_X-M$ relation changes, becoming smaller especially at the lowest masses. This effect is also redshift dependent, being more extreme in the lowest and in the highest redshift bins. It shows that at the least favorable redshifts for strong lensing, the X-ray luminosities of the strong lensing clusters tend to be larger than for the general cluster population, especially if the lens mass is relatively small. This seems to suggest that some cluster property rather than the mass plays an important role for boosting the lensing cross sections of these small lenses, which also influences their X-ray luminosity. As discussed in \cite{TO04.1} and in \cite{FE07.1} mergers are likely to be a explanation for the effect we observe here.  

\section{Cluster dynamical state}
\label{sct:eq}

In this section, we use indicators of the virial and of the hydrostatic equilibria in clusters to investigate wether there is a correlation between strong lensing and dynamical activity in the lenses.

\subsection{Virial equilibrium}

The virial equilibrium is perhaps the most natural choice when trying to quantify the dynamical state of a bound structure. In the case of self-gravitating systems, this is quantified by the competition between the
total potential energy and (twice) the internal kinetic energy. 

In agreement with the scalar virial theorem, and following \cite{SH06.1}, we introduce the parameter $\beta$ for the simulated clusters, defined as
\begin{equation}
\beta \equiv 1 + \frac{2T-S}{U} = 1 - \frac{2T-S}{|U|},
\end{equation}
where $T$ is the internal kinetic energy, $U$ the potential energy, and $S$ a surface pressure term that arises from considering the structure as contained in a limited region (in this case the finite sphere). If $N_{\rm vir}$ is the total number of particles contained in the sphere of radius $R_{\rm vir}$, then the kinetic energy is evaluated as
\begin{equation}\label{eqn:kin}
T = \frac{1}{2} \sum_{i=1}^{N_{\rm vir}} m_i \|{\bf v}_i\|^2,
\end{equation}
and the potential energy is given by 
\begin{equation}\label{eqn:pot}
U = -\frac{G}{2} \sum_{i\ne j = 1}^{N_{\rm vir}} \sum_{j=1}^{N_{\rm vir}} \frac{m_i m_j}{\|{\bf r}_i-
{\bf r}_j\|}.
\end{equation}
In Eqs. (\ref{eqn:kin}) and (\ref{eqn:pot}) $m_i$, ${\bf r}_i$ and ${\bf v}_i$ are the mass, the position with respect to the center of mass and velocity with respect to the center of motion of the $i-$th particle, respectively.

\begin{figure}[t!]
\begin{center}
 \includegraphics[width=1.0\hsize]{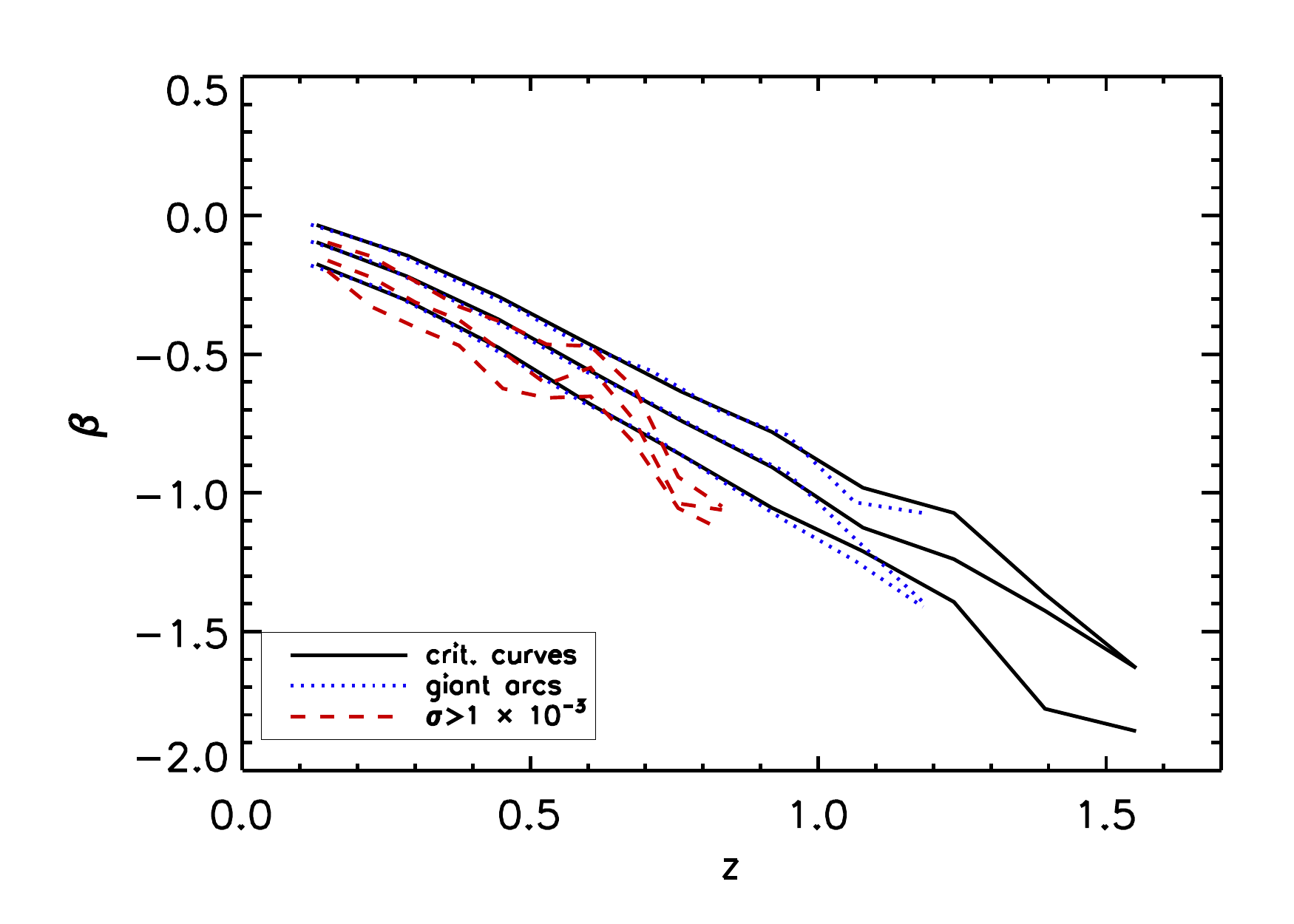}
  \includegraphics[width=1.0\hsize]{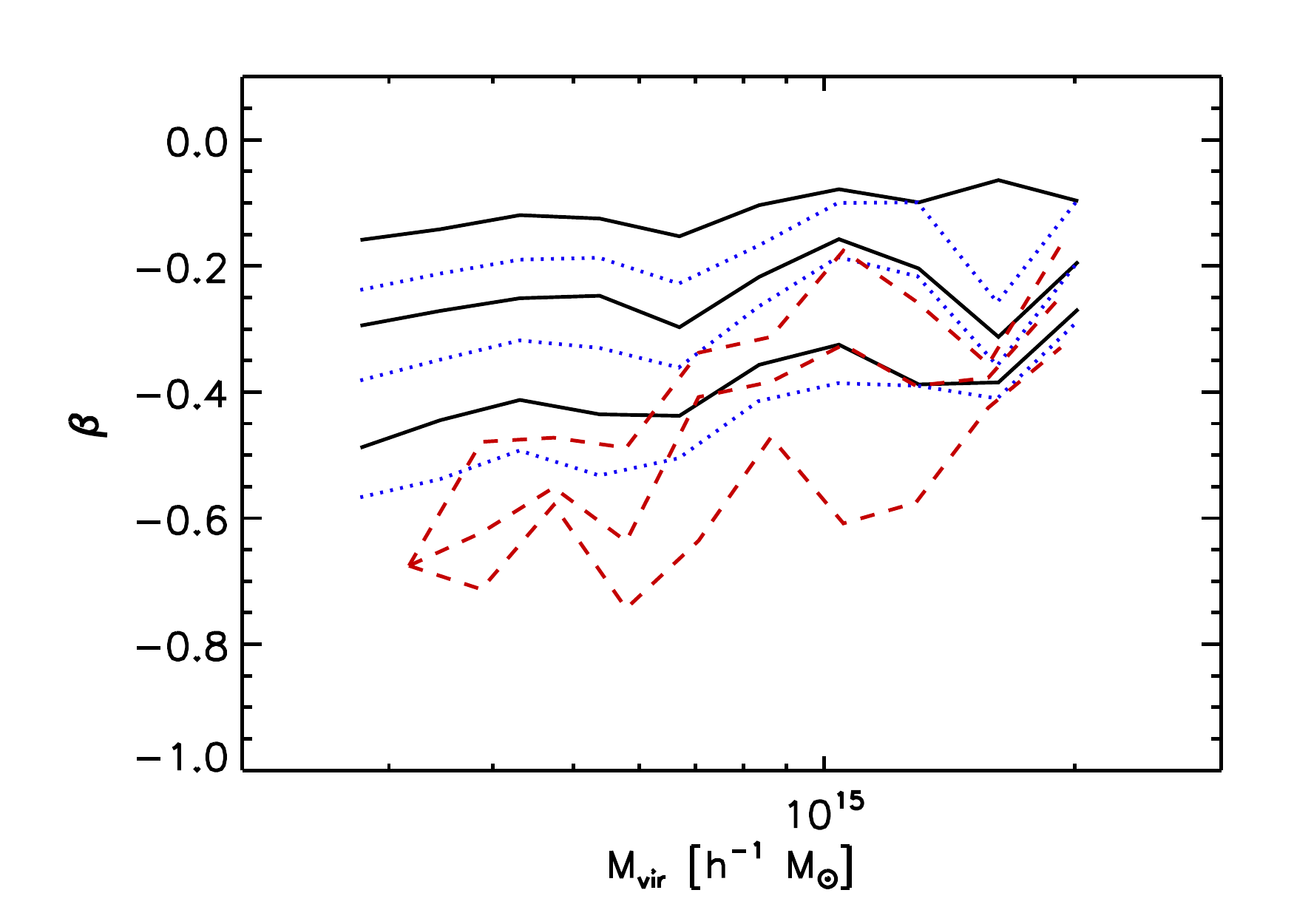}
\end{center}
\caption{Top panel: the median value and the $25\%$ and $75\%$ percentiles of the distributions of the parameter $\beta$ as a function of redshift. Solid, dotted, and dashed line indicate the results for clusters with critical lines, with non-vanishing cross section for giant arcs, and with cross section $\sigma>10^{-3}\;h^{-2}$Mpc$^2$. Bottom panel: the median value and the $25\%$ and $75\%$ percentiles of the distributions of the parameter $\beta$ as a function of the cluster mass.}
\label{fig:virZ}
\end{figure}

The surface term, $S$,  is computed by selecting the outermost $20\%$ of the particles that are inside $R_{\rm vir}$. Let us call $Q_{0.8}$ this set of particles. If $R_{0.8}$ and $R_{0.9}$ are respectively the innermost and the median distances from the cluster center of the particles within $Q_{0.8}$, then 
\begin{equation}
S = \frac{R_{0.9}^3}{R_{\rm vir}^3-R_{0.8}^3} \sum_{i \in Q_{0.8}} m_i
\|{\bf v}_i\|^2,
\end{equation}
where the sum is now extended to all the particles between $R_{0.8}$ and
$R_{\rm vir}$. 

For an ideal cluster in perfect virial equilibrium, $\beta = 0$. On the other hand, whenever the cluster is dynamically active, like accreting mass or merging with substructures, $\beta < 0$, because the dynamics of the structure is dominated by the kinetic energy. According to the hierachical structure formation, we expect that structures at high redshift have $\beta$ on average more negative than at low redshift, the lowest mass objects having the smallest $\beta$. At the typical redshift for lensing clusters ($z \sim 0.3-0.5$) the mass range considered here is well beyond the characteristic collapsing mass ($M_* = $ few $\times \:10^{12} h^{-1} M_\odot$). Thus, all our clusters are in the process of formation and have negative $\beta$ parameters. 
As expected, we find a significant evolution of $\beta$ with redshift. In  the top panel of Fig.~\ref{fig:virZ} we show  the median (together with $25\%$ and $75\%$ percentiles) values of $\beta$ as a function of redshift. Different line styles indicate the results for clusters with critical lines, for clusters which have non-vanishing cross section for giant arcs, and for clusters with large lensing cross sections ($\sigma>10^{-3}\;h^{-2}$Mpc$^2$, as usual). The $\beta$ parameter is close to zero at low redshift, and it becomes increasingly more  negative going to  high redshift. In particular, the value of $\beta$ decreases from $\sim 0$ to $\sim -1$  from $z = 0.1$ to $z = 1$. Note that the most efficient lensing clusters, i.e. the clusters with large lensing cross sections, are characterized by smaller $\beta$ parameters. The bottom panel shows the median and the $25\%$ and $75\%$ percentiles of $\beta$ as a function of the cluster mass. The differences between the cluster sub-samples become more pronounced at the small masses. This indicates that the least massive clusters with large lensing cross sections tend to be farther from the virial equilibrium than clusters with small lensing cross sections. These dynamically active clusters also have higher X-ray luminosity, explaining the flattening of the $L_X-M$ relation shown in Fig.~\ref{fig:lumM} towards the small masses.

\begin{figure*}[ht!]
\begin{center}
  \includegraphics[width=0.49\hsize]{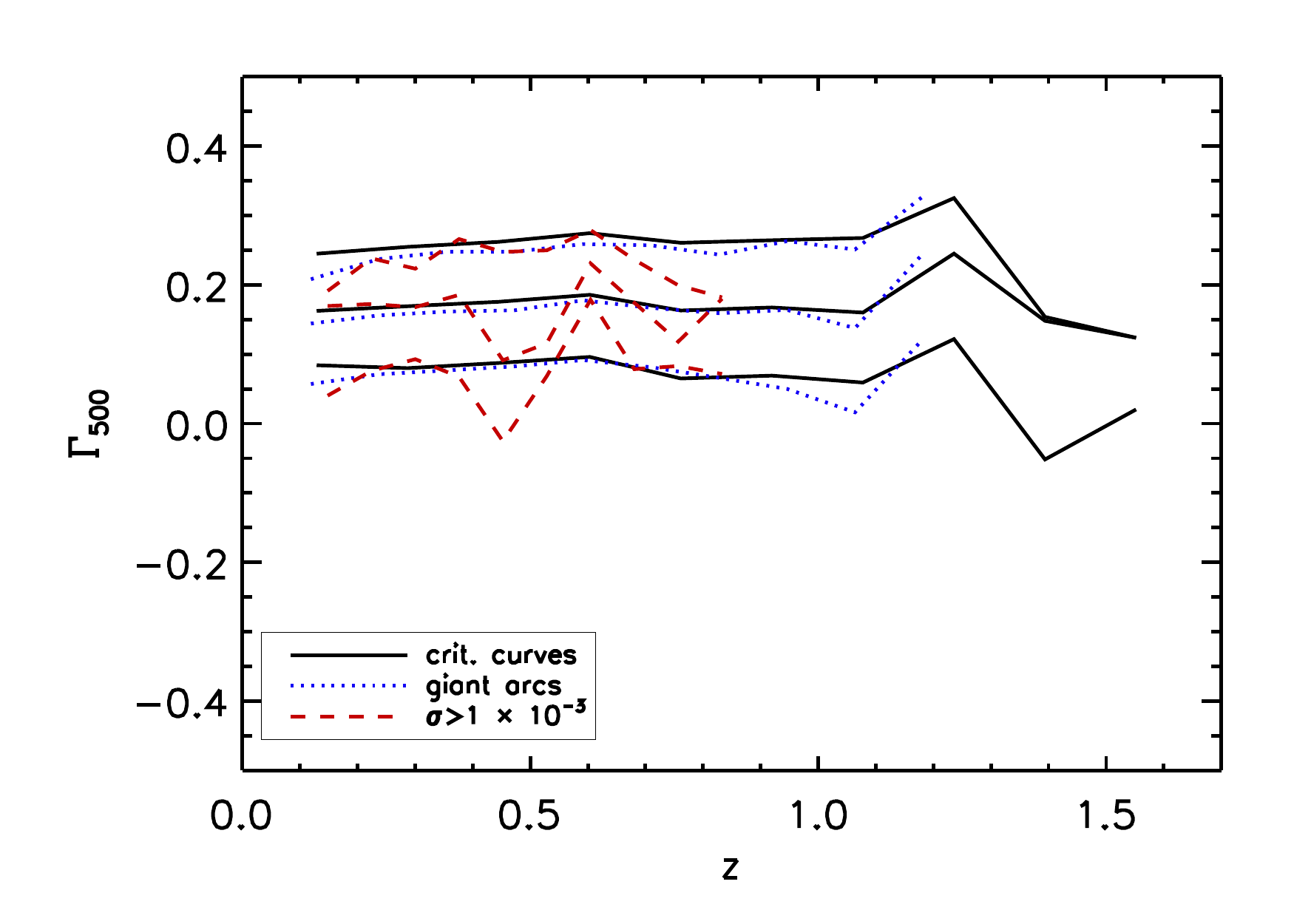}
  \includegraphics[width=0.49\hsize]{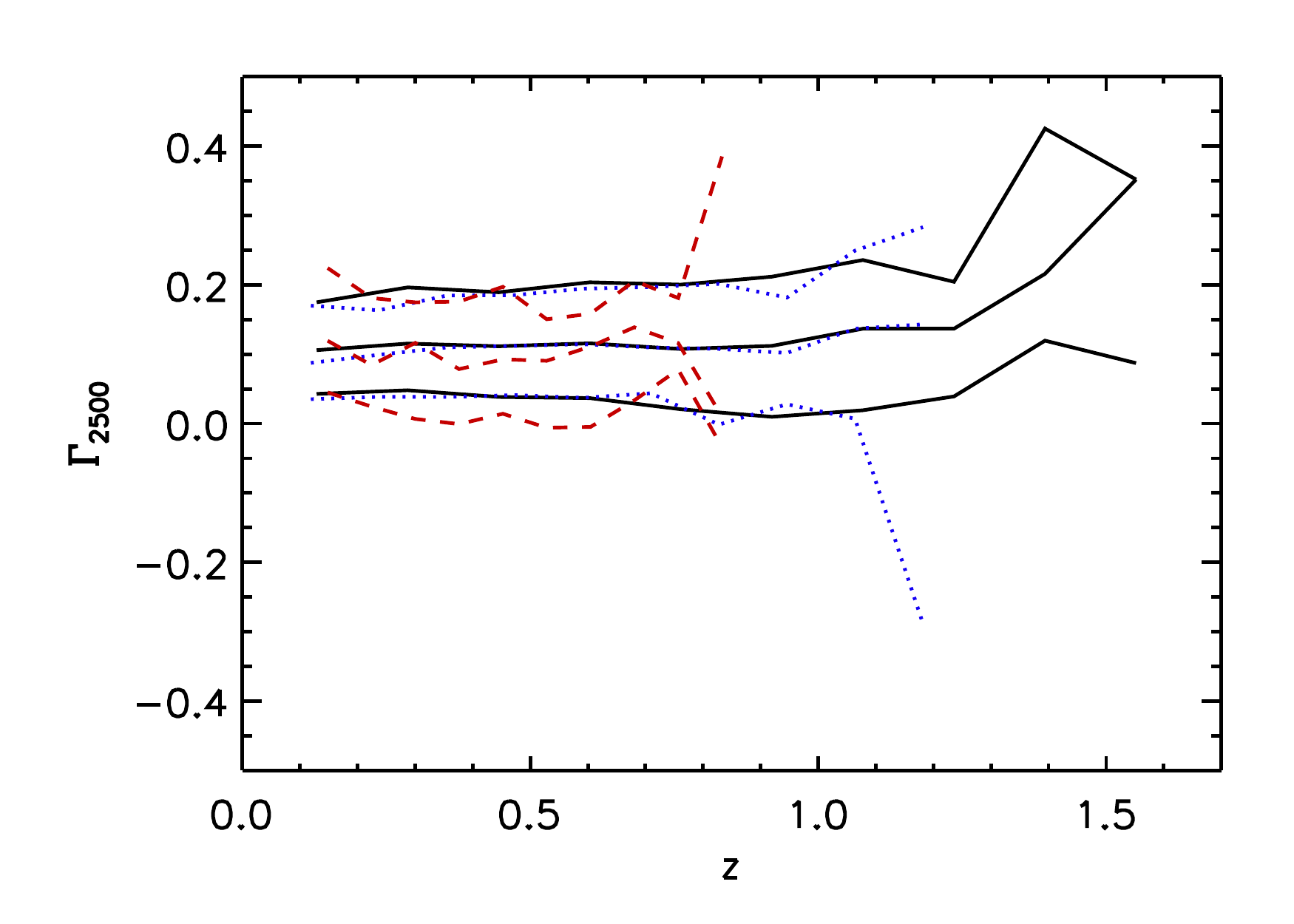}
\end{center}
\caption{The median value and the $25\%$ and $75\%$ percentiles of the distributions of the parameter $\Gamma$ as a function of redshift. Solid, dotted, and dashed line indicate the results for clusters with critical lines, with non-vanishing cross section for giant arcs, and with cross section $\sigma>10^{-3}\;h^{-2}$Mpc$^2$. From left to right, the results are shown for $\Gamma$ measured at $R_{500}$ and $R_{2500}$, respectively.}
\label{fig:hydZ}
\end{figure*}

\subsection{Hydrostatic equilibrium}

A different method for quantifying the equilibrium state of a cluster consists of determining the hydrostatic equilibrium of the gas filling its potential well. If the system is in equilibrium the pressure and the gravitational forces at a given position are exactly counter-acting. 

Given the hydrostatic equilibrium equation, $\nabla \Phi = -\nabla P_\mathrm{g}/\rho_\mathrm{g}$, we define the \emph{hydrostatic equilibrium parameter} $\Gamma$ at a given distance from the cluster center as
\begin{equation}\label{eqn:gamma}
\Gamma(r) = 1 - \frac{W(r)}{4\pi G M(r)},
\end{equation}
where $M(r)$ is the total mass in a sphere of radius $\mathbb{S}_r$, and
\begin{equation}
W(r) \equiv \int_{\partial \mathbb{S}_r} \left| \frac{\nabla P_\mathrm{g}}{\rho_\mathrm{g}} \right| dA.
\end{equation}
Note that, while $\Phi$ is the total potential of the cluster, $P_\mathrm{g}$ and $\rho_\mathrm{g}$ are the pressure and density of the gas component.

If hydrostatic equilibrium holds in the gas shell of radius $r$, then $\Gamma(r) = 0$. Depending on whether the gas is compressed or is not thermalized, for example due to merging with substructures or infall of material from the external regions, $\Gamma$ becomes negative or positive. As shown by \cite{RA04.1}, the gas is generally not at rest inside the cluster potential wells. Non-negligible subsonic bulk motions contribute to the total pressure support of the gas. This typically leads to underestimate the total mass of the system under the assumption of hydrostatic equilibrium \citep[see e.g.][]{2003MNRAS.346..731A,RA06.1,2008ApJ...674..728R,2007ApJ...655...98N,2009MNRAS.394..479A,ME09.1}. The effect is the largest at the largest distances from the cluster center, where the infall of matter is more pronounced. 
We evaluate the hydrostatic equilibrium parameter $\Gamma$ at two different cluster-centric distances, namely $R_{500}$, and $R_{2500}$.
In Figure \ref{fig:hydZ} we show the redshift evolution of $\Gamma$. The median (with $25\%$ and $75\%$ percentiles of the distribution)  in each redshift bin are shown  for the subsamples of critical clusters, clusters capable of producing giant arcs, and clusters with lensing cross section larger than $10^{-3}h^{-2}$Mpc$^2$ (solid, dotted, and dashed lines, respectively). The curves are rather flat, differently from what found for the $\beta$ parameter.
\begin{figure*}[t!]
\begin{center}
  \includegraphics[width=0.49\hsize]{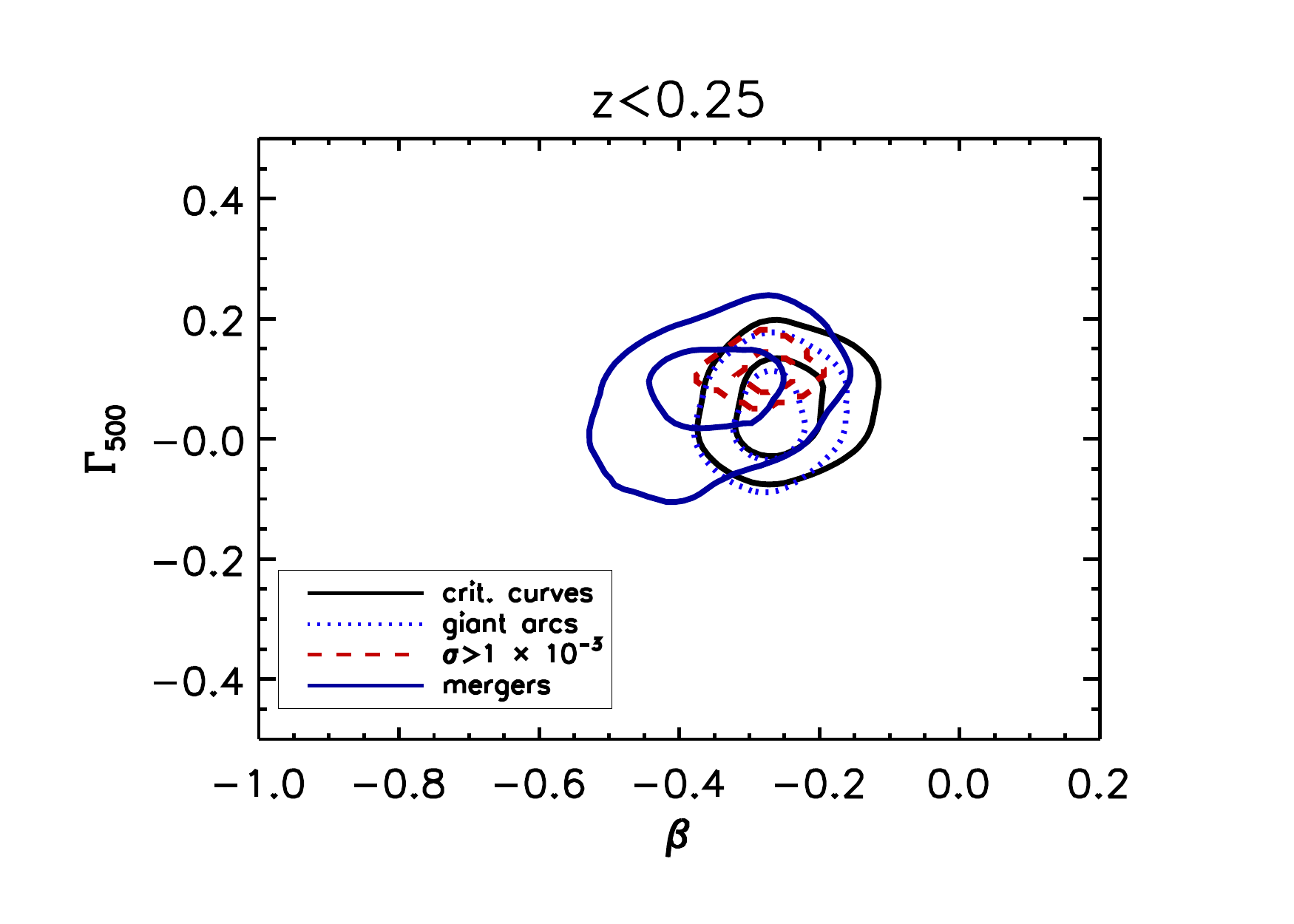}
  \includegraphics[width=0.49\hsize]{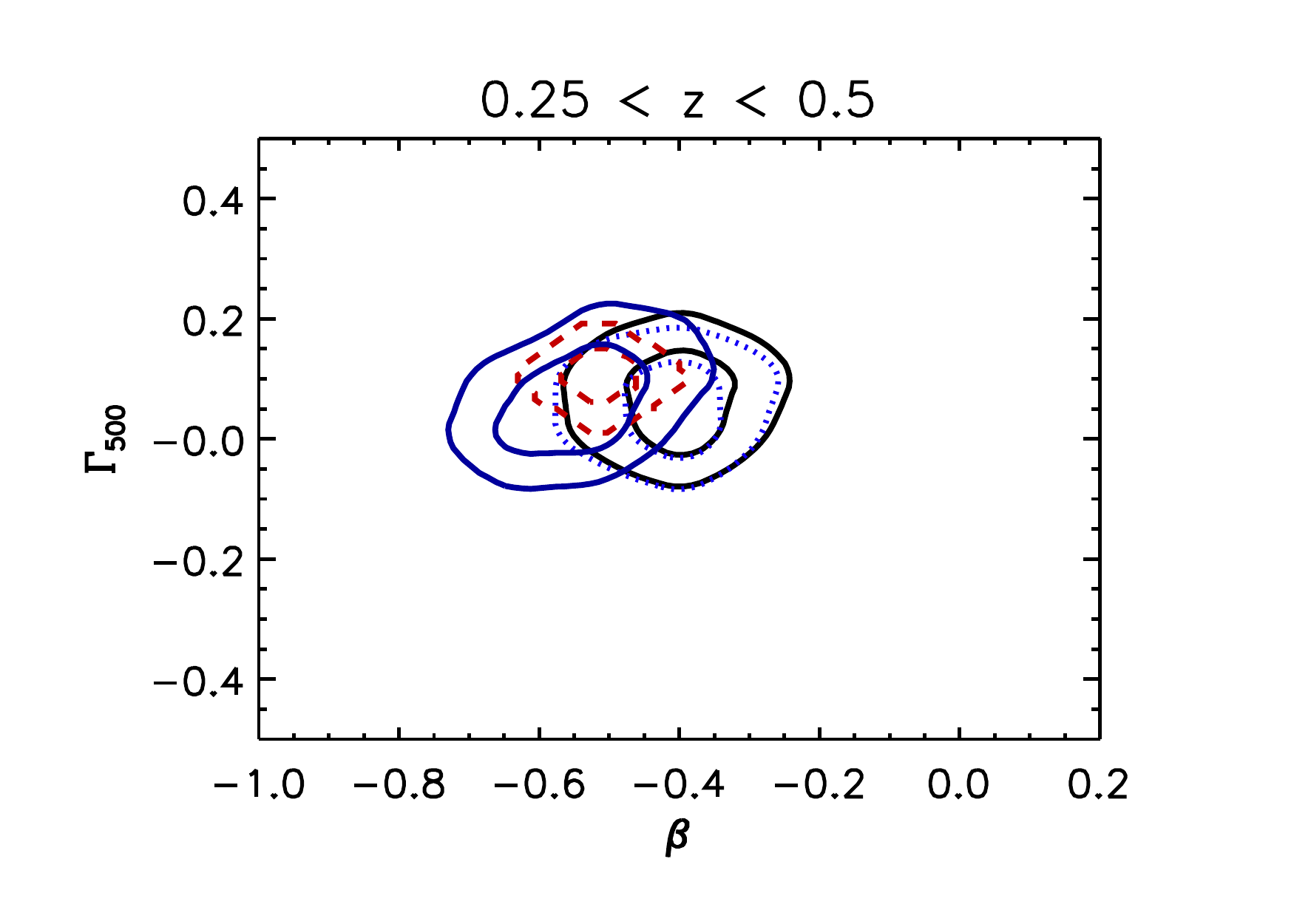}
  \includegraphics[width=0.49\hsize]{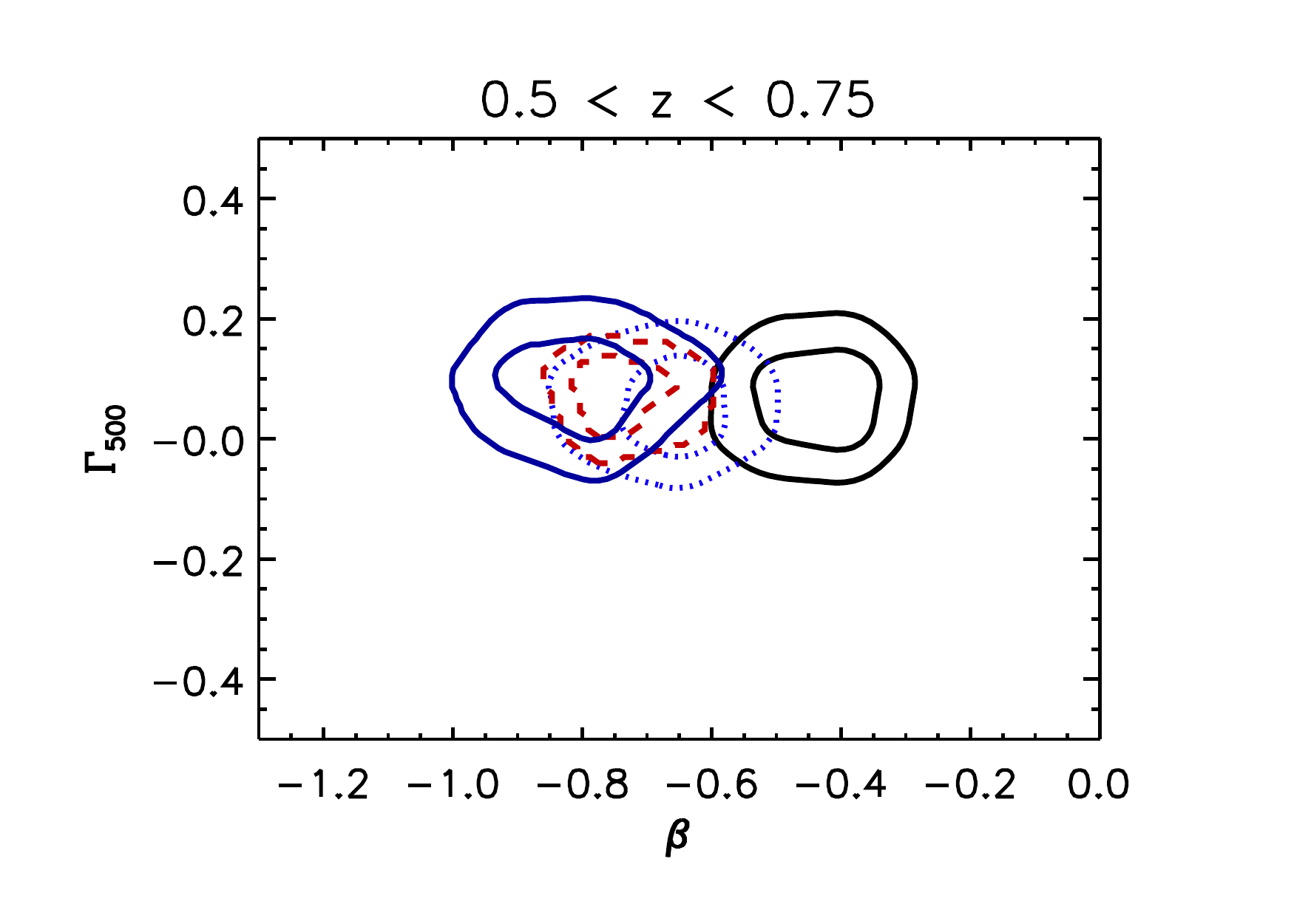}
  \includegraphics[width=0.49\hsize]{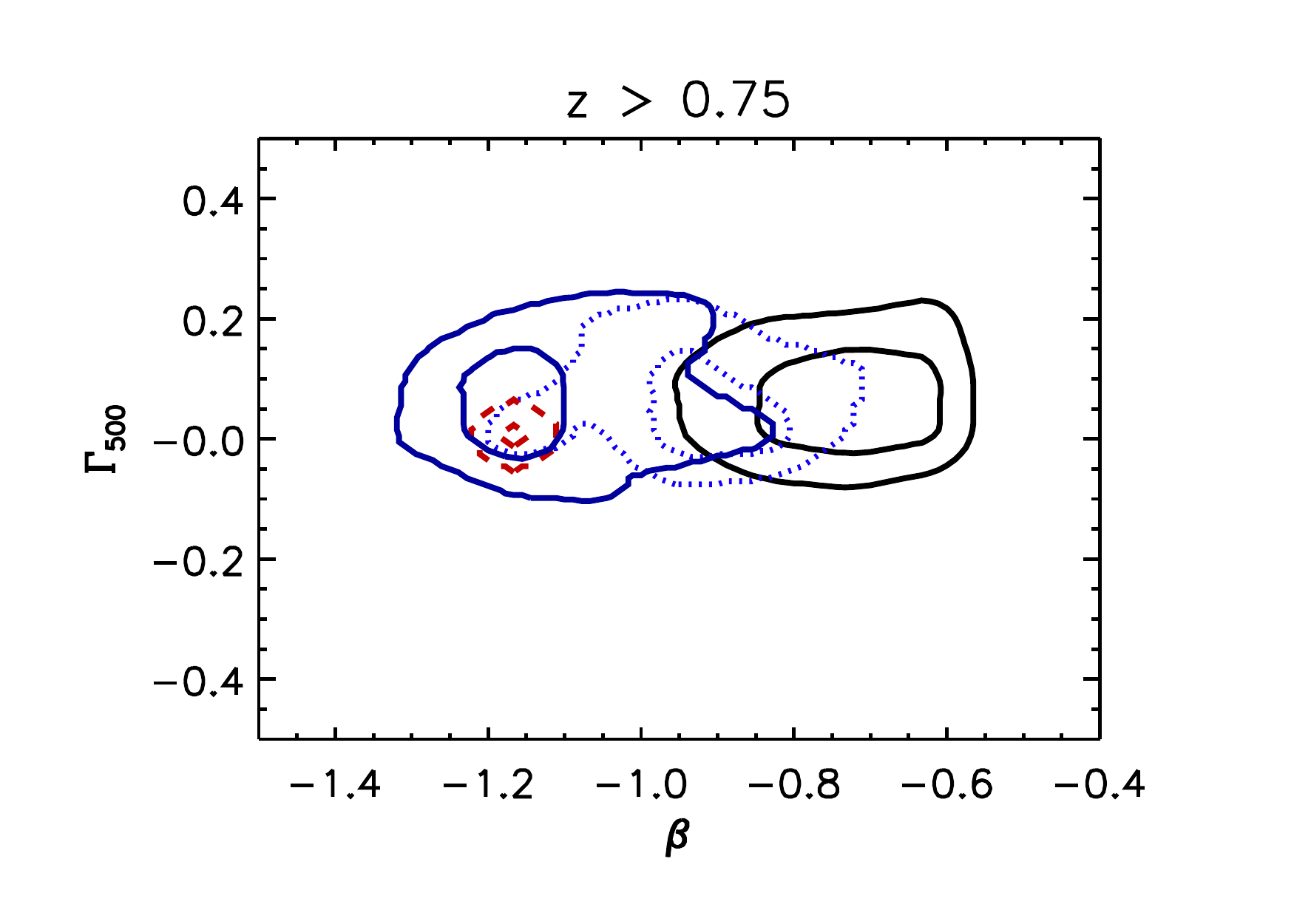}
\end{center}
\caption{The distributions of lensing clusters with increasing ability to produce large distortions in the $\beta-\Gamma_{500}$ plane. The contours indicate the levels corresponding to $90\%$ and $50\%$ of the distribution peaks. The black solid contours show the distribution of all clusters which have critical lines for sources at $z_{\rm s}=2$. The blue-dotted and the red-dashed contours show the distributions of clusters with minimal lensing cross sections for giant arcs $\sigma=0$ and $\sigma=10^{-3}h^{-2}$Mpc$^2$, respectively. Finally, the blue-solid contours indicate the distribution of clusters identified as ``mergers", i.e. for which a variation of mass $>30\%$ between consecutive snapshots has been measured in the projected region of radius $R_{2500}$ around the cluster center.}
\label{fig:virHyd}
\end{figure*}
The results at different radii are consistent with the previous finding of \cite{RA04.1}: the excess of pressure support due to gas bulk motions amounts to $\sim 15-20\%$ at $R_{\rm 500}$, and decreases towards the center. We do not see a clear dependence of $\Gamma$ on the lensing cross section. Note however that, while $\beta$ is a {\em global} indicator of departure from virial equilibrium, i.e. it is sensitive to any excess of kinetic energy within the virial radius, $\Gamma$ is only a {\em local} indicator of hydrostatic equilibrium, i.e. it can only be used to measure a local departure from hydrostatic equilibrium. If, as we believe, mergers play an important role for strong lensing, this parameter is less efficient to capture them. 

It is interesting to see where the clusters in our sample are located in the $\beta-\Gamma$ plane. This is shown in Fig.~\ref{fig:virHyd} for $\Gamma$ measured at $R_{500}$ and for four different redshift bins. Similar results would be seen using $\Gamma_{2500}$ instead of $\Gamma_{500}$. The contours in each panel show the distributions of clusters in different sub-samples: critical lenses (black solid lines), clusters with non-vanishing cross sections for giant arcs (dotted lines), and clusters with large lensing cross sections (dashed lines). The inner and the outer contours indicate the $90\%$ and the $50\%$ of the distribution peaks. As the lens strength increases, the distributions shift along the $\Gamma$ axes towards smaller $\beta$s. The separations between the distribution peaks become increasingly larger with redshift, indicating that the dynamical activity of clusters, as highlighted by the departure of these structures from virial equilibrium, has a strong impact on their ability to produce large distortions especially at high redshift. This is expected since the dynamical activity in clusters grows by going back in time.

\begin{figure}[t!]
\begin{center}
  \includegraphics[width=\hsize]{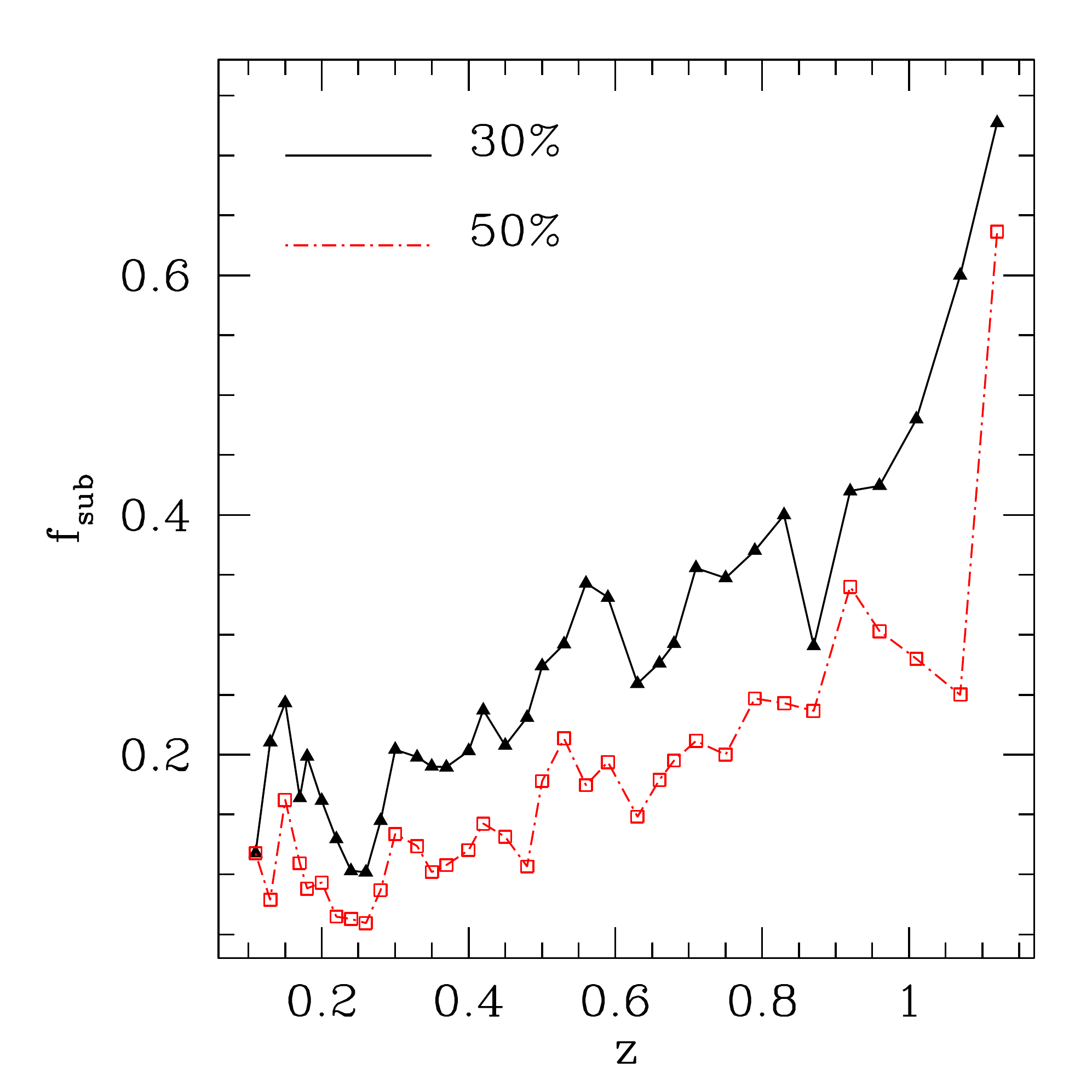}
\end{center}
\caption{The fraction $f_\mathrm{sub}$ of structures that, at each redshift step, change the value of the mass inside at least a rectangular volume (see text for details) of at least $30\%$ (black filled triangles) and $50\%$ (red empty squares) with respect to the previous step, as labelled in the plot.}
\label{fig:strZ}
\end{figure}

\subsection{Mass fluctuations in the cluster core}\label{sct:fluct}

In order to verify once more our interpretation of the results discussed above, we perform an additional test to verify that the clusters with increasingly negative values of $\beta$ are indeed clusters with infalling substructures. 
Considering  the potential impact of cluster mergers on strong lensing, we shall take into account not only those sub-structures that merge with the main cluster clump and start orbiting around it at small distance from its center, but also those sub-clumps that orbit at large distances but which transit close to cluster core perpendicularly to the line of sight. 
Thus, for each cluster projection we reconstruct the mass accretion history by measuring the projected mass within $R_{2500}$ at different epochs.  Sudden jumps in the curves $M_{p,2500}(z)$ signal the passage of some substructure across the cluster central region. We analyze these curves as follows. First, we smooth the curves by removing the peaks with a boxcar smoothing algorithm. This provides us a smooth accretion mass history $M_{\rm smooth}(z)$. Then, we compare the un-smoothed curve $M_{p,2500}(z)$ to the smoothed one, and we identify as mergers those events where  $M_{p,2500}(z)>K \times M_{\rm smooth}(z)$. For this analysis we choose $K=1.3$ and $K=1.5$.

In Fig.~\ref{fig:strZ} we show the redshift evolution of the fraction $f_\mathrm{sub}$ of clusters exhibiting these temporal mass variations compared to the total sample. As expected, the fraction of clusters with large projected mass variations in the core tends to increase with increasing redshift. The fraction of clusters where the excess of mass within $R_{2500}$ is at least $30\%$ varies from $\sim 0.1$ at $z\simeq 0.1$ to up to $\sim 0.7$ at $z\gtrsim 1$.

In Fig.~\ref{fig:virHyd}, the distribution of the "merging" clusters in the $\beta-\Gamma$ plane is given by the blue solid contours. Clearly, merging clusters are typically objects with extremely negative $\beta$ parameters. Their $\Gamma$ parameters do not significantly differ from those of clusters with critical lines, confirming that this parameter is not a good indicator of substructure accretion. Interestingly, the distributions of "merging" clusters in the four redshift bins nicely overlay the distributions of clusters with large lensing cross sections for giant arcs. This confirms our interpretation that the most efficient strong lenses in the {\sc MareNostrum Universe} are dynamically active clusters, with substructures located near the cluster core. Thanks to their contribution of the shear and to the overall convergence, these substructures boost the strong lensing cross sections.

\begin{figure*}[lt!]
\begin{center}
  \includegraphics[width=0.49\hsize]{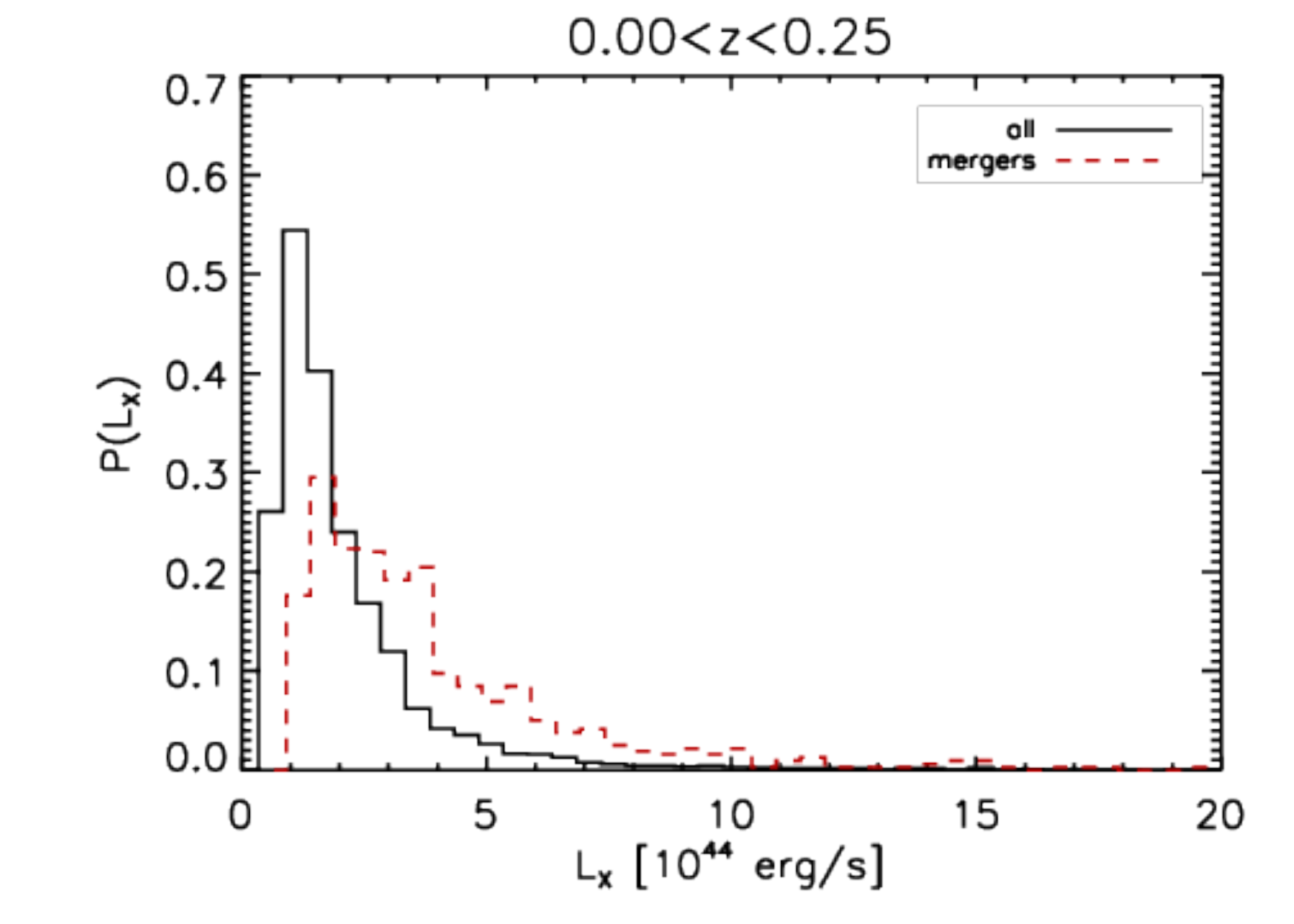}
  \includegraphics[width=0.49\hsize]{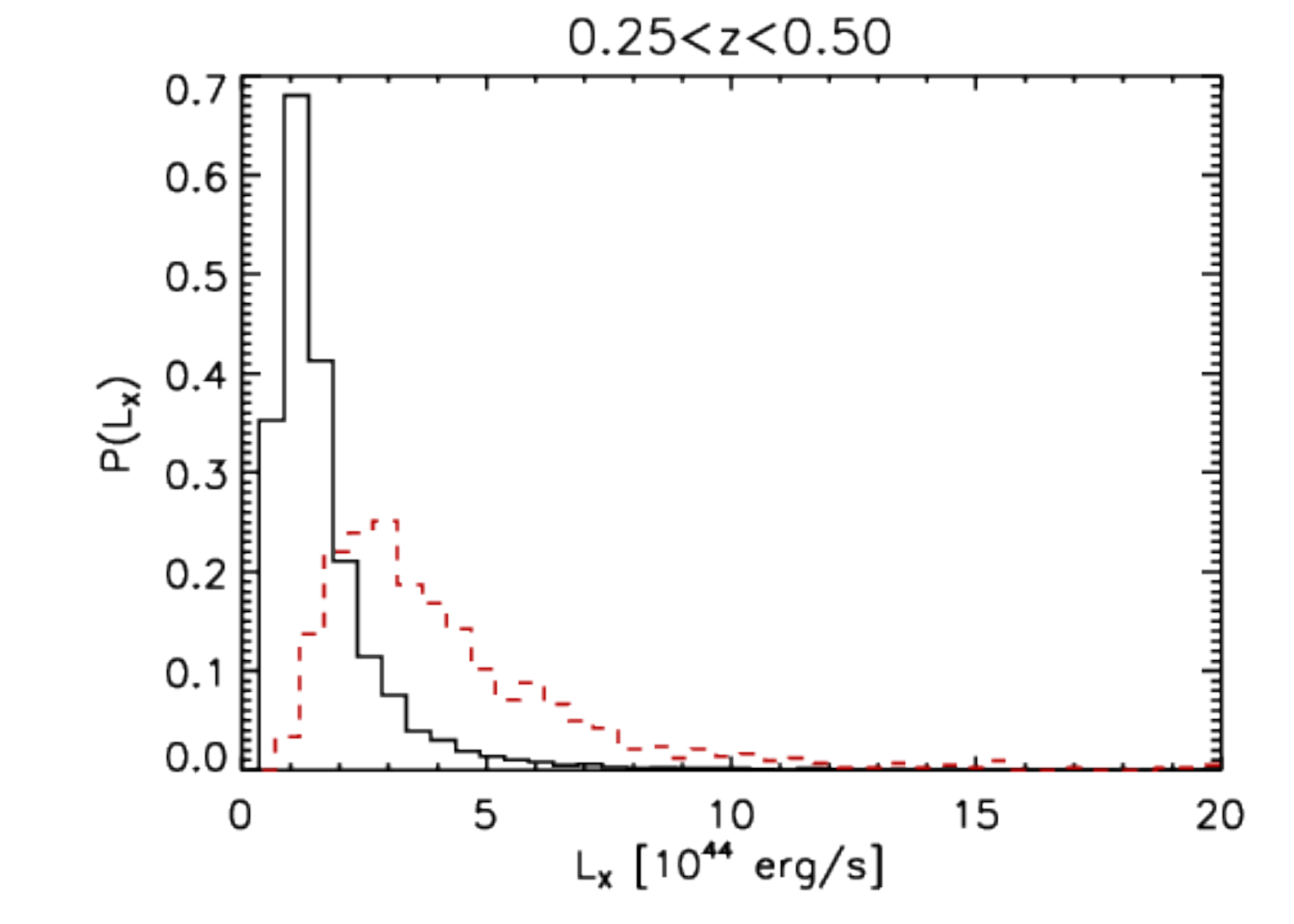}
  \includegraphics[width=0.49\hsize]{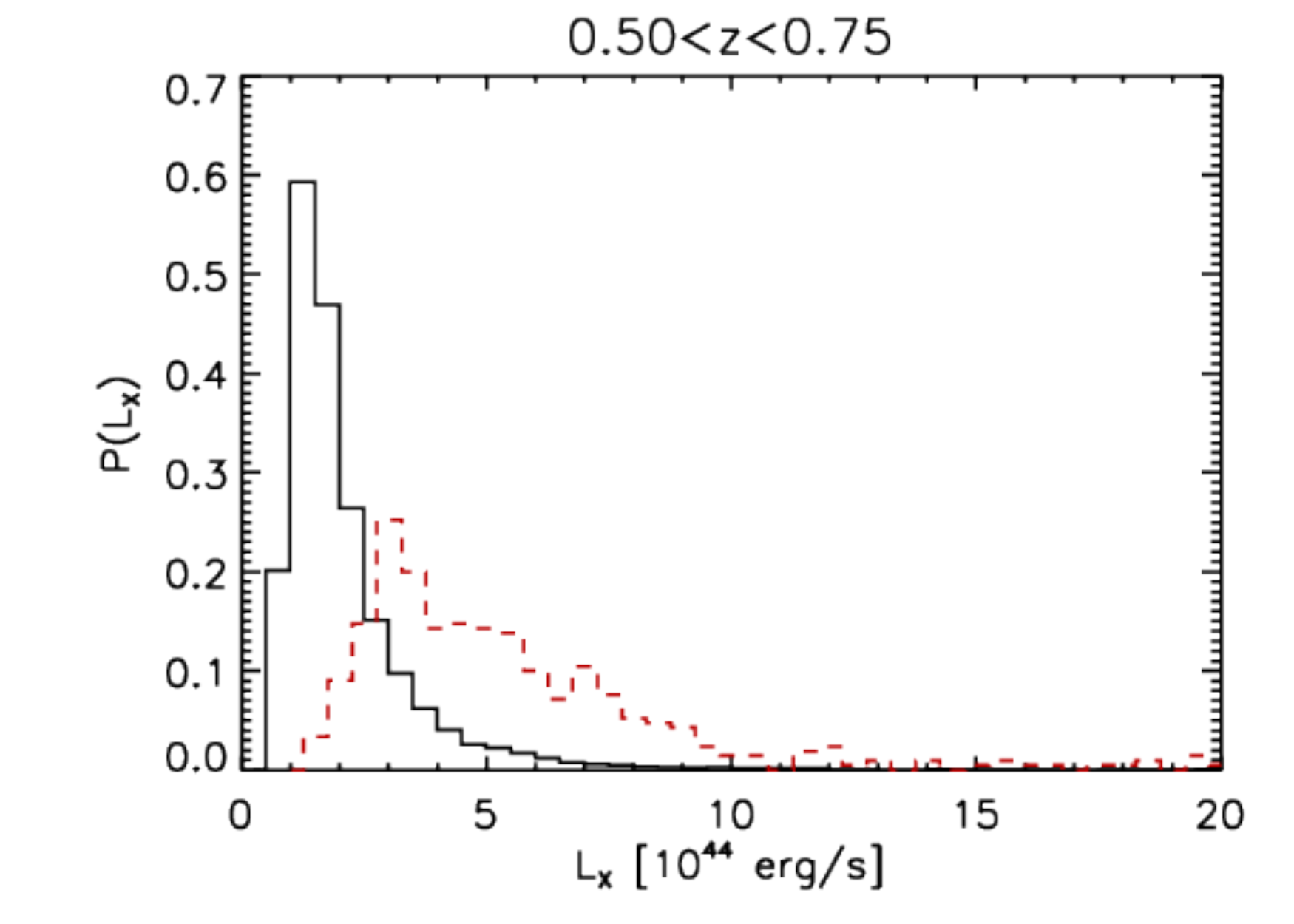}
  \includegraphics[width=0.49\hsize]{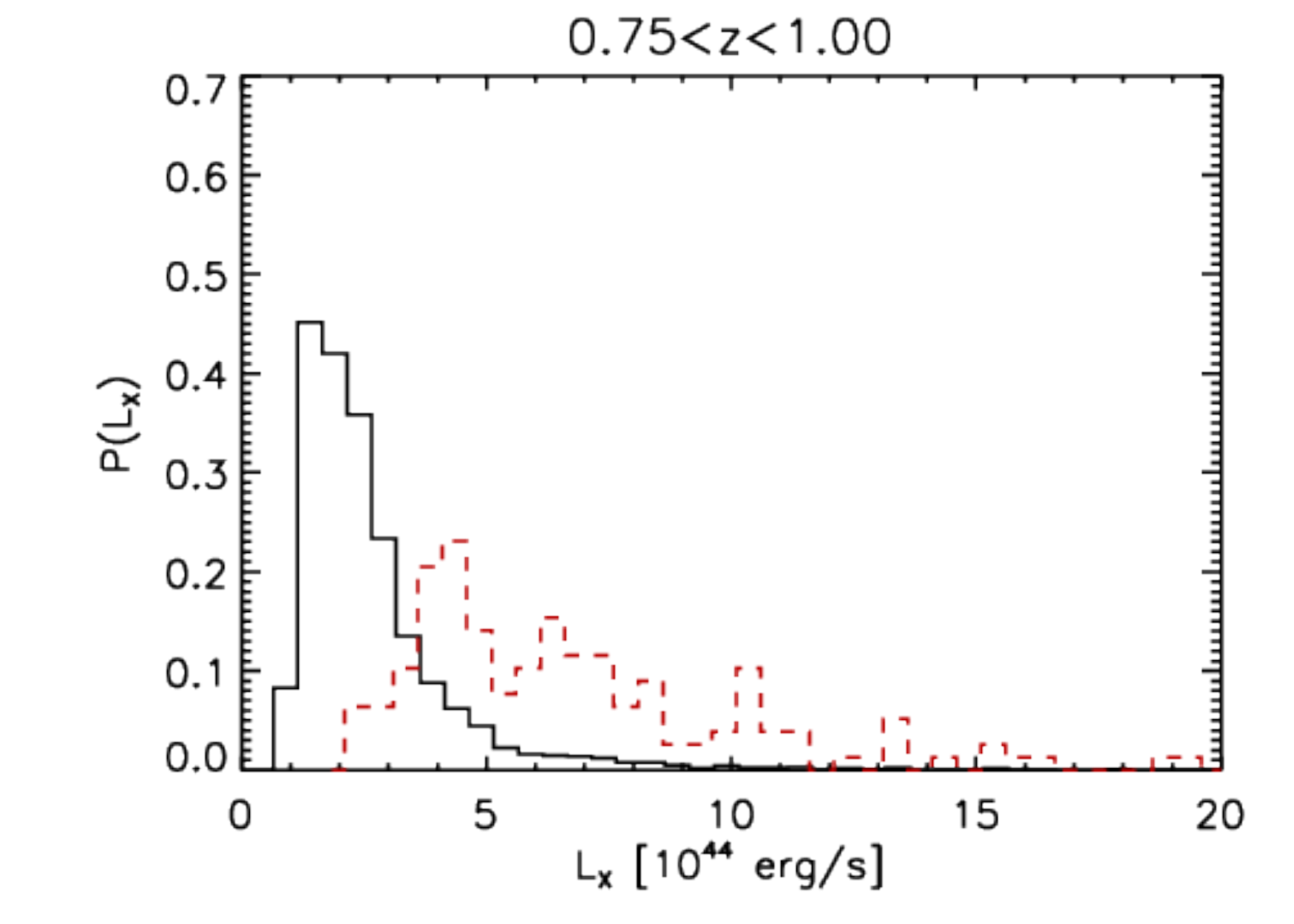}
\end{center}
\caption{The probability distributions of X-ray luminosities of all clusters in the {\sc MareNostrum Universe} (solid histograms) are compared to those of merging clusters (dashed histograms). Each panel refers to different redshift bins, as specified in the title above each figure.}
\label{fig:lxmerge}
\end{figure*}

\begin{figure*}[lt!]
\begin{center}
  \includegraphics[width=0.49\hsize]{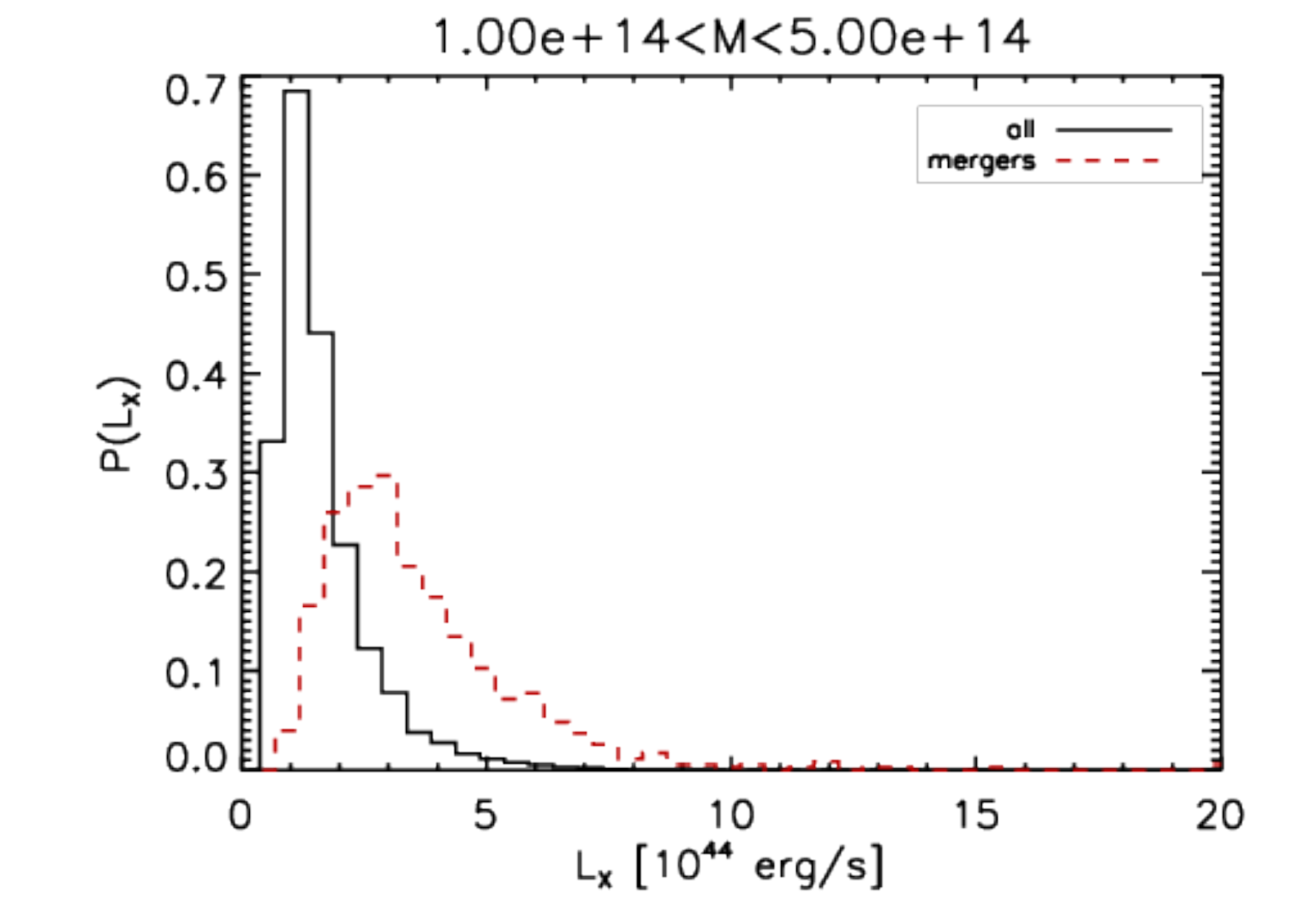}
  \includegraphics[width=0.49\hsize]{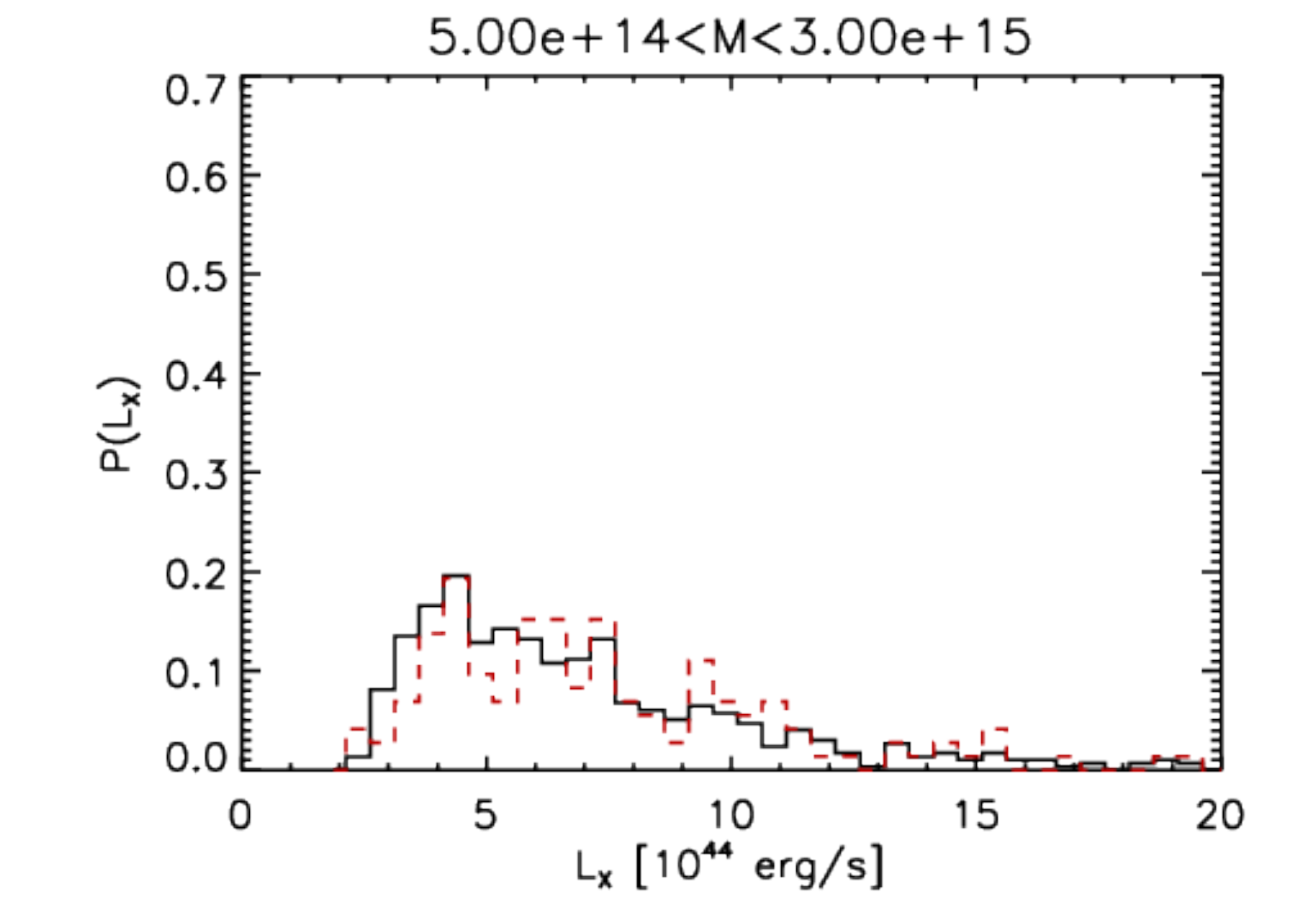}
\end{center}
\caption{As in Fig.~\ref{fig:lxmerge} but showing the X-ray luminosity distributions of general and merging clusters in the redshift range $0.25<z<0.5$. In the left panel we select the clusters with mass $10^{14}h^{-1}M_\odot<M_{\rm vir}\leq 5\times 10^{14}h^{-1}M_\odot$. The distributions in the right panel refer to clusters with mass $5\times 10^{14}h^{-1}M_\odot<M_{\rm vir}\leq 3\times 10^{15}h^{-1}M_\odot$ (right panel).}
\label{fig:lxmerge_mass}
\end{figure*}

{  In Fig.~\ref{fig:lxmerge} we compare the distributions of the X-ray luminosities of all clusters in the {\sc MareNostrum Universe} (solid histograms) and of the sub-sample of merging clusters (dashed histograms). Clearly, the clusters ongoing a merging phase have higher X-ray luminosities compared to the general cluster population \citep[see also][]{2004MNRAS.352..508R}. The differences between the distributions increase with redshift, indicating that the impact of mergers is larger at higher redshifts, where clusters have smaller masses. In Fig.~\ref{fig:lxmerge_mass} we show the X-ray luminosity distributions of general and merging clusters in the redshift range $0.25 < z \leq 0.5$. In the left panel, we select the clusters with masses  $10^{14}h^{-1}M_\odot<M_{\rm vir}\leq 5\times 10^{14}h^{-1}M_\odot$. Among these relatively small mass clusters, those ongoing a merging phase are significantly more X-ray luminous. In the right panel, we re-plot the distributions after selecting only the most massive clusters ($5\times 10^{14}h^{-1}M_\odot<M_{\rm vir}\leq 3\times 10^{15}h^{-1}M_\odot$). In this case, the X-ray luminosity distributions are almost identical, indicating that most of the differences between merging and general clusters appear at low masses. The results shown here help explaining the behavior of the luminosity-mass relations displayed in Fig.~\ref{fig:lumM}. As the lens strength increases, the strong lensing cluster population tends to be dominated by merging clusters, which are characterized by higher X-ray luminosities. For this reason, the $L_X-M$ relation tilts at the smallest masses, because clusters with high lensing efficiency and small mass are typically merging objects.} 

\section{Summary and conclusions}

In this paper, we have investigated the properties of $\sim 50000$ strong lensing clusters at different redshifts in the {\sc MareNostrum Universe} cosmological simulation. Projecting each of these clusters along three orthogonal lines of sight, we have considered almost $150000$ lens realizations in total. With such a big number of objects, we can statistically characterize the population of strong lensing clusters much better than it was done in the past. Moreover, the {\sc MareNostrum Universe} includes gas, thus it allows to investigate the correlation between strong lensing and X-ray observables. We have classified the strong lenses into two categories, namely clusters which are critical for sources at $z_{\rm s}=2$ and clusters which can induce large distortions in the images of these sources, i.e. form giant arcs. We have explored several structural properties of the strong lensing clusters, like the masse, the shapes, the concentrations, the X-ray luminosity, and the dynamical activity.

Our main results can be summarized as follows:
\begin{itemize}

	\item strong lensing clusters are typically massive objects. Their masses vary over two orders of magnitude. The minimal mass for strong lensing depends on both the redshift of  the lenses and of the sources. For sources at $z_{\rm s}=2$, we find that clusters can develop critical lines down to masses of $\sim 10^{13}\;h^{-1}\;M_\odot$. Requiring that clusters are also able to produce giant arcs, increases the mass limit by almost one order of magnitude. The lensing cross section further depends on the mass, so that we can estimate that the minimal mass required for a cluster, in order that the expected number of giant arcs in a deep HST observation is $\sim 1$, is $\sim 2 \times 10^{14}\;h^{-1}\;M_\odot$;
	
	\item the three-dimensional shape of strong lensing clusters does not seem to be significantly different from that of the general cluster population. However, strong lensing clusters tend to have their major axes oriented along the line of sight. This ``orientation bias" is larger for clusters which produce giant arcs than for clusters which only possess critical lines, and it becomes larger by increasing the lensing cross section;
	
	\item due to the orientation bias and to the fact that their halos are generally well described by prolate triaxial models, strong lensing clusters tend to appear rounder when projected on the sky. However, zooming over their central regions, we noticed that their projected mass maps are described by rather elongated distributions, evidencing the presence of substructures projected near cluster cores;
	
	\item the concentrations measured by fitting the density profiles of strong lensing clusters stacked in mass and redshift bins do not differ significantly from the concentrations of the general cluster population. Nevertheless, due to the orientation bias, the concentrations of the same objects inferred from the analysis of the projected density profiles are generally larger than in 3D. For clusters with large lensing cross sections for giant arcs, the 2D-concentrations can be larger by more than a factor of two. These results may provide a viable explanation of the large concentrations reported for some of the strongest lenses observed so far;
	
	\item the X-ray luminosity-mass relation of strong lensing clusters is likely to differ from that of  the general cluster population, especially at the lowest masses. We found that at a fixed mass, strong lenses with increasingly larger cross sections for lensing have higher X-ray luminosities, indicating that some process occurring in these objects enhances both the X-ray luminosity and the strong lensing cross section;
	
	\item the strong lensing efficiency is certainly correlated with the dynamical activity in clusters. We found that clusters with large lensing cross sections are characterized by a systematic departure from virial equilibrium. A similar departure from virial equilibrium is found for clusters which are in the process of accreting substructures, which accidentally transit across their cores perpendicularly to the line of sight.	 		
\end{itemize}

In conclusion, our results show that strong lensing clusters are likely to be a very peculiar class of objects, characterized by several selection biases, which need to be properly taken into account in many applications. For example, due to the orientation bias, it is very likely that the 3D-masses inferred from strong lensing models of observed clusters are biased high. In arc statistics studies, the statistical modeling of the strong lensing cluster population need to include mergers, triaxiality, and asymmetries in the projected mass distributions, as also suggested by previous studies. Due to the intrinsic difficulties at modeling analytically all of these effects, numerical simulations again seem to be the only viable way to describe strong lensing clusters. In this sense, large statistical samples of numerically simulated lenses, as that extracted from the {\sc MareNostrum Universe}, are fundamental tools for interpreting the current strong lensing observations. A major effort is now necessary to clarify the existing inconsistencies between the properties of simulated and observed galaxy clusters, especially in the central regions.  

\acknowledgements{This work was supported by the HPC-Europa2 Transnational Access programme. We acknowledge financial contributions from contracts ASI-INAF I/023/05/0, ASI-INAF I/088/06/0 and INFN PD51. We warmly thank L. Moscardini and M. Bartelmann for their considerable input to this paper, M. Roncarelli for aid in the computation of X-ray luminosities and E. Puchwein for providing us the code for the computation of hydrostatic equilibrium parameters. The {\sc MareNostrum Universe} simulation has been done at the BSC-CNS
(Spain) and analyzed at NIC J\"ulich (Germany). SG acknowledges the
support of the European Science Foundation through the ASTROSIM
Exchange Visits Program. MM acknowledges the support from the HPC Europa program and thanks the Institut f\"ur Theoretische Astrophysik of the University of Heidelberg for the hospitality during the preparation of this work. GY would like to thank also the Spanish  Ministry of Science and Technology for financial support under
project numbers  FPA2006-01105  and AYA2006-15492-C03.}

\bibliography{./master}

\begin{thebibliography}{100}
\expandafter\ifx\csname natexlab\endcsname\relax\def\natexlab#1{#1}\fi

\bibitem[{{Ameglio} {et~al.}(2009){Ameglio}, {Borgani}, {Pierpaoli}, {Dolag},
  {Ettori}, \& {Morandi}}]{2009MNRAS.394..479A}
{Ameglio}, S., {Borgani}, S., {Pierpaoli}, E., {et~al.} 2009, \mnras, 394, 479

\bibitem[{{Arnaud}(1996)}]{AR96.1}
{Arnaud}, K.~A. 1996, in Astronomical Society of the Pacific Conference Series,
  Vol. 101, Astronomical Data Analysis Software and Systems V, ed. G.~H.
  {Jacoby} \& J.~{Barnes}, 17--+

\bibitem[{{Ascasibar} {et~al.}(2003){Ascasibar}, {Yepes}, {M{\"u}ller}, \&
  {Gottl{\"o}ber}}]{2003MNRAS.346..731A}
{Ascasibar}, Y., {Yepes}, G., {M{\"u}ller}, V., \& {Gottl{\"o}ber}, S. 2003,
  \mnras, 346, 731

\bibitem[{Bartelmann {et~al.}(1998)Bartelmann, Huss, Colberg, Jenkins, \&
  Pearce}]{BA98.2}
Bartelmann, M., Huss, A., Colberg, J., Jenkins, A., \& Pearce, F. 1998, A\&A,
  330, 1

\bibitem[{Bartelmann \& Meneghetti(2004)}]{BA04.1}
Bartelmann, M. \& Meneghetti, M. 2004, A\&A, 418, 413

\bibitem[{Bartelmann \& Schneider(2001)}]{BA01.1}
Bartelmann, M. \& Schneider, P. 2001, Physics Reports, 340, 291

\bibitem[{Bartelmann \& White(2003)}]{BA03.4}
Bartelmann, M. \& White, S. 2003, A\&A, 407, 845

\bibitem[{{Beckwith} {et~al.}(2006){Beckwith}, {Stiavelli}, {Koekemoer},
  {Caldwell}, {Ferguson}, {Hook}, {Lucas}, {Bergeron}, {Corbin}, {Jogee},
  {Panagia}, {Robberto}, {Royle}, {Somerville}, \&
  {Sosey}}]{2006AJ....132.1729B}
{Beckwith}, S.~V.~W., {Stiavelli}, M., {Koekemoer}, A.~M., {et~al.} 2006, \aj,
  132, 1729

\bibitem[{{Benjamin} {et~al.}(2007){Benjamin}, {Heymans}, {Semboloni}, {van
  Waerbeke}, {Hoekstra}, {Erben}, {Gladders}, {Hetterscheidt}, {Mellier}, \&
  {Yee}}]{2007MNRAS.381..702B}
{Benjamin}, J., {Heymans}, C., {Semboloni}, E., {et~al.} 2007, \mnras, 381, 702

\bibitem[{{Borgani} {et~al.}(2008){Borgani}, {Diaferio}, {Dolag}, \&
  {Schindler}}]{2008SSRv..134..269B}
{Borgani}, S., {Diaferio}, A., {Dolag}, K., \& {Schindler}, S. 2008, Space
  Science Reviews, 134, 269

\bibitem[{{Borgani} {et~al.}(2004){Borgani}, {Murante}, {Springel}, {Diaferio},
  {Dolag}, {Moscardini}, {Tormen}, {Tornatore}, \&
  {Tozzi}}]{2004MNRAS.348.1078B}
{Borgani}, S., {Murante}, G., {Springel}, V., {et~al.} 2004, \mnras, 348, 1078

\bibitem[{{Brada{\v c}} {et~al.}(2005){Brada{\v c}}, {Erben}, {Schneider},
  {Hildebrandt}, {Lombardi}, {Schirmer}, {Miralles}, {Clowe}, \&
  {Schindler}}]{2005A&A...437...49B}
{Brada{\v c}}, M., {Erben}, T., {Schneider}, P., {et~al.} 2005, \aap, 437, 49

\bibitem[{{Broadhurst} {et~al.}(2008){Broadhurst}, {Umetsu}, {Medezinski},
  {Oguri}, \& {Rephaeli}}]{2008ApJ...685L...9B}
{Broadhurst}, T., {Umetsu}, K., {Medezinski}, E., {Oguri}, M., \& {Rephaeli},
  Y. 2008, \apjl, 685, L9

\bibitem[{{Bryan} \& {Norman}(1998)}]{1998ApJ...495...80B}
{Bryan}, G.~L. \& {Norman}, M.~L. 1998, \apj, 495, 80

\bibitem[{{Cacciato} {et~al.}(2006){Cacciato}, {Bartelmann}, {Meneghetti}, \&
  {Moscardini}}]{2006A&A...458..349C}
{Cacciato}, M., {Bartelmann}, M., {Meneghetti}, M., \& {Moscardini}, L. 2006,
  \aap, 458, 349

\bibitem[{{Clowe} {et~al.}(2006){Clowe}, {Schneider}, {Arag{\'o}n-Salamanca},
  {Bremer}, {De Lucia}, {Halliday}, {Jablonka}, {Milvang-Jensen}, {Pell{\'o}},
  {Poggianti}, {Rudnick}, {Saglia}, {Simard}, {White}, \&
  {Zaritsky}}]{2006A&A...451..395C}
{Clowe}, D., {Schneider}, P., {Arag{\'o}n-Salamanca}, A., {et~al.} 2006, \aap,
  451, 395

\bibitem[{{Coe} {et~al.}(2006){Coe}, {Ben{\'{\i}}tez}, {S{\'a}nchez}, {Jee},
  {Bouwens}, \& {Ford}}]{2006AJ....132..926C}
{Coe}, D., {Ben{\'{\i}}tez}, N., {S{\'a}nchez}, S.~F., {et~al.} 2006, \aj, 132,
  926

\bibitem[{{Dahle}(2006)}]{2006ApJ...653..954D}
{Dahle}, H. 2006, \apj, 653, 954

\bibitem[{{Diego} {et~al.}(2005){Diego}, {Sandvik}, {Protopapas}, {Tegmark},
  {Ben{\'{\i}}tez}, \& {Broadhurst}}]{2005MNRAS.362.1247D}
{Diego}, J.~M., {Sandvik}, H.~B., {Protopapas}, P., {et~al.} 2005, \mnras, 362,
  1247

\bibitem[{Dolag {et~al.}(2004)Dolag, Jubelgas, Springel, Borgani, \&
  Rasia}]{DO04.1}
Dolag, K., Jubelgas, M., Springel, V., Borgani, S., \& Rasia, E. 2004, ApJL,
  606, L97

\bibitem[{{Fedeli} \& {Bartelmann}(2007)}]{FE07.1}
{Fedeli}, C. \& {Bartelmann}, M. 2007, \aap, 474, 355

\bibitem[{{Fedeli} {et~al.}(2007{\natexlab{a}}){Fedeli}, {Bartelmann},
  {Meneghetti}, \& {Moscardini}}]{2007A&A...473..715F}
{Fedeli}, C., {Bartelmann}, M., {Meneghetti}, M., \& {Moscardini}, L.
  2007{\natexlab{a}}, \aap, 473, 715

\bibitem[{{Fedeli} {et~al.}(2007{\natexlab{b}}){Fedeli}, {Bartelmann},
  {Meneghetti}, \& {Moscardini}}]{FE07.2}
{Fedeli}, C., {Bartelmann}, M., {Meneghetti}, M., \& {Moscardini}, L.
  2007{\natexlab{b}}, \aap, 473, 715

\bibitem[{Fedeli {et~al.}(2006)Fedeli, Meneghetti, Bartelmann, Dolag, \&
  Moscardini}]{FE06.1}
Fedeli, C., Meneghetti, M., Bartelmann, M., Dolag, K., \& Moscardini, L. 2006,
  A\&A, 447, 419

\bibitem[{Flores {et~al.}(2000)Flores, Maller, \& Primack}]{FL00.1}
Flores, R., Maller, A., \& Primack, J. 2000, ApJ, 535, 555

\bibitem[{{Fu} {et~al.}(2008){Fu}, {Semboloni}, {Hoekstra}, {Kilbinger}, {van
  Waerbeke}, {Tereno}, {Mellier}, {Heymans}, {Coupon}, {Benabed}, {Benjamin},
  {Bertin}, {Dor{\'e}}, {Hudson}, {Ilbert}, {Maoli}, {Marmo}, {McCracken}, \&
  {M{\'e}nard}}]{2008A&A...479....9F}
{Fu}, L., {Semboloni}, E., {Hoekstra}, H., {et~al.} 2008, \aap, 479, 9

\bibitem[{Fukazawa {et~al.}(1998)Fukazawa, Makishima, Tamura, Ezawa, Xu, Ikebe,
  Kikuchi, \& Ohashi}]{FU98.1}
Fukazawa, Y., Makishima, K., Tamura, T., {et~al.} 1998, PASJ, 50, 187

\bibitem[{{Gao} {et~al.}(2009){Gao}, {Jing}, {Mao}, {Li}, \&
  {Kong}}]{2009arXiv0910.4013G}
{Gao}, G.~J., {Jing}, Y.~P., {Mao}, S., {Li}, G.~L., \& {Kong}, X. 2009, ArXiv
  e-prints

\bibitem[{{Gavazzi}(2005)}]{2005A&A...443..793G}
{Gavazzi}, R. 2005, \aap, 443, 793

\bibitem[{{Gavazzi} {et~al.}(2003){Gavazzi}, {Fort}, {Mellier}, {Pell{\'o}}, \&
  {Dantel-Fort}}]{2003A&A...403...11G}
{Gavazzi}, R., {Fort}, B., {Mellier}, Y., {Pell{\'o}}, R., \& {Dantel-Fort}, M.
  2003, \aap, 403, 11

\bibitem[{Gladders {et~al.}(2003)Gladders, Hoekstra, Yee, Hall, \&
  Barrientos}]{GL03.1}
Gladders, M., Hoekstra, H., Yee, H., Hall, P., \& Barrientos, L. 2003, ApJ,
  593, 48

\bibitem[{{Gottl{\"o}ber} \& {Yepes}(2007)}]{2007ApJ...664..117G}
{Gottl{\"o}ber}, S. \& {Yepes}, G. 2007, \apj, 664, 117

\bibitem[{{Gottl\"ober} {et~al.}(2008){Gottl\"ober}, {Yepes}, {Wagner}, \&
  {Sevilla}}]{2006astro.ph..8289G}
{Gottl\"ober}, S., {Yepes}, G., {Wagner}, C., \& {Sevilla}, R. 2008, in
  Proceedings of the XLIst Rencontres de Moriond, XXVIth Astrophysics Moriond
  Meeting, Vol.~1, From dark halos to light, ed. S.~{Maurogordato}, J.~{Tran
  Than Van}, \& L.~{Tresse}, 309

\bibitem[{{Gunn} \& {Gott}(1972)}]{gunngott72}
{Gunn}, J.~E. \& {Gott}, J.~R.~I. 1972, \apj, 176, 1

\bibitem[{{Hennawi} {et~al.}(2007){Hennawi}, {Dalal}, {Bode}, \&
  {Ostriker}}]{HE07.1}
{Hennawi}, J.~F., {Dalal}, N., {Bode}, P., \& {Ostriker}, J.~P. 2007, \apj,
  654, 714

\bibitem[{{Hennawi} {et~al.}(2008){Hennawi}, {Gladders}, {Oguri}, {Dalal},
  {Koester}, {Natarajan}, {Strauss}, {Inada}, {Kayo}, {Lin}, {Lampeitl},
  {Annis}, {Bahcall}, \& {Schneider}}]{2008AJ....135..664H}
{Hennawi}, J.~F., {Gladders}, M.~D., {Oguri}, M., {et~al.} 2008, \aj, 135, 664

\bibitem[{{Hilbert} {et~al.}(2008){Hilbert}, {White}, {Hartlap}, \&
  {Schneider}}]{2008MNRAS.386.1845H}
{Hilbert}, S., {White}, S.~D.~M., {Hartlap}, J., \& {Schneider}, P. 2008,
  \mnras, 386, 1845

\bibitem[{Hockney \& Eastwood(1988)}]{HO88.1}
Hockney, R. \& Eastwood, J. 1988, Computer simulation using particles (Bristol:
  Hilger, 1988)

\bibitem[{{Hoekstra}(2007)}]{2007MNRAS.379..317H}
{Hoekstra}, H. 2007, \mnras, 379, 317

\bibitem[{{Jelinsky} \& {SNAP Collaboration}(2006)}]{2006AAS...209.9809J}
{Jelinsky}, P. \& {SNAP Collaboration}. 2006, in Bulletin of the American
  Astronomical Society, Vol.~38, Bulletin of the American Astronomical Society,
  1039--+

\bibitem[{Jing \& Suto(2000)}]{JI00.2}
Jing, Y. \& Suto, Y. 2000, ApJL, 529, L69

\bibitem[{Jing \& Suto(2002)}]{JI02.1}
Jing, Y. \& Suto, Y. 2002, ApJ, 574, 538

\bibitem[{{Kaiser}(1986)}]{1986MNRAS.222..323K}
{Kaiser}, N. 1986, \mnras, 222, 323

\bibitem[{{Kaiser}(2007)}]{2007AAS...210.5102K}
{Kaiser}, N. 2007, in Bulletin of the American Astronomical Society, Vol.~38,
  Bulletin of the American Astronomical Society, 163--+

\bibitem[{Keeton(2001)}]{KE01.2}
Keeton, C. 2001, ApJ, 562, 160

\bibitem[{{Klypin} {et~al.}(1999){Klypin}, {Gottl{\"o}ber}, {Kravtsov}, \&
  {Khokhlov}}]{Klypin99}
{Klypin}, A., {Gottl{\"o}ber}, S., {Kravtsov}, A.~V., \& {Khokhlov}, A.~M.
  1999, \apj, 516, 530

\bibitem[{{Kneib} {et~al.}(2003){Kneib}, {Hudelot}, {Ellis}, {Treu}, {Smith},
  {Marshall}, {Czoske}, {Smail}, \& {Natarajan}}]{2003ApJ...598..804K}
{Kneib}, J.-P., {Hudelot}, P., {Ellis}, R.~S., {et~al.} 2003, \apj, 598, 804

\bibitem[{{Komatsu} {et~al.}(2009){Komatsu}, {Dunkley}, {Nolta}, {Bennett},
  {Gold}, {Hinshaw}, {Jarosik}, {Larson}, {Limon}, {Page}, {Spergel},
  {Halpern}, {Hill}, {Kogut}, {Meyer}, {Tucker}, {Weiland}, {Wollack}, \&
  {Wright}}]{wmap5}
{Komatsu}, E., {Dunkley}, J., {Nolta}, M.~R., {et~al.} 2009, \apjs, 180, 330

\bibitem[{Li {et~al.}(2005)Li, Mao, Jing, Bartelmann, Kang, \&
  Meneghetti}]{LI05.1}
Li, G., Mao, S., Jing, Y., {et~al.} 2005, ApJ, 635, 795L

\bibitem[{{Liedahl} {et~al.}(1995){Liedahl}, {Osterheld}, \&
  {Goldstein}}]{LI95.2}
{Liedahl}, D.~A., {Osterheld}, A.~L., \& {Goldstein}, W.~H. 1995, \apjl, 438,
  L115

\bibitem[{{Limousin} {et~al.}(2007){Limousin}, {Richard}, {Jullo}, {Kneib},
  {Fort}, {Soucail}, {El{\'{\i}}asd{\'o}ttir}, {Natarajan}, {Ellis}, {Smail},
  {Czoske}, {Smith}, {Hudelot}, {Bardeau}, {Ebeling}, {Egami}, \&
  {Knudsen}}]{2007ApJ...668..643L}
{Limousin}, M., {Richard}, J., {Jullo}, E., {et~al.} 2007, \apj, 668, 643

\bibitem[{{Limousin} {et~al.}(2008){Limousin}, {Richard}, {Kneib}, {Brink},
  {Pell{\'o}}, {Jullo}, {Tu}, {Sommer-Larsen}, {Egami}, {Micha{\l}owski},
  {Cabanac}, \& {Stark}}]{2008A&A...489...23L}
{Limousin}, M., {Richard}, J., {Kneib}, J.-P., {et~al.} 2008, \aap, 489, 23

\bibitem[{Luppino {et~al.}(1999)Luppino, Gioia, Hammer, {Le F{\` e}vre}, \&
  Annis}]{LU99.1}
Luppino, G., Gioia, I., Hammer, F., {Le F{\` e}vre}, O., \& Annis, J. 1999,
  A\&AS, 136, 117

\bibitem[{{Mantz} {et~al.}(2008){Mantz}, {Allen}, {Ebeling}, \&
  {Rapetti}}]{2008MNRAS.387.1179M}
{Mantz}, A., {Allen}, S.~W., {Ebeling}, H., \& {Rapetti}, D. 2008, \mnras, 387,
  1179

\bibitem[{{Mao} {et~al.}(2004){Mao}, {Jing}, {Ostriker}, \&
  {Weller}}]{2004ApJ...604L...5M}
{Mao}, S., {Jing}, Y., {Ostriker}, J.~P., \& {Weller}, J. 2004, \apjl, 604, L5

\bibitem[{{Markevitch} {et~al.}(2004){Markevitch}, {Gonzalez}, {Clowe},
  {Vikhlinin}, {Forman}, {Jones}, {Murray}, \& {Tucker}}]{2004ApJ...606..819M}
{Markevitch}, M., {Gonzalez}, A.~H., {Clowe}, D., {et~al.} 2004, \apj, 606, 819

\bibitem[{{Massey} {et~al.}(2007){Massey}, {Rhodes}, {Leauthaud}, {Capak},
  {Ellis}, {Koekemoer}, {R{\'e}fr{\'e}gier}, {Scoville}, {Taylor}, {Albert},
  {Berg{\'e}}, {Heymans}, {Johnston}, {Kneib}, {Mellier}, {Mobasher},
  {Semboloni}, {Shopbell}, {Tasca}, \& {Van Waerbeke}}]{2007ApJS..172..239M}
{Massey}, R., {Rhodes}, J., {Leauthaud}, A., {et~al.} 2007, \apjs, 172, 239

\bibitem[{{Mead} {et~al.}(2010){Mead}, {King}, {Sijacki}, {Leonard},
  {Puchwein}, \& {McCarthy}}]{2010arXiv1001.2281M}
{Mead}, J.~M.~G., {King}, L.~J., {Sijacki}, D., {et~al.} 2010, ArXiv e-prints

\bibitem[{{Meneghetti} {et~al.}(2007{\natexlab{a}}){Meneghetti}, {Argazzi},
  {Pace}, {Moscardini}, {Dolag}, {Bartelmann}, {Li}, \& {Oguri}}]{ME07.1}
{Meneghetti}, M., {Argazzi}, R., {Pace}, F., {et~al.} 2007{\natexlab{a}}, \aap,
  461, 25

\bibitem[{Meneghetti {et~al.}(2005)Meneghetti, Bartelmann, Dolag, Moscardini,
  Perrotta, Baccigalupi, \& Tormen}]{ME05.1}
Meneghetti, M., Bartelmann, M., Dolag, K., {et~al.} 2005, A\&A, 442, 413

\bibitem[{{Meneghetti} {et~al.}(2007{\natexlab{b}}){Meneghetti}, {Bartelmann},
  {Jenkins}, \& {Frenk}}]{2007MNRAS.381..171M}
{Meneghetti}, M., {Bartelmann}, M., {Jenkins}, A., \& {Frenk}, C.
  2007{\natexlab{b}}, \mnras, 381, 171

\bibitem[{Meneghetti {et~al.}(2003{\natexlab{a}})Meneghetti, Bartelmann, \&
  Moscardini}]{ME03.2}
Meneghetti, M., Bartelmann, M., \& Moscardini, L. 2003{\natexlab{a}}, MNRAS,
  346, 67

\bibitem[{Meneghetti {et~al.}(2003{\natexlab{b}})Meneghetti, Bartelmann, \&
  Moscardini}]{ME03.1}
Meneghetti, M., Bartelmann, M., \& Moscardini, L. 2003{\natexlab{b}}, MNRAS,
  340, 105

\bibitem[{Meneghetti {et~al.}(2000)Meneghetti, Bolzonella, Bartelmann,
  Moscardini, \& Tormen}]{ME00.1}
Meneghetti, M., Bolzonella, M., Bartelmann, M., Moscardini, L., \& Tormen, G.
  2000, MNRAS, 314, 338

\bibitem[{{Meneghetti} {et~al.}(2005){Meneghetti}, {Jain}, {Bartelmann}, \&
  {Dolag}}]{2005MNRAS.362.1301M}
{Meneghetti}, M., {Jain}, B., {Bartelmann}, M., \& {Dolag}, K. 2005, \mnras,
  362, 1301

\bibitem[{{Meneghetti} {et~al.}(2008){Meneghetti}, {Melchior}, {Grazian}, {De
  Lucia}, {Dolag}, {Bartelmann}, {Heymans}, {Moscardini}, \&
  {Radovich}}]{2008A&A...482..403M}
{Meneghetti}, M., {Melchior}, P., {Grazian}, A., {et~al.} 2008, \aap, 482, 403

\bibitem[{{Meneghetti} {et~al.}(2009){Meneghetti}, {Rasia}, {Merten},
  {Bellagamba}, {Ettori}, {Mazzotta}, \& {Dolag}}]{ME09.1}
{Meneghetti}, M., {Rasia}, E., {Merten}, J., {et~al.} 2009, ArXiv e-prints

\bibitem[{Meneghetti {et~al.}(2001)Meneghetti, Yoshida, Bartelmann, Moscardini,
  Springel, Tormen, \& White}]{ME01.1}
Meneghetti, M., Yoshida, N., Bartelmann, M., {et~al.} 2001, MNRAS,, 325, 435

\bibitem[{{Merten} {et~al.}(2009){Merten}, {Cacciato}, {Meneghetti}, {Mignone},
  \& {Bartelmann}}]{2009A&A...500..681M}
{Merten}, J., {Cacciato}, M., {Meneghetti}, M., {Mignone}, C., \& {Bartelmann},
  M. 2009, \aap, 500, 681

\bibitem[{{Mewe} {et~al.}(1985){Mewe}, {Gronenschild}, \& {van den
  Oord}}]{ME85.1}
{Mewe}, R., {Gronenschild}, E.~H.~B.~M., \& {van den Oord}, G.~H.~J. 1985,
  \aaps, 62, 197

\bibitem[{{Mewe} {et~al.}(1986){Mewe}, {Lemen}, \& {van den Oord}}]{ME86.1}
{Mewe}, R., {Lemen}, J.~R., \& {van den Oord}, G.~H.~J. 1986, \aaps, 65, 511

\bibitem[{{Nagai} {et~al.}(2007){Nagai}, {Vikhlinin}, \&
  {Kravtsov}}]{2007ApJ...655...98N}
{Nagai}, D., {Vikhlinin}, A., \& {Kravtsov}, A.~V. 2007, \apj, 655, 98

\bibitem[{Navarro {et~al.}(1997)Navarro, Frenk, \& White}]{NA97.1}
Navarro, J., Frenk, C., \& White, S. 1997, ApJ, 490, 493

\bibitem[{{Oguri} \& {Blandford}(2009)}]{2009MNRAS.392..930O}
{Oguri}, M. \& {Blandford}, R.~D. 2009, \mnras, 392, 930

\bibitem[{{Oguri} {et~al.}(2009){Oguri}, {Hennawi}, {Gladders}, {Dahle},
  {Natarajan}, {Dalal}, {Koester}, {Sharon}, \&
  {Bayliss}}]{2009ApJ...699.1038O}
{Oguri}, M., {Hennawi}, J.~F., {Gladders}, M.~D., {et~al.} 2009, \apj, 699,
  1038

\bibitem[{Oguri {et~al.}(2003)Oguri, Lee, \& Suto}]{OG03.1}
Oguri, M., Lee, J., \& Suto, Y. 2003, ApJ, 599, 7

\bibitem[{{Oguri} {et~al.}(2005){Oguri}, {Takada}, {Umetsu}, \&
  {Broadhurst}}]{2005ApJ...632..841O}
{Oguri}, M., {Takada}, M., {Umetsu}, K., \& {Broadhurst}, T. 2005, \apj, 632,
  841

\bibitem[{Puchwein {et~al.}(2005)Puchwein, Bartelmann, Dolag, \&
  Meneghetti}]{PU05.1}
Puchwein, E., Bartelmann, M., Dolag, K., \& Meneghetti, M. 2005, A\&A in press;
  preprint astro-ph/0504206

\bibitem[{{Puchwein} \& {Hilbert}(2009)}]{2009arXiv0904.0253P}
{Puchwein}, E. \& {Hilbert}, S. 2009, ArXiv e-prints

\bibitem[{{Rasia} {et~al.}(2006){Rasia}, {Ettori}, {Moscardini}, {Mazzotta},
  {Borgani}, {Dolag}, {Tormen}, {Cheng}, \& {Diaferio}}]{RA06.1}
{Rasia}, E., {Ettori}, S., {Moscardini}, L., {et~al.} 2006, \mnras, 369, 2013

\bibitem[{{Rasia} {et~al.}(2008){Rasia}, {Mazzotta}, {Bourdin}, {Borgani},
  {Tornatore}, {Ettori}, {Dolag}, \& {Moscardini}}]{2008ApJ...674..728R}
{Rasia}, E., {Mazzotta}, P., {Bourdin}, H., {et~al.} 2008, \apj, 674, 728

\bibitem[{Rasia {et~al.}(2004)Rasia, Tormen, \& Moscardini}]{RA04.1}
Rasia, E., Tormen, G., \& Moscardini, L. 2004, MNRAS, 351, 237

\bibitem[{{Refregier} {et~al.}(2008){Refregier}, {Douspis}, \& {the DUNE
  collaboration}}]{2008arXiv0807.4036R}
{Refregier}, A., {Douspis}, M., \& {the DUNE collaboration}. 2008, ArXiv
  e-prints

\bibitem[{{Rowley} {et~al.}(2004){Rowley}, {Thomas}, \&
  {Kay}}]{2004MNRAS.352..508R}
{Rowley}, D.~R., {Thomas}, P.~A., \& {Kay}, S.~T. 2004, \mnras, 352, 508

\bibitem[{{Sand} {et~al.}(2005){Sand}, {Treu}, {Ellis}, \&
  {Smith}}]{2005ApJ...627...32S}
{Sand}, D.~J., {Treu}, T., {Ellis}, R.~S., \& {Smith}, G.~P. 2005, \apj, 627,
  32

\bibitem[{{Sand} {et~al.}(2004){Sand}, {Treu}, {Smith}, \&
  {Ellis}}]{2004ApJ...604...88S}
{Sand}, D.~J., {Treu}, T., {Smith}, G.~P., \& {Ellis}, R.~S. 2004, \apj, 604,
  88

\bibitem[{Schindler(1999)}]{SC99.2}
Schindler, S. 1999, A\&A, 349, 435

\bibitem[{{Sereno} {et~al.}(2010){Sereno}, {Jetzer}, \&
  {Lubini}}]{2010arXiv1001.1696S}
{Sereno}, M., {Jetzer}, P., \& {Lubini}, M. 2010, ArXiv e-prints

\bibitem[{{Shaw} {et~al.}(2006){Shaw}, {Weller}, {Ostriker}, \&
  {Bode}}]{SH06.1}
{Shaw}, L.~D., {Weller}, J., {Ostriker}, J.~P., \& {Bode}, P. 2006, \apj, 646,
  815

\bibitem[{Sheth \& Tormen(2002)}]{SH02.1}
Sheth, R. \& Tormen, G. 2002, MNRAS, 329, 61

\bibitem[{{Short} {et~al.}(2010){Short}, {Thomas}, {Young}, {Pearce},
  {Jenkins}, \& {Muanwong}}]{2010arXiv1002.4539S}
{Short}, C.~J., {Thomas}, P.~A., {Young}, O.~E., {et~al.} 2010, ArXiv e-prints

\bibitem[{{Soucail} {et~al.}(2004){Soucail}, {Kneib}, \&
  {Golse}}]{2004A&A...417L..33S}
{Soucail}, G., {Kneib}, J.-P., \& {Golse}, G. 2004, \aap, 417, L33

\bibitem[{{Spergel} {et~al.}(2007){Spergel}, {Bean}, {Dor{\'e}}, {Nolta},
  {Bennett}, {Dunkley}, {Hinshaw}, {Jarosik}, {Komatsu}, {Page}, {Peiris},
  {Verde}, {Halpern}, {Hill}, {Kogut}, {Limon}, {Meyer}, {Odegard}, {Tucker},
  {Weiland}, {Wollack}, \& {Wright}}]{wmap3}
{Spergel}, D.~N., {Bean}, R., {Dor{\'e}}, O., {et~al.} 2007, \apjs, 170, 377

\bibitem[{{Springel}(2005)}]{SP05.1}
{Springel}, V. 2005, \mnras, 364, 1105

\bibitem[{Torri {et~al.}(2004)Torri, Meneghetti, Bartelmann, Moscardini, Rasia,
  \& Tormen}]{TO04.1}
Torri, E., Meneghetti, M., Bartelmann, M., {et~al.} 2004, MNRAS, 349, 476

\bibitem[{Wambsganss {et~al.}(1998)Wambsganss, Cen, \& Ostriker}]{WA98.2}
Wambsganss, J., Cen, R., \& Ostriker, J. 1998, ApJ, 494, 29

\bibitem[{{Wambsganss} {et~al.}(2008){Wambsganss}, {Ostriker}, \&
  {Bode}}]{2008ApJ...676..753W}
{Wambsganss}, J., {Ostriker}, J.~P., \& {Bode}, P. 2008, \apj, 676, 753

\bibitem[{{Wittman} {et~al.}(2006){Wittman}, {Jain}, {Jarvis}, {Knox},
  {Margoniner}, {Takada}, {Tyson}, {Zhan}, \& {LSST Weak Lensing Science
  Collaboration}}]{2006AAS...209.8610W}
{Wittman}, D.~M., {Jain}, B., {Jarvis}, M., {et~al.} 2006, in Bulletin of the
  American Astronomical Society, Vol.~38, Bulletin of the American Astronomical
  Society, 1019--+

\bibitem[{{Yepes} {et~al.}(2007){Yepes}, {Sevilla}, {Gottl{\"o}ber}, \&
  {Silk}}]{YSGS07}
{Yepes}, G., {Sevilla}, R., {Gottl{\"o}ber}, S., \& {Silk}, J. 2007, \apjl,
  666, L61

\bibitem[{Zaritsky \& Gonzalez(2003)}]{ZA03.1}
Zaritsky, D. \& Gonzalez, A. 2003, ApJ, 584, 691

\end{thebibliography}
\bibliographystyle{aa}

\appendix
\section{The most efficient lens in the {\sc MareNostrum Universe}}
\label{sect:app1}
\begin{figure*}[t!]
\begin{center}
  \includegraphics[width=1.0\hsize]{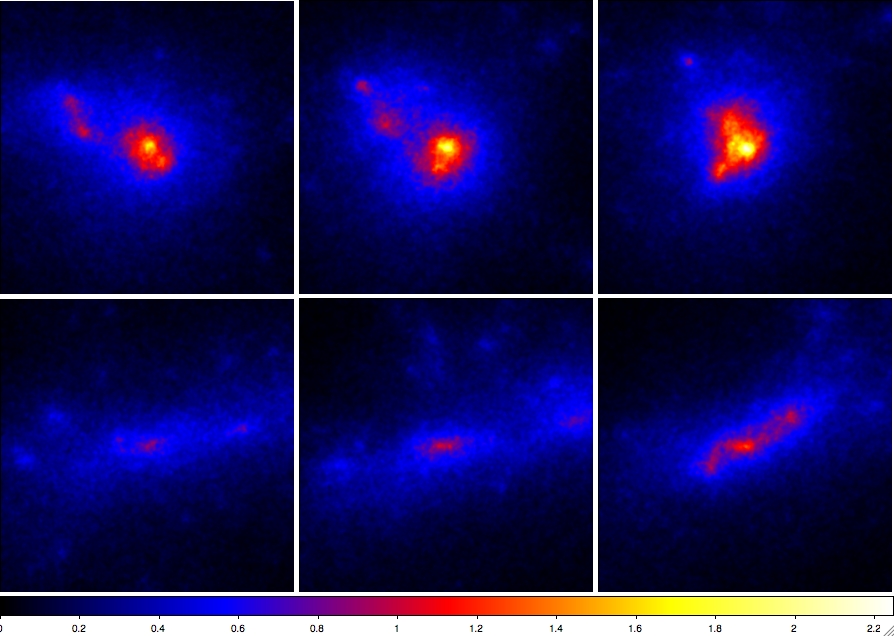}
\end{center}
\caption{The most efficient lens in the {\sc MareNostrum Universe}. In the top row of panels we show the convergence maps of this cluster at three different epochs, namely at redshifts $z_1=0.42$, $z_2=0.45$, and $z_3=0.48$ starting from the left. The highest efficiency for lensing, which corresponds to a cross section for giant arcs of $\sigma=1.92\times10^{-2}h^{-2}$Mpc$^{2}$, is reached at $z_3$. At this epoch, the lensing cross section is approximately two times larger than at $z_1$. The bottom row of panels shows the same cluster sequence but along a line of sight perpendicular to that of  the upper panels. The side-length of the panels on the left is $633$ arcsec, while the remaining panels have sizes of $423$ arcsec.}
\label{fig:bigmerg}
\end{figure*}

We have ranked the clusters in our sample by their lensing cross section, creating a list of the 10 most efficient lenses between $z=0$ and $z=2$. It turns out that all of them are clusters which are elongated along the line of sight and which have significant substructures projected near the center. In the upper panels of Fig.~\ref{fig:bigmerg} we show an example, given by the most efficient lens we have identified in the simulations. From left to right, shown are the convergence maps at three different epochs, namely $z_1=0.42$, $z_2=0.45$, and $z_3=0.48$.  The bottom panels show the cluster at the same epochs as in the upper panels, but along a different line of sight, which is perpendicular to the previous one. The scale of the figures in the left panels is $633$ arcsec. In the remaining panels it is $423$ arcsec. We can see that:
\begin{enumerate}
	\item in the upper panels the cluster appears rounder and denser than in the bottom panels. Clearly, the cluster is elongated along the line of sight in the upper plots. In these projections, the lensing cross section is larger by more than an order of magnitude compared to the projections shown in the bottom panels;
	\item the concentrations measured by fitting the surface density profiles of the cluster in the upper panels is significantly larger than that obtained by fitting the 3D-density profiles. For example, at $z_3$ the 3D-concentration is $4.11$, while that inferred from the projected mass distribution is $8.29$; 
	\item going back in time, few substructures approach the cluster center, and, at $z_3$, they seem to cross the very inner region of the cluster, boosting the lensing cross section significantly. At this epoch the lensing cross section is $\sigma=1.92\times10^{-2}h^{-2}$Mpc$^{2}$. Between $z_3$ and $z_1$,  it drops by almost a factor of two. At the epoch of maximal lensing efficiency the cluster is characterized by rather extreme values of the $\beta$ and of the $\Gamma_{500}$ parameters, which are equal to $-0.74$ and $0.36$, respectively;
	\item being the center of the cluster more clumpy, the measured ellipticity is larger in the central than in the external region. For example, at $z_3$ the ellipticities measured within $0.1\times R_{200}$ and $R_{200}$ are $0.23$ and $0.07$, respectively.    
\end{enumerate}
All this shows that the most powerful strong lenses in the universe are likely to be a very special class of objects, characterized by several peculiarities, which we need to properly take into account in order to statistically model them. 

{  \cite{2009MNRAS.392..930O} (OB hereafter) used semi-analytic methods based on triaxial NFW halos for calculating the probability distributions of several properties of the clusters producing the largest critical lines in the universe. The size of the critical lines is quantified by means of the Einstein radius (see Eq. 18 of OB). Their calculations include scatter in concentrations and axial ratios which are calibrated with numerical simulations. The length of the critical lines is strongly correlated with the lensing cross section (Meneghetti et al. in prep.), given that the latter is an area surrounding the caustics, which are mapped on the critical lines via the lens equation. Thus, we expect that the clusters with the most extended critical lines will have the largest cross sections.  OB  also considered a WMAP1 normalized cosmology and adopted several source redshifts for drawing their distributions. In particular, they used $z_s=1$ and $z_s=3$, which encompass the redshift used in our simulations ($z_s=2$). We attempt now a comparison with their results. 

We begin with  the distributions of the cluster orientations and triaxial shapes. They find that in $\sim 90\%$ of their Montecarlo realizations, the cluster with the largest Einstein radius has  
$\theta \lesssim 26$ degrees and the median of their distribution is between 11 and 14 degrees. The cluster exhibiting the largest cross section for $z_s=2$ in the {\sc MareNostrum Universe} has its major axis forming an angle of 23 degrees with the line of sight. This cluster has a minor to major axis ratio of 0.21 which is also well in agreement with the distributions found by OB, whose medians for $z_s=1$ and $z_s=3$ are $0.32^{+0.1}_{-0.1}$ and $0.23^{+0.11}_{-0.07}$, respectively.

Considering a WMAP1 normalized cosmology, OB report that the typical redshift for the lens with the largest Einstein radius is $z=0.28^{+0.11}_{-0.09}$ for $z_s=1$ and $z=0.47^{+0.25}_{-0.17}$ for $z_s=3$. This is also compatible with the redshift of the cluster discussed above ($z=0.48$). The distribution of the cluster  masses found by OB has a median $M_{\rm vir}=2.49^{+0.95}_{-0.92}\times 10^{15}h^{-1}M_\odot$ for $z_s=1$ and $M_{\rm vir}=1.98^{+1.30}_{-1.08}\times 10^{15}h^{-1}M_\odot$ for $z_s=3$. The strongest lens in the {\sc MareNostrum Universe} has a mass of $1.85\times 10^{15}h^{-1}M_\odot$, which agrees with the results of OB at $1\sigma$ level.

As discussed above, the projected ellipticity of the cluster shown in the upper right panel of Fig.~\ref{fig:bigmerg} varies between 0.23 in the center and 0.07 in the external region. OB find that the cluster with the largest Einstein radius has a typical projected ellipticity $<0.3$. Given that they use simple triaxial mass distributions to model their clusters, the projected ellipticity does not change with radius as we find by analyzing numerically simulated clusters. Nevertheless, our and OB's results seem consistent once again. Similarly, there is a very good agreement also between the 2D-concentrations: while the most efficient lens discussed in this Section has a projected concentration of $8.29$, OB find that a distribution with median between $8.91^{+2.64}_{-2.11}$ and $7.03^{+2.27}_{-1.93}$ is expected in a WMAP1 cosmology for sources between $z_s=1$ and $z_s=3$.

In conclusions, the most efficient lens in the {\sc MareNostrum Universe} has properties which match the expectations for the cluster with the largest Einstein radius in the Universe, as derived by OB modeling galaxy clusters with triaxial halos, whose structural properties are calibrated using numerical simulations different from the  {\sc MareNostrum Universe}. This can be considered a valid ``sanity check" for our results. 
}

\end{document}